%% file: REVISION_CLEAN_IS_Context_Dependent.tex
\documentclass[aap]{imsart}
\RequirePackage[colorlinks,citecolor=blue,urlcolor=blue]{hyperref}

\usepackage[utf8]{inputenc}
\usepackage[english]{babel}
\usepackage{amsmath,amssymb,amsthm,verbatim,mathtools,bbm,mathrsfs,float,subcaption}
\usepackage[numbers,sort&compress]{natbib}
\usepackage{algorithm,algpseudocode}
\usepackage{enumitem}
\usepackage{url}
\usepackage{xfrac}

\usepackage{tikz} 
\usetikzlibrary{automata}
\usetikzlibrary{calc,decorations.pathreplacing, arrows.meta, positioning}

\usepackage{float}
\usepackage[md]{titlesec}
\usepackage{ulem}[normalem]
\usepackage{soul,cancel}

\newcommand{\me}{\mathrm{e}}
\newcommand{\rank}[1]{\operatorname{rank}(#1)}

\DeclareMathOperator{\scrA}{\mathscr{A}}

\DeclareMathOperator{\E}{\mathbb{E}}

\DeclareMathOperator{\Prob}{\mathbb{P}}
\DeclareMathOperator{\tP}{\text{P}}
\DeclareMathOperator{\Pa}{\mathcal{P}}
\DeclareMathOperator{\scrP}{\mathscr{P}}

\DeclareMathOperator{\Sca}{\calS}

\DeclareMathOperator{\KL}{\text{D}_{\text{KL}}}

\DeclareMathOperator{\tT}{\text{T}}

\DeclareMathOperator{\SymS}{\text{Sym}(\calS)}

\DeclareMathOperator{\tC}{\text{C}}
\DeclareMathOperator{\tG}{\text{G}}
\DeclareMathOperator{\tA}{\text{A}}

\newcommand{\iidsim}{\stackrel{\mbox{\tiny{iid}}}{\sim}}
\newcommand{\ind}{\mathbbm{1}}
\newcommand{\defeq}{\coloneqq}

\newtheorem{theorem}{Theorem}
\newtheorem{proposition}{Proposition}
\newtheorem{lemma}{Lemma}
\newtheorem{definition}{Definition}
\newtheorem*{definition*}{Definition}
\newtheorem*{notation*}{Notation}

\newtheorem{remark}{Remark}
\theoremstyle{plain}

\newcommand{\bb}{\mathbf{b}}
\newcommand{\s}{\mathbf{s}}
\newcommand{\bt}{\mathbf{t}}
\newcommand{\x}{\mathbf{x}}
\newcommand{\tx}{\tilde{x}}
\newcommand{\tp}{\tilde{p}}
\newcommand{\tpsi}{\tilde{\psi}}
\newcommand{\y}{\mathbf{y}}

\newcommand{\C}{\mathcal{C}}

\newcommand{\I}{\mathcal{I}}

\newcommand{\calS}{\mathcal{S}}

\newcommand{\bigO}{\mathcal{O}}

\newcommand{\R}{\mathbf{R}}

\newcommand{\Q}{\mathbf{Q}}
\newcommand{\tQ}{\tilde{\mathbf{Q}}}
\newcommand{\tPhi}{\tilde{\Phi}}
\newcommand{\tg}{\tilde{\gamma}}
\newcommand{\gmin}{\gamma_{\min}}
\newcommand{\gmax}{\gamma_{\max}}

\newcommand{\tgmin}{\tg_{\min}}
\newcommand{\tgmax}{\tg_{\max}}

\makeatletter
\newtheorem{repeatthm@}{Lemma}

\makeatother

\usepackage{thm-restate}

\makeatletter
\newcommand{\mytag}[2]{%
  \text{#1}%
  \@bsphack
  \begingroup
    \@onelevel@sanitize\@currentlabelname
    \edef\@currentlabelname{%
      \expandafter\strip@period\@currentlabelname\relax.\relax\@@@%
    }%
    \protected@write\@auxout{}{%
      \string\newlabel{#2}{%
        {#1}%
        {\thepage}%
        {\@currentlabelname}%
        {\@currentHref}{}%
      }%
    }%
  \endgroup
  \@esphack
}
\makeatother

\begin{document}

\begin{frontmatter}

\title{Importance Sampling Approximation of Sequence Evolution Models with Site-Dependence}
%
\runtitle{Importance Sampling for Site-Dependent Evolution}

\begin{aug}
\author[A]{\fnms{Joseph}~\snm{Mathews}\ead[label=e1]{joseph.mathews@duke.edu}}
\and
\author[B]{\fnms{Scott C.}~\snm{Schmidler}\ead[label=e2]{scott.schmidler@duke.edu}\orcid{0000-0000-0000-0000}}
\address[A]{Department of Statistical Science, Duke University \printead[presep={,\ }]{e1}}

\address[B]{Department of Statistical Science,
Duke University \printead[presep={,\ }]{e2}}
\end{aug}

%





\begin{abstract}
We consider models for molecular sequence evolution in which the transition rates at each site depend on the local sequence context, giving rise to a time-inhomogeneous Markov process in which sites evolve under a complex dependency structure.  We introduce a randomized approximation algorithm for the marginal sequence likelihood under these models using importance sampling, and provide matching order upper and lower bounds on the finite sample approximation error.  Given two sequences of length $n$ with $r$ observed mutations, we show that for practical regimes of $r/n$, the complexity of the importance sampler does not grow exponentially in $n$, but rather in $r$, making the algorithm practical for many applied problems. We demonstrate the use of our techniques to obtain problem-specific complexity bounds for a well-known dependent-site model from the phylogenetics literature. 
%
%
\end{abstract}



\end{frontmatter}

\section{Introduction}
\label{sec:Intro}

\subsection{Problem:} Consider a sequence of random variables $\x = (x_1,x_2,\ldots,x_n)$ each taking values in some finite alphabet $\scrA \ni x_i$ of size $a := |\scrA|$, and evolving according to a continuous-time Markov process, thus giving  rise to a stochastic process $\x_t = (x_1(t),\ldots,x_n(t))$ taking values in $\scrA^n$ for every $t \in [0,\infty)$.  For example, $\x$ may be a DNA sequence with $\scrA = \{\tA,\tG,\tC,\tT\}$.  We consider models in which the time evolution of $\x_t$ is given by a time-\textit{inhomogeneous} CTMC with $a^n\times a^n$ rate matrix $\tQ$ specified by a \textit{sequence context-dependent} rate function
\begin{align}
\label{eqn:CD Rates}
\tg_i(b; \tx_i) = \gamma_i(b; x_i)\phi(b; \tx_i) \quad \text{ for } \; b \in \scrA \setminus x_i,
\end{align}
where  $\gamma_i: \scrA^2 \rightarrow (0,\infty)$ is the context-independent rate
at which $x_i \in \scrA$ mutates to $b$,
$\phi: \scrA^{k+1} \rightarrow (0,\infty)$ is a context-dependent multiplier, and
\begin{align*}
 \tx_i = (x_{i_1},\ldots,x_{i_{k/2}},x_i,x_{i_{k/2 + 1}},\ldots,x_{i_k})
\end{align*}
denotes the sequence \textit{context} of site $x_i$.  When $k = 0$ we have a standard sequence evolution model, with the sites evolving independently, each according to a CTMC with $a\times a$ rate matrix $\Q$ specified by $\gamma_i$ (see e.g. \citet{Yang:2014}). Examples of $\Q$ include the well-known Jukes-Cantor (JC69) \cite{Jukes:1969} and generalized time reversible (GTR) \cite{Tavare:1986} models.

Let $\y = (y_1,y_2,\ldots,y_n) = \x_T$ denote an observation of the process at time $T$. A fundamental quantity in phylogenetics is the 
probability that $\x$ transitions to $\y$ in time $T$:
\begin{align}
p_{(T,\tQ)}(\y \mid \x) := 
\text{Pr}(\x_{T} = \y \mid \x_{0} = \x) = (e^{T\tQ})_{\x,\y}.
\label{eqn:DSMmarglik}
\end{align}
Calculation of $p_{(T,\tQ)}(\y \mid \x)$ arises in evaluating the likelihood function for a wide variety of evolutionary inference problems: the reconstruction of phylogenetic tree topologies \cite{Felenstein:1985,MrBayes:2012,Felenstein:1973}; the estimation of divergence times (branch lengths) \cite{Sanderson:1997,Thorne:1998,Kishino:2001}, mutation model parameters \cite{Rodriguez:1990,Yang:1994}, and selection coefficients \cite{Halpern:1998,Yang:2008}; and the reconstruction of ancestral sequences \cite{Pagel:2004,Yang:1995}, to name just a few. Most models and software packages in common use invoke the \textit{independent site model} (ISM) assumption for  
computational tractability; namely, 
$k=0$ and the processes $x_i(t)$ and $x_j(t)$ are assumed to evolve independently for $i\neq j$.  This enables straightforward calculation of $p_{(T,\tQ)}(\y \mid \x)$ since under an ISM it factors site-by-site:
\begin{align}
\label{eqn:typical MargLik}
p_{(T,\tQ)}(\y \mid \x) = \prod^n_{i=1} p_{(T,\Q)}(y_i \mid x_i) =  \prod^n_{i=1} (e^{T\Q})_{x_i,y_i}.
\end{align}
%
However, independent-site models fail to capture certain important features of biological evolution known to give rise to dependence among sites; examples include CpG di-nucleotide mutability \cite{Pederson:1998}, enzyme-driven somatic hypermutation in B-cell affinity maturation \cite{Wiehe:2018,Mathews:2023b}, and structural constraints in RNA and proteins \cite{Robinson:2003}. In the latter case, dependencies arise between sites that are not necessarily contiguous in the DNA sequence, allowing for long-range dependence. 
A variety of \textit{dependent} site models (DSMs) have been proposed to relax this independence assumption \cite{Robinson:2003,Siepel:2004,Jensen:2000,Jensen:2001,Larson:2020,Hwang:2004,VonHaeseler:1998,Christensen:2005,Arndt:2005,Lunter:2004}
but exact computation of $p_{(T,\tQ)}(\y \mid \x)$ under models with site dependence is generally intractable since the corresponding likelihood no longer factors. 

Markov chain Monte Carlo (MCMC) algorithms \cite{Jensen:2000,Robinson:2003, Hwang:2004,Rodrigue:2005,Rodrigue:2006,Hobolth:2009,HobolthThorne:2014,Li:2024} address this by sampling unobserved sequence evolution paths from $\x$ to $\y$, and $p_{(T,\tQ)}(\y \mid \x)$ can be obtained by marginalization over all such paths (see equation \eqref{eqn:Transition Probability} in Section~\ref{sec: background}). 
However, MCMC algorithms suffer from several undesirable properties that limit their practical utility in many applications of DSMs (e.g. phylogenetic tree reconstruction), which often require repeated calculation of marginal likelihoods during iterative sampling (Bayesian MCMC) or optimization (MLE). These undesirable properties include their iterative nature and the necessity of assessing convergence empirically \cite{Gelman:1992,Cowles:1996} along with the challenges of doing so in high dimensional spaces \cite{Brooks:1998,Vandewerken:2017}, the lack of simple parallelization  \cite{Vandewerken:2013}, and the difficulty of obtaining theoretical guarantees on the accuracy of the resulting estimators, the latter requiring rare and difficult-to-obtain quantitative mixing-time bounds \cite{Rosenthal:1995,Jones:2001}. Moreover, estimation of marginal likelihoods directly from MCMC output is notoriously difficult \cite{Wolpert:2012,Fourment:2020,Chib:1995}. In situations requiring repeated evaluation within an optimization or sampling loop, we instead desire  a highly computationally efficient estimator, ideally with bounded approximation error.

Here we propose an approach to approximating $p_{(T,\tQ)}(\y \mid \x)$ under an arbitrary DSM using
importance sampling, with instrumental distribution based on an ISM. A practical benefit of this approach is that it allows practitioners to leverage the vast software resources developed by the phylogenetics community over several decades \cite{Li:2025}. Because calculation of the transition probability is straightforward  under an ISM, and endpoint-conditioned path sampling under an ISM is feasible (see Section~\ref{sec: Methods IS}), we readily obtain an importance sampling estimator using an ISM as an instrumental distribution.

Our main result is a bound on the $\chi^2$-divergence between the corresponding independent and dependent site models, which in turn provides an upper bound on the number of samples needed to approximate $p_{(T,\tQ)}(\y \mid \x)$ under the DSM with $\epsilon$-relative error. A common shortcoming of importance sampling is that the sample complexity typically grows exponentially in the problem dimension. Here, we show that  the required sample size depends on the sequence length $n$ only through the number $r$ of observed mutations, although it grows exponentially in $r$.  Nevertheless, we demonstrate that our approach yields  practical sample sizes on a popular DSM known as the CpG model \cite{Hobolth:2008}. We also provide a lower bound on the KL-divergence of the ISM and DSM models, providing conditions under which our upper bounds are near-optimal by appealing to results of \citet{Chaterjee:2018}.


The main technical difficulty in establishing our results is that the weights (density ratio) between the ISM and DSM models $\mu$ and $\pi$ are not uniformly bounded, because there is no limit on the  number of possible unobserved jumps along any endpoint-conditioned path from $\x$ to $\y$. As a result, error bounds  are not readily obtained using standard probability concentration tail bounds.
Instead we directly bound the $\chi^2$-divergence by obtaining a bound on the moment generating function (MGF) of the total number of such unobserved mutations.
We then take advantage of a connection between
the space of paths from $\x$ to $\y$ and the symmetric group $\SymS$ on $\mathcal{S} \defeq \{i: x_i \neq y_i \}$ corresponding to the \textit{orderings} in which the $r$ observed mutations occur, and establish an inequality relating the $L^2$ norms of the density ratios (importance weights) on these two spaces. The proof techniques and MGF bounds developed here also provide a foundation for establishing finite sample error bounds for sequential Monte Carlo (SMC) estimators of the marginal sequence likelihood \cite{Mathews:2025}.


The remainder of this paper is organized as follows: Section~\ref{sec: background} establishes notation and the problem setup; Section~\ref{sec:MainResults} presents our main results; Section~\ref{sec:IS Results} establishes an upper bound on the $\chi^2$-divergence and a lower bound on the KL-divergence between the ISM and DSM models, and provides an application to the CpG model; Section~\ref{sec:Conclusion} summarizes our results and discusses future directions. Technical results are deferred to the appendices: 
Appendix~\ref{Appdx:Importance Sampling Results} provides supporting details used to show the upper bound on the $\chi^2$-divergence between the ISM and DSM;  Appendix~\ref{Appdx:Importance Sampling Lower Bound} provides supporting details for establishing the lower bound on KL-divergence from the ISM to DSM.

\section{Background and Notation}
\label{sec: background}

\subsection{Notation}
Consider a sequence $\x_t = (x_1(t),\ldots,x_n(t))$ evolving according to the model \eqref{eqn:CD Rates}.  Let 
$\tQ$ be the $a^n \times a^n$ rate matrix defined by the context-dependent rates \eqref{eqn:CD Rates} of the Markov process operating on the space of all sequences:
\begin{align}
\label{eqn: tilde Q expanded definition}
\tQ_{\x,\x^{\prime}} = 
\begin{cases}
\tg_{i}(b; \tilde{x}_{i}) & \text{ for } \text{d}_{\text{H}}(\x,\x^{\prime}) = 1 \text{ and }x^{\prime}_{i} = b \neq x_{i}   \\
-\tg(\cdot; \x) & \text{ for } \text{d}_{\text{H}}(\x,\x^{\prime}) = 0 \\
0 & \text{ for } \text{d}_{\text{H}}(\x,\x^{\prime}) > 1,
\end{cases}  
\end{align}
where $\tg_i(\cdot; \tx_i) = \sum_{b \neq x_i} \tg_i(b; \tx_i)$ 
denotes the rate at which site $i$ exits state $x_i$, and $\tg(\cdot ; \x) =  \sum^n_{i=1} \tg_i(\cdot; \tx_i)$ the total rate at which sequence $\x$ mutates, and $\text{d}_{\text{H}}(\x,\x')$ is the Hamming distance between two sequences $\x,\x' \in \mathscr{A}^n$. We adopt the standard assumption that multiple substitutions cannot occur simultaneously, a natural one for most sequence evolution models.

We are interested in the calculation of probabilities of the form \eqref{eqn:DSMmarglik}, where
$(e^{T\tQ})_{\x,\y}$ denotes the element of the matrix $e^{T\tQ}$ corresponding to the sequences $\x$ and $\y$. However, direct computation of $e^{T\tQ}$ is intractable as $\rank{\tQ}$ grows exponentially in $n$. Alternatively, we can write $p_{(T,\tQ)}(\y\mid\x)$ as a marginalization over latent \textit{paths} that start in $\x$ and end in $\y$ at time $T$. Specifically, let
%
\begin{align*}
\Pa =   (m,t^1,\ldots,t^m,s^1,\ldots,s^m,b^1,\ldots,b^m )
\end{align*}
denote a path of \textit{length} $m$,
where $m \in \{0,1,\ldots \}$ is the number of mutations occurring along the path, $t^1,\ldots,t^m \in \mathbb{R}_+$ are the times of the mutation events satisfying $ t_0 = 0 < t^1 < \ldots < t^m < T$, $s^1,\ldots,s^m \in \{1,\ldots,n\}$ the sites at which the mutations occur, and $b^1,\ldots,b^m \in \scrA$ are the values of the base changes. At times, we will make the length of the path explicit by writing $\Pa^l$ and letting $\scrP^l$ denote the set of all length $l$ paths. Let 
\begin{alignat*}{2}   
\x^j    &:=&\; \x(t^j)     &= \x\bigl(j; s^1,\ldots,s^m,b^1,\ldots,b^m,m\bigr)\\
\tx^j_i &:=&\; \tx_i(t^j)  &= \tx_i\bigl(j; s^1,\ldots,s^m,b^1,\ldots,b^m,m\bigr)
\end{alignat*}
%
denote the sequence and context at site $i$, respectively, following the $j^{\text{th}}$ jump along a given path, i.e. in the interval $t\in[t_j,t_{j+1})$, and let $\Delta^t(j) := t^{j+1} - t^j$ with $\Delta^t(m) := T - t^m$ be the inter-arrival times between jumps. Let $\scrP = \cup^{\infty}_{l=0} \scrP^l$ denote the set of all such paths. Let $\nu^l := \nu^l_t \otimes \nu_s^l \otimes \nu^l_b$, where $\nu^l_t$ denotes the Lebesgue measure on $[0,T]^l$, and $\nu_s^l$ and $\nu^l_b$ the counting measures on $\{1,\ldots,n\}^l$ and $\scrA^l$, respectively, and define the measure $\nu(d\Pa) := \sum^{\infty}_{l=0} \ind_l(dm) \nu^l(d\bt^l, d\s^l, d\bb^l)$. Then we can write 
\begin{align}
\label{eqn:Transition Probability}
p_{(T,\tQ)}(\y \mid  \x)  = \sum^{\infty}_{l=0}\int_{\scrP^l} \tP_{(T,\tQ)}(\y, \Pa^{l} \mid \x) \nu^l(d\Pa^l) = \int_{\scrP} \tP_{(T,\tQ)}(\y, \Pa \mid \x) \nu(d\Pa) ,
\end{align}
%
where the conditional joint density of a path from $\x$ ending in $\y$ is given by 
\begin{align}
\label{eqn:Path Density}
\tP_{(T,\tQ)}(\y,\Pa \mid \x) := \left[\prod^{m(\Pa)}_{j=1} \tg_{s^j}(b^j; \tx^{j-1}_{s^j})
\me^{-\Delta^t(j-1) \tg(\cdot; \x^{j-1})}
 \right] \me^{-\Delta^t(m)\tg(\cdot; \y) } \ind_{\x^m = \y}(\Pa),
\end{align}
if the times satisfy the ordering constraint $0<t^1<\ldots<t^m<T$, and zero otherwise. We let $r = \text{d}_{\text{H}}(\x,\y)$ denote the Hamming distance between $\x$ and $\y$, and $\calS = \{i : y_i \neq x_i \}$ denote the set of \textit{observed} mutated sites. Note that \eqref{eqn:Path Density} is zero unless the sites $ \{s^1,\ldots,s^m\} \supseteq \mathcal{S}$, 
so each endpoint conditioned path $\Pa$ contains $r$ \textit{required} jumps and $m(\Pa) - r$ \textit{extra} jumps.

It follows from \eqref{eqn:Transition Probability} that $p_{(T,\tQ)}(\y \mid \x)$ can be approximated by Monte Carlo integration by sampling from the \textit{endpoint-conditioned} distribution
%
\begin{align*}
\pi(\Pa \mid \x,\y) :=  \frac{\tP_{(T,\tQ)}(\y, \Pa \mid \x)}{p_{(T,\tQ)}(\y \mid \x)},
\end{align*}
where we have suppressed the dependence of $\pi$ on $T$ for brevity; hereafter $T$ will be assumed fixed.
%
(At times we will abuse notation by denoting both a probability measure and its density with respect to $\nu$ by the same symbol; it will also often be convenient to leave conditioning on $\x$ and $\y$ implicit, writing $\pi(\Pa)$ and $\mu(\Pa)$ in place of $\pi(\Pa \mid \x,\y)$ and $\mu(\Pa \mid \x,\y)$.)
However, generating samples from the joint distribution $\pi(\Pa \mid \x, \y)$ is not straightforward. For a single site,
paths can be sampled efficiently by specialized algorithms \cite{Hobolth:2009} taking advantage of rate matrix calculations. Hence under an \textit{independent} site model (ISM) paths can be sampled on a site-by-site basis. However, under site dependence, constructing the rate matrix of the joint process (of size $a^n$) is intractable for even moderate $n$.

\subsection{Importance Sampling}\label{sec: Methods IS}
We define the endpoint-conditioned distribution under the ISM as follows. Let $\Q_i= (\gamma_i(y; x))$ for $x,y\in \scrA$ be an $a \times a$ rate matrix corresponding to the CTMC at site $i$ (see \eqref{eqn:CD Rates}, with $\phi \equiv 1$). Consider the endpoint-conditioned distribution
\begin{align}
\label{Eqn:ISM}
\mu(\Pa \mid \x, \y) \propto \tP_{(T,\Q)}(\y, \Pa \mid \x),
\end{align}
with rate matrix $\Q^{(n)} = \mathbf{I}_a \otimes \Q^{(n-1)} + \Q_n \otimes \mathbf{I}_{a^{n-1}}$
where $\Q^{(1)} = \Q_1$ and 
$\mathbf{I}_a$ is the $a$-dimensional identity matrix. The density $\tP_{(T,\Q)}(\y, \Pa \mid \x)$ is given by \eqref{eqn:Path Density} but with $\phi \equiv 1$. In this case the joint density \eqref{eqn:Path Density} can be factored by site. Let 
\begin{align}\label{eqn: ith path def}
\Pa_i = (m_i,t^1_i,\ldots,t^{m_i}_i,b^1_i,\ldots,b^{m_i}_i)
\end{align}
denote the path at site $i$ defined by $\Pa$. That is, 
$(t^1_i,\ldots,t^{m_i}_i,b^1_i,\ldots,b^{m_i}_i) = \{(t^j,b^j) \in \Pa: s^j = i \}$ and $m_i = \sum^m_{j=1} \ind(s^j = i)$ is the number of jumps at site $i$. We let $\scrP^l_i$ denote the set of all length $l$ paths, and $\scrP_i = \cup^{\infty}_{l=0} \scrP^l_i$ be the set of all paths, at the $i$th site. Define $\Delta_i^t(j) := t_i^{j+1} - t_i^j$ with $\Delta_i^t(m_i) := T - t_i^{m_i}$. The joint density of a path at site $i$ that begins at $x_i$ and ends at $y_i$ is given by 
\begin{align}
\label{eqn:Path Density ISM}
 \tP_{(T,\Q_i)}(y_i,\Pa_i \mid x_i) = \left[\prod^{m_i}_{j=1} \gamma_i(b^j_i; b^{j-1}_i))
\me^{-\Delta_i^t(j-1) \gamma_i(\cdot; b^{j-1}_i)}
 \right] \me^{-\Delta_i^t(m_i)\gamma_i(\cdot; y_i)} \ind_{ \{b^{m_i}_i = y_i \} }(\Pa_i).
\end{align}
Computing the transition probability \eqref{eqn:Transition Probability} under the ISM is straightforward:
\begin{align}
\label{eqn: Norm Constant Mu}
p_{(T,\Q)}(\y \mid \x) := \int_{\scrP} \tP_{(T,\Q)}(\y , \Pa \mid \x) \nu(d\Pa) &= \prod^n_{i=1} 
 \int_{\scrP_i} \tP_{(T,\Q_i)}(y_i,\Pa_i \mid x_i) \nu_i(d\Pa_i) \\
&= \prod^n_{i=1} (e^{T\Q_i})_{x_i,y_i},
\end{align}
where $\nu_i(d\Pa_i) := \sum^{\infty}_{l=0} \ind_{dm_i}(l) \nu_i^l(d\bt^l_i, d\bb^l_i) $ for $\nu^l_i = \nu^l_t \otimes \nu^l_b$.

Similarly, letting $\mu_i(\Pa_i \mid x_i, y_i) \propto \tP_{(T,\Q_i)}(y_i,\Pa_i \mid x_i)$ denote the endpoint-conditioned measure for site $i$, we have $\mu =  \mu_1 \times \ldots \times \mu_n$ under the ISM.
As noted previously, sampling paths from the endpoint-conditioned measure $\mu$ under the ISM is also straightforward \cite{Hobolth:2009}. 
We will use samples from $\mu$ to define an importance sampling estimator of $p_{(T,\tQ)}(\y \mid \x)$ under the DSM \eqref{eqn:Transition Probability}.  Define the \textit{importance weight} of a path $\Pa$ by
\begin{align}
\label{eqn:Importance Weights}
w(\Pa) := \tP_{(T,\tQ)}(\y,\Pa \mid \x)/ \tP_{(T,\Q)}(\y,\Pa \mid \x),
\end{align}
and suppose $\Pa^{(1)},\ldots,\Pa^{(N)} \iidsim \mu$. Then an unbiased estimator of $p_{(T,\tQ)}(\y \mid \x)$ is given by
\begin{align}
\label{eqn:IS Estimator}
\hat{p}_{(T,\tQ)}(\y \mid \x) := \frac{p_{(T,\Q)}(\y \mid \x)}{N} \sum^N_{i=1} \frac{\tP_{(T,\tQ)}(\y, \Pa^{(i)} \mid \x)}{\tP_{(T,\Q)}(\y, \Pa^{(i)} \mid \x)} = \bar{w}(\Pa^{(1:N)}) p_{(T,\Q)}(\y \mid \x)
\end{align}
for
\begin{equation}
\bar{w}(\Pa^{(1:N)}) := \frac{1}{N} \sum^N_{i=1} \frac{\tP_{(T,\tQ)}(\y,\Pa^{(i)} \mid \x)}{\tP_{(T,\Q)}(\y,\Pa^{(i)} \mid \x)}.
\end{equation}
\noindent Our goal is to quantify the number of samples $N$  required to guarantee that the estimator \eqref{eqn:IS Estimator}  has relative error at most $\epsilon \in (0,1)$. To do so, we bound the $\chi^2$-divergence of $\mu$ from $\pi$, denoted 
$\chi^2(\pi \mid \mid \mu) = \text{Var}_\mu(\pi/\mu)  =L^2(\pi ,\mu) - 1$, where 
\begin{align*}
L^2(\pi, \mu) := \int_{\scrP}\left(\frac{\pi(\Pa \mid \x, \y)}{\mu(\Pa \mid \x, \y)}  \right)^2 \mu(d\Pa \mid \x, \y)
\end{align*}
%
is the squared $L^2(\mu)$ norm of $\pi/\mu$.
The variance of the estimator \eqref{eqn:IS Estimator} is given 
by 
$p_{(T,\Q)}^2(\y \mid \x)\text{Var}(\bar{w}) = N^{-1} p_{(T,\tQ)}(\y \mid \x)^2\chi^2(\pi \mid \mid \mu)$, and
by Chebychev's inequality
\begin{align}
\label{eqn:Concentration}
\Prob_{\mu}\left( \big| \hat{p}_{(T,\tQ)}(\y \mid \x) - p_{(T,\tQ)}(\y \mid \x) \big| > \epsilon \, p_{(T,\tQ)}(\y \mid \x) \right)  \leq  \frac{\chi^2(\pi \mid \mid \mu)}{N \epsilon^2} .
\end{align}
Therefore, $N = N(\epsilon) = \bigO(\epsilon^{-2}L^2(\pi, \mu))$
samples from $\mu$ suffice. Let
\begin{align}\label{eqn: N star definition}
N^{\star}(\epsilon) = \inf\{N: \Prob_{\mu}\big( | \hat{p}_{(T,\tQ)}(\y \mid \x) - p_{(T,\tQ)}(\y \mid \x) | > \epsilon \, p_{(T,\tQ)}(\y \mid \x)\big) \geq 3/4 \}
\end{align}
be the smallest sample size required to obtain an $\epsilon$-relative error estimate of $p_{(T,\tQ)}(\y \mid \x)$ with probability at least $3/4$, then we have immediately from \eqref{eqn:Concentration} that
\begin{align*}
    N^\star \leq \frac{4}{3}  \frac{L^2(\pi,\mu)}{\epsilon^2}.
\end{align*}
(The error tolerance probability $1/4$ can be reduced to $1 - \delta$ for $\delta \in (0, 1/4)$ by running the importance sampling algorithm $\mathcal{O}(\log(1/\delta))$ times and taking the median of the estimates; see e.g. \cite{Jerrum:1986}). Further improvements 
using sharper concentration bounds (e.g., Chernoff) seem not to be available, as the MGF of $w(\Pa)$ need not exist even for simple models; see Appendix~\ref{sec: Example MGF}.

Note that the $L^2$ distance
is related to the (squared) coefficient of variation of the importance weights under $\mu$:
\begin{align}
\label{eqn:ChiSq ISM}
L^2(\pi, \mu)  =  \frac{\int_{\mathscr{\Pa}} w^2(\Pa)   \mu(d\Pa \mid \x,\y) }{\left(\int_{\mathscr{\Pa}} w(\Pa)   \mu(d\Pa \mid \x, \y)  \right)^2} = \E_{\mu}[w^2(\Pa)]/ \left(\E_{\mu}[w(\Pa)] \right)^2.
\end{align}
%
In addition, by Jensen's inequality we have
\begin{align}
\label{eqn:Divergence Comparison}
\KL(\pi \mid \mid \mu) = \E_{\pi}[\log(\pi/\mu)] \leq \log\left(\E_{\pi}[\pi/\mu] \right) = \log(L^2(\pi,\mu)).
\end{align}
\citet{Chaterjee:2018} proved upper and lower bounds on the mean absolute error for general importance sampling estimators in terms of $\KL(\pi \mid \mid \mu)$,
showing that $N^{\star}$ necessarily grows exponentially in $\KL(\pi \mid \mid \mu)$. We apply this result by providing a lower bound on $\KL(\pi \mid \mid \mu)$ to obtain a lower bound on $N^*$ of matching order to our upper bound on the right-hand side of \eqref{eqn:Divergence Comparison}, showing that this bound is tight.

\section{Main Results}
\label{sec:MainResults}
We now state the main results of the paper; supporting results are established in the following sections. 
%

Our first main result concerns the sample complexity of the importance sampler.  It is obtained by establishing a bound on the $\chi^2$-divergence between the endpoint-conditioned distribution over paths under a DSM ($\pi$) and the same distribution under the corresponding ISM ($\mu$), which will grow as a function of the problem size (sequence length) $n$.  As we will see, the number of observed mutations $r =\text{d}_{\text{H}}(\x,\y)$ and the time interval $T$ play important roles; thus we must also consider the relative growth of $r(n)$ and $T(n)$ as $n$ increases. Luckily, there is a natural interval of interest for $T$ determined by $n$ and $r$, centered at $r/n$ \cite{Mathews2:2025}. Hence we will adopt the following assumption, the justification for which is provided below:
\begin{restatable}{assump}{rTAssumption}
\label{assump: T assumption}
The time interval $T(n) = \mathcal{O}(\frac{r(n)}{n})$ and the observed mutation count $r(n) \leq n^{\frac{1}{2}}$.
\end{restatable}
%
%
%
In what follows, we often write $r$ and $T$ instead of $r(n)$ and $T(n)$ for brevity, except where we wish to emphasize the dependence explicitly.

We can now state our first theorem:
\begin{theorem}
\label{thm:Main Thm Bound}
Let $\phi_{\star} = \phi_{\max} / \phi_{\min}$ denote the ratio of the largest and smallest context-dependent multipliers. For endpoint-conditioned independent-site process $\mu$ and dependent-site process $\pi$ defined in Section~\ref{sec: background}, there exists a constant $c \in (0, \infty)$ such that
\begin{align*} 
\chi^2(\pi \mid \mid \mu) \leq  \phi^{2r}_{\star} \exp\left(c(k,\phi_{\star}) \cdot (rT + (n-r) T^2)  \right) - 1,
\end{align*}
for $c(k,\phi_{\star}) = \bigO(\phi_{\star}^{4(k+1)})$ and consequently under Assumption~\ref{assump: T assumption}, the importance sampling estimator \eqref{eqn:IS Estimator} with
\begin{align*}
N^{\star}(\epsilon) =  \bigO(\phi^{2r}_{\star} e^{c^{\prime}(k,\phi_{\star})}\epsilon^{-2})
\end{align*}
for $\epsilon \in (0,1)$ and constant $c^{\prime}(k,\phi_{\star}) = \bigO(\phi_{\star}^{4(k+1)})$ provides a randomized approximation scheme for $p_{(T,\tQ)}(\y \mid \x)$.
\end{theorem}
The proof of Theorem~\ref{thm:Main Thm Bound} is given in  Section~\ref{sec: upper bound proof}. Theorem~\ref{thm:Main Thm Bound}  says that, for $T = \bigO(r/n)$, the required sample size (and hence runtime) of the importance sampler to approximate the marginal sequence likelihood is at most exponential in the number of \textit{observed} mutations $r$ when $r^2 \leq n$, bounding the complexity of the algorithm independent of the sequence length $n$. Theorem~\ref{thm:Main Thm Bound} thus establishes that the importance sampler is substantially more efficient than, for example, rejection sampling or matrix exponentiation when $r \ll n$ and $T = \bigO(r/n)$. 
Establishing this bound is non-trivial since the mutation count $m(\Pa)$ is unbounded, and a key ingredient of the proof is a bound on the moment generating function (MGF) of $m(\Pa)$ under $\mu$.
The assumption $r(n) = \bigO(n^\frac{1}{2})$ is reasonable in situations where $\x$ and $\y$ are close in evolutionary distance  (share a recent common ancestor, or correspond to highly conserved regions of the genome). This assumption ensures that the mutational process is far from  equilibrium so that $T$ can be reliably estimated \cite{Challis:2012,Mossel:2003,Mossel:2004} (for example, under the JC69 model \cite{Jukes:1969}, the MLE of $T$ does not exist if $r(n) > 3n/4$).  In many applications (e.g. genomic sequences) $n$ is very large, but the fraction of sites with observed mutations is low.

The efficiency of the importance sampler depends critically on the assumption that $T = \bigO(r/n)$, i.e. that $T$ not be too far from $r/n$. 
Because $T$ and mutation rates are not simultaneously identifiable, rate matrices are commonly scaled to one expected substitution per site per unit time, making $r/n$ -- a well known measure of genetic distance often called the \textit{p-distance} -- a natural estimate of $T$. However under DSMs, estimators of $T$ such as the maximum likelihood estimate (MLE) or posterior mean are not available in closed form and require iterative optimization or MCMC sampling, with the marginal likelihood evaluated at each iteration.  \citet{Mathews2:2025} show that the likelihood decays exponentially for values of $T$ far from $r/n$-- and the posterior distribution of $T$ concentrates close to $r/n$, under any 
reasonable prior distribution -- such that larger values of $T$ can be safely omitted from consideration in such algorithms without compromising their accuracy. 

Our next result, established in Section~\ref{sec: LB}, shows that the upper bound in Theorem~\ref{thm:Main Thm Bound} is tight by establishing a matching \textit{lower bound}. We first define the \textit{CpG model}, a simple DSM
that has been applied to account for low observed CG frequencies across codon boundaries in lentiviral genes \cite{Jensen:2000} and mammalian genomes \cite{Hwang:2004}. 
\begin{definition} The {\em CpG model} is defined by context-dependent rates
\begin{align}
\label{eqn:CpG Island Rates}
\tg_i(b; \tx_i) = \gamma_i(b; x_i) \lambda^{\ind_{\text{CG}}(x_{i - 1},x_i) + \ind_{\text{CG}}(x_i,x_{i + 1})},
\end{align}
where $\lambda \in (0,\infty)$ is a constant reflecting the relative bias against formation of CG pairs across codon boundaries \cite{Lunter:2004, Arndt:2005, Christensen:2005}.
\end{definition}
We can now state our second main result:
\begin{theorem}
\label{thm: main thm IS LB}
There exist $T^{\star} = \bigO(r/n)$ and sequences $\x^{\star}$ and $\y^{\star}$ satisfying Assumption~\ref{assump: T assumption}, such that under the CpG model with rates \eqref{eqn:CpG Island Rates}:
\begin{align*}
\text{D}_{\text{KL}}(\pi \mid \mid \mu) = \Theta(r).
\end{align*}
\end{theorem}
The proof of Theorem~\ref{thm: main thm IS LB} is given in Section~\ref{sec: LB}, with supporting details in Appendix~\ref{Appdx:Importance Sampling Lower Bound}.
\citet{Chaterjee:2018} provide conditions under which $N^{\star}(\epsilon)$ must grow exponentially in $\text{D}_{\text{KL}}(\pi \mid \mid \mu)$. Hence, their result together with Theorem~\ref{thm: main thm IS LB} indicates that exponential growth in $r$ is also \textit{necessary} whenever $\log(\pi(\Pa)/\mu(\Pa))$ concentrates around $\text{D}_{\text{KL}}(\pi \mid \mid \mu)$ under $\pi$. However, despite the exponential growth of the lower bound provided by Theorem~\ref{thm: main thm IS LB}, we show in Section~\ref{sec:Seq Bounds} that practical values of $N^{\star}(\epsilon)$ may still be obtained even in this worst case.


%
\section{Proofs of Main Results}
\label{sec:IS Results}
In this section we provide bounds for the importance sampler which give rise to Theorems~\ref{thm:Main Thm Bound} and \ref{thm: main thm IS LB}. We begin by  providing an overview of the proof strategy in Section~\ref{sec: approach}. The proofs of the upper and lower bounds are given in Sections~\ref{sec: upper bound proof} and \ref{sec: LB}, respectively. Finally, in Section~\ref{sec:Seq Bounds} we show how practical estimates can be obtained for a CpG model.

\subsection{Approach}
\label{sec: approach}
Theorems~\ref{thm:Main Thm Bound} and Theorem~\ref{thm: main thm IS LB} are proved by establishing a mapping
between $\scrP$ and $\SymS$, the symmetric group on $\calS$, the set of observed mutation sites. Specifically, let $\s(\Pa) = (s^1,\ldots,s^{m(\Pa)})$ be the set of sites mutated on the path $\Pa$ and $\bb(\Pa) = (b^1,\ldots,b^{m(\Pa)})$ the corresponding base changes, and for $\Pa \in \scrP$ define
\begin{align}
\label{eqn: Phi general}
\Phi(\s(\Pa), \bb(\Pa)) :=
\prod^{m(\Pa)}_{j=1}\phi_{s^j}(b^j; \tx^{j-1}_{s^j}).
\end{align}
Recall that at least $r = |\calS| = \text{d}_{\text{H}}(\x,\y)$ jumps must occur among any path satisfying the endpoint constraint $\x(T) = \y$. Hence, for any path $\Pa^r$ of length exactly $r$, we have 
$\s(\Pa^r) \in \SymS$ and
\begin{align}
\label{eqn: Phi length r}
\Phi(\s(\Pa^r), \bb(\Pa^r)) = \Phi(\s(\Pa^r))   =  \prod^r_{j=1}\phi_{s^j}(y_{s^j}; \tx^{j-1}_{s^j} ).
\end{align}
In other words, paths of length $r$ correspond to permutations of the order of the observed mutations, and $\Phi$ can be viewed as a function defined over $\text{Sym}(\calS)$ instead of $\scrP$. At times we will simply write $\Phi(\s^r)$ for $\s^r \in \SymS$. Normalizing \eqref{eqn: Phi length r} yields a probability measure on $\SymS$:
\begin{align}
\label{eqn: Symmetric Group Distribution}
\tPhi(\s^r) := \frac{\Phi(\s^r)}{\sum_{\tilde{\s}^r \in \text{Sym}(\calS)} \Phi(\tilde{\s}^r)} = \frac{\Phi(\s^r)}{Z^{r}_{\Phi}},
\end{align}
where $Z^r_{\Phi} \defeq \sum_{\tilde{\s}^r \in \text{Sym}(\calS)} \Phi(\tilde{\s}^r)$ is the normalizing constant of $\tPhi$.
%
Under the ISM ($\phi \equiv 1$) \eqref{eqn: Symmetric Group Distribution} is the uniform measure on $\SymS$, which we denote as $U_{\calS}$. We will show that
\begin{align}
\label{eqn:Approach upper bound}
L^2(\pi,\mu) \leq \exp\left(rT c + (n-r)T^2 c^{\prime} \right) L^2(\tPhi, U_{\Sca}),
\end{align}
for constants $c,c^{\prime} \in (0,\infty)$. Theorem~\ref{thm:Main Thm Bound} will follow as a special case of this result. The exponential term on the right-hand side of \eqref{eqn:Approach upper bound} can be interpreted as the increase in variability of the importance weights due to the $m(\Pa) - r$ \textit{extra} jumps. Indeed, establishing \eqref{eqn:Approach upper bound} is non-trivial precisely because the number of extra jumps is unbounded. We address this by establishing a key bound on the MGF of $m(\Pa)$ under $\mu$ in Appendix~\ref{Appdx:Importance Sampling Results}.

The connection between $L^2(\pi,\mu)$ and $L^2(\tPhi, U_{\Sca})$ is made by observing that $\scrP$ can be partitioned into the following equivalence classes. Define an equivalence relation $\sim$ on $\scrP$ such that $\Pa \sim \Pa^{\prime}$ if the two paths agree on the order in which the \textit{final} jumps among sites in $\calS$ occur.
Let $\scrP_{[i]}$ denote the $i$th equivalence class and  $\s^r_i \in \SymS$ the permutation associated with $\scrP_{[i]}$. For example, in the  case of length $r$ paths we obtain
\begin{align*}
\scrP_{[i]}^r :=  \scrP^r \cap \scrP_{[i]} = \{\Pa: m(\Pa) = r, (s^1,\ldots,s^r) = \s^r_i, (b^1,\ldots,b^r) = (y_{s^1},\ldots,y_{s^r})  \},
\end{align*}
%
with paths in $\scrP^r \cap \scrP_{[i]}$ differing only by jump \textit{times}, with jump order given by $\s^r_i$. Then we can express an integral over the full path space as a summation of integrals over the $r!$ equivalence classes:
\begin{align}
\label{eqn: approach expansion}
\int_{\scrP} w^2(\Pa) \mu(d\Pa) & =
\left[1 + \frac{\sum_{m > r}\sum^{r!}_{i=1}\int_{\scrP_{[i]}^m} w^2(\Pa) \mu(d\Pa)}{\sum^{r!}_{i=1} \int_{\scrP_{[i]}^r} w^2(\Pa^r) \mu(d\Pa^r)} \right] \sum^{r!}_{i=1}\int_{\scrP_{[i]}^r} w^2(\Pa^r) \mu(d\Pa^r).
\end{align}
We show in Appendix~\ref{Appdx: weight bound}
that under Assumption~\ref{assump: T assumption}  there exist constants $0< c < c^{\prime}$ such that with probability one under $\mu$:
\begin{align*}
c \cdot \Phi(\s(\Pa^r)) \leq w(\Pa^r) \leq c^{\prime} \cdot \Phi(\s(\Pa^r)),
\end{align*}
that is, the weights for any length $r$ paths sharing the same order of the observed mutations lie within a constant factor.
In this sense,
the variance of $w(\Pa^{r})$  across all length $r$ paths is dominated by the variance of $\Phi(\s(\Pa^r))$ across the $r!$ possible orderings in which the $r$ required jumps occur. The bracketed term in \eqref{eqn: approach expansion} can be viewed as the ``inflation'' in this variance due to the presence of extra jumps.
Controlling this inflation is critical and, as mentioned above, we do so by bounding the MGF of $m(\Pa)$ under $\mu$. Specifically, we find a constant $\theta \in (0,\infty)$ and bound the bracketed term from above by an expectation $1 + \E_{\mu}[\theta^{m(\Pa) -r}]$.
This upper bound can then be factored site-by-site due to site independence:
\begin{align*}
    \E_{\mu}[\theta^{m(\Pa) -r}] = \int_{\mathscr{P}} \theta^{m(\Pa) - r} \mu(d\Pa) = \theta^{-r} \prod^{n}_{i=1} \int_{\mathscr{P}_{i}} \theta^{m(\Pa_i)} \mu(d\Pa_i) = \theta^{-r} \prod^{n}_{i=1} \E_{\mu_i}[\theta^{m(\Pa_i)}].
\end{align*}
This factorization is critical to our analysis, as it allows the bracketed term to be bounded above in terms of the MGF of $m(\Pa_i)$, the number of jumps at a \textit{single} site.

We then give a specific choice of sequences $(\x^{\star},\y^{\star})$ which, evolving under the CpG model, serves to establish a corresponding \textit{lower bound}:
\begin{align*}
\text{D}_{\text{KL}}(\pi \mid \mid \mu) = \Omega(\text{D}_{\text{KL}}(\tPhi \mid \mid U_{\Sca}) ).
\end{align*}
Applying Jensen's inequality and combining these upper and lower bounds then yields $\log(L^2(\pi,\mu)) = \Theta(\log(L^2(\tPhi,U_{\calS})))$.

\subsection{Proof of Theorem~\ref{thm:Main Thm Bound}}
\label{sec: upper bound proof}
We prove a general upper bound on $\chi^2(\pi \mid \mid \mu)$, from which  Theorem~\ref{thm:Main Thm Bound} follows as a special case. Some supporting details are deferred to Appendix~\ref{Appdx:Importance Sampling Results}.
Denote the largest and smallest context-dependent rate multipliers by
\begin{align*}
\phi_{\max} = \max_{\tx \in \scrA^{k+1}}\max_{y \in \scrA \setminus \{x \}} \phi(y; \tx ) \quad\quad \phi_{\min} = \min_{\tx \in \scrA^{k+1}}\min_{y \in \scrA \setminus \{x\}} \phi(y; \tx),
\end{align*}
where the outer $\max$ and $\min$ are over possible size $k$ contexts $\tx = (x_1,\ldots,x,\ldots,x_k)$. Let 
\begin{align*}
\gmax = \max_{i \in \{1,\ldots,n\}}\max_{x \in \scrA}\max_{y \in \scrA \setminus \{x\}} \gamma_i(y;x) \quad\quad \gmin = \min_{i \in \{1,\ldots,n\}}\min_{x \in \scrA}\min_{y \in \scrA \setminus \{x\}}\gamma_i(y;x)
\end{align*}
be the largest and smallest ISM rates across all sites, and similarly let $\tg_{\min}$ and $\tg_{\max}$ be bounds on the largest and smallest rates under the DSM.
%

Let
\begin{align*}
\Delta^{\tg}(j)  := \Delta^{\tg}(j; \Pa) &=  \tg(\cdot; \x^j) - \tg(\cdot; \x^{j-1}) \\
\Delta^{\gamma}(j) := \Delta^{\gamma}(j; \Pa) &=  \gamma(\cdot; \x^j) - \gamma(\cdot; \x^{j-1})
\end{align*}
denote the change in sequence mutation rate following the $j$th jump along path $\Pa$.
Observe that we can rewrite the exponential term in the path density  \eqref{eqn:Path Density} as
\begin{align}
\label{eqn: delta t to t conversion}
-\sum^{m(\Pa)}_{j=1}\Delta^t(j-1) \tg(\cdot; \x^{j-1})  -\Delta^t(m)\tg(\cdot; \y)  = \sum^{m(\Pa)}_{j=1}t^j \Delta^{\tg}(j)  
-T\tg(\cdot; \y) .
\end{align}
Therefore, for $\Pa$ drawn from $\mu$, the importance weight \eqref{eqn:Importance Weights} can be written
\begin{align}
\label{eqn:Likelihood Ratio}
w(\Pa) = 
\Phi(\s(\Pa),\bb(\Pa)) 
e^{-T (\tg(\cdot; \y) - \gamma(\cdot; \y)) }
\me^{\sum^{m(\Pa)}_{j=1} t^j (\Delta^{\tg}(j)  - \Delta^{\gamma}(j) )},
\end{align}
and since $e^{-T (\tg(\cdot; \y) - \gamma(\cdot; \y)) }$ is constant we have
\begin{align}
\label{eqn:L2 cancellation}
L^2(\pi, \mu) = 
\E_{\mu}[w^2(\Pa)]/ \left(\E_{\mu}[w(\Pa)] \right)^2 = 
\E_{\mu}[w_0^2(\Pa)]/ \left(\E_{\mu}[w_0(\Pa)] \right)^2,
\end{align}
where
\begin{align*}
w_0(\Pa) =
\Phi(\s(\Pa), \bb(\Pa))
\me^{\tilde{\psi}(\s(\Pa),\mathbf{b}(\Pa))  - \psi(\s(\Pa),\mathbf{b}(\Pa))}
\end{align*}
for
\begin{align*}
\tpsi(\s(\Pa),\bb(\Pa)) \defeq \sum^{m(\Pa)}_{j=1} t^j \Delta^{\tg}(j) \qquad\text{ and }\qquad \psi(\s(\Pa),\bb(\Pa)) \defeq \sum^{m(\Pa)}_{j=1} t^j \Delta^{\gamma}(j).
\end{align*}
We first bound $w_0(\Pa)$ from above and below by a function that depends only on $m(\Pa)$, $\s(\Pa)$, and $\bb(\Pa)$. To do so we use the following result, proven in Appendix~\ref{Appdx:Importance Sampling Results}, which bounds the change in the sequence mutation rates uniformly.
\begin{restatable}{lemma}{LemmaOne}
\label{lemma: Uniform Bounds on Delta}
Let $q = a-1$
and define
\begin{align*}
\delta := q(\gmax - \gmin)
\qquad \text{and} \qquad 
\tilde{\delta} := q(k+1) (\tgmax - \tgmin).
\end{align*}
%
Then the following statements hold for the random variables $\Delta^{\tg}(j)$ and $\Delta^{\gamma}(j)$: 
\begin{enumerate}
\item $\Prob_{\mu}(|\tpsi(\s(\Pa),\bb(\Pa))| \leq m(\Pa) T\tilde{\delta} ) = 1 $
\item $\Prob_{\mu}(|\psi(\s(\Pa),\bb(\Pa))| \leq m(\Pa)T\delta ) = 1 $
\item $\Prob_{\mu}(|\tpsi(\s(\Pa),\bb(\Pa)) - \psi(\s(\Pa),\bb(\Pa))| \leq m(\Pa)T(\tilde{\delta}+\delta)) = 1 $.
\end{enumerate}
\end{restatable}
%
\noindent Using Lemma~\ref{lemma: Uniform Bounds on Delta} we  obtain the following bound on $w_0(\Pa)$, again for $\Pa$ drawn from $\mu$.
\begin{restatable}{lemma}{LemmaTwo}
\label{lemma:uniform bound w}
\begin{align}
\label{eqn:uniform bound w statement}
\Prob_{\mu}\left(\Phi(\s(\Pa), \bb(\Pa)) e^{- m(\Pa)T (\tilde{\delta} + \delta)} 
\leq w_0(\Pa) \leq \Phi(\s(\Pa), \bb(\Pa)) e^{  m(\Pa) T (\tilde{\delta} + \delta)} \right) = 1.
\end{align}
%
\end{restatable}
%
\noindent 
The proof follows immediately by Lemma~\ref{lemma: Uniform Bounds on Delta} and is given in Appendix~\ref{Appdx:Importance Sampling Results}. We now apply Lemmas~\ref{lemma: Uniform Bounds on Delta} and~\ref{lemma:uniform bound w} to obtain the following lower bound on $\E_{\mu}[w_0(\Pa)]$ in the denominator of \eqref{eqn:L2 cancellation}. Note that the bound involves 
\begin{equation}
p_r = \Prob_{\mu}(m(\Pa) = r \mid \x,\y),
\label{Eqn:Pr}
\end{equation}
the probability of exactly $r$ mutations occurring in the transition from $\x$ to $\y$ under the ISM; we provide an explicit lower bound on this quantity in Lemma~\ref{lemma:expectation and pr bound} below.
\begin{lemma}
\label{lemma:Denominator of chi square}
The following lower bound holds:
\begin{align*}
\E_{\mu}[w_0(\Pa)] \geq e^{-rT(\tilde{\delta} + 3\delta)} p_r  \frac{Z^r_{\Phi}}{r!}.
\end{align*}
\end{lemma}
To prove Lemma~\ref{lemma:Denominator of chi square}, recall that $\scrP$ is the set of all paths from $\x$ to $\y$, and observe that the expectation of a measurable function $f: \scrP \rightarrow \mathbb{R}$ can be expressed as a summation over the equivalence classes $\scrP_{[1]},\ldots,\scrP_{[r!]}$ defined in Section~\ref{sec: approach}: 
\begin{align*}
\E_{\mu}[f(\Pa)] = \sum^{r!}_{v=1} \int_{\scrP_{[v]}} f(\Pa) \mu(d\Pa).
\end{align*}
In particular, since $\Phi(\Pa) = \Phi(\s(\Pa),\bb(\Pa))$ depends only on the site and base changes for $\Phi(\s(\Pa),\bb(\Pa))$  defined in \eqref{eqn: Phi general}, $\Phi$ is constant over $\scrP_{[v]} \cap \scrP^r$ (since for any $\Pa^r \in \scrP_{[v]} \cap \scrP^r$ we have $\bb(\Pa^r) = (y_{\s^r_{v}(1)},\ldots, y_{\s^r_{v}(r)} ) := \y_{\s^r_{v}}$ and $\s(\Pa^r) = \s^r_{v}$) and therefore since $\Phi > 0$:
\begin{align}
\E_{\mu}[\Phi(\Pa)] \geq \E_{\mu}[\Phi(\Pa)\ind(m(\Pa)=r)] &= 
\sum^{r!}_{v=1} \int_{\scrP_{[v]} \cap \scrP^r} \Phi(\s(\Pa),\bb(\Pa)) \mu(d\Pa) \nonumber\\
& = \sum^{r!}_{v=1} \Phi(\s^r_{v}, \y_{\s^r_{v}}) \int_{\scrP_{[v]} \cap \scrP^r} \mu(d\Pa) \nonumber \\
&= p_r\sum^{r!}_{v=1} \Phi(\s^r_{v}, \y_{\s^r_{v}}) \mu_{\mid r}(\scrP_{[v]}),
\label{eqn: lemma lower bound proof overview}
\end{align}
where $\mu_{\mid r}(B)$ denotes the restriction of the ISM to paths of length $r$, defined as $\mu_{\mid r}(B) := \mu(B \cap \scrP^r)/\mu(\scrP^r)$ for $B \subset \scrP$.
Lemma~\ref{lemma:Denominator of chi square} will follow by combining Lemmas~\ref{lemma: Uniform Bounds on Delta} and~\ref{lemma:uniform bound w}:
\begin{proof}
As in \eqref{eqn: lemma lower bound proof overview}, we write $\E_{\mu}[w_0(\Pa)] = \sum^{r!}_{v=1} \int_{\scrP_{[v]}}w_0(\Pa) \mu(d\Pa)$ and apply Lemma~\ref{lemma:uniform bound w} to obtain 
\begin{equation}
\E_{\mu}[w_0(\Pa)] 
\geq 
e^{-rT(\tilde{\delta} + \delta)} 
\E_{\mu}[\Phi(\Pa)] \geq  e^{-rT(\tilde{\delta} + \delta)}  p_r\sum^{r!}_{v=1} \Phi(\s^r_{v}, \y_{\s^r_{v}}) \mu_{\mid r}(\scrP_{[v]}).
\label{eqn:Denominator chi square proof 1}
\end{equation}
By Lemma~\ref{lemma: Uniform Bounds on Delta} the restricted density $\mu_{\mid r}(\Pa^r) \geq e^{-r(2\delta T + \log T)}$
since
\begin{align*}
\mu_{\mid r}(\Pa^r) & =  \frac{\prod^r_{j=1}\gamma_{s^j}(b^j; x_{s^j}^{j-1})
\me^{\sum^r_{j=1} t^j \Delta^{\gamma}(j)}}{\int_{\scrP^r}\prod^r_{j=1}\gamma_{s^j}(b^j; x_{s^j}^{j-1})
\me^{\sum^r_{j=1} t^j \Delta^{\gamma}(j) } \nu(d\Pa)}\\
& = \frac{
\me^{\sum^r_{j=1} t^j  \Delta^{\gamma}(j) }}{\int_{\scrP^r}
\me^{\sum^r_{j=1} t^j\Delta^{\gamma}(j) } \nu(d\Pa)} \geq  \frac{e^{-2rT\delta}}{\int_{\scrP^r} \nu(d\Pa)} = \frac{e^{-2rT\delta}}{ r! \cdot \frac{T^r}{r!} } .
\end{align*}
%
The second equality follows since $\prod^r_{j=1}\gamma(b^j; x^{j-1})$ is invariant under permutations due to site independence, and the last equality uses $\int_{0<t^1 < \ldots < t^r < T} dt^1\ldots dt^r = T^r/r!$. It follows that 
\begin{align}
\label{eqn:Denominator chi square proof 2 2}
\mu_{\mid r}(\scrP_{[v]}) = \int_{\scrP_{[v]} \cap \scrP^r} \mu_{\mid r}(\Pa) d\nu(\Pa) \geq T^{-r}e^{-2rT\delta} \int_{0 < t^{1} < \ldots t^r < T} dt^{1}\ldots dt^r = \frac{e^{-2rT\delta}}{r!}.
\end{align}
%
%
%
The stated bound follows by combining \eqref{eqn:Denominator chi square proof 1} and \eqref{eqn:Denominator chi square proof 2 2} since $\Phi(\s^r_{v}, \y_{\s^r_{v}}) \equiv \Phi(\s^r_{v})$:
\begin{align*}
\E[w_0(\Pa)]  \geq e^{-rT(\tilde{\delta}+3\delta)} p_r  \frac{Z^r_{\Phi}}{r!}.
\end{align*}
\end{proof}
Lemma~\ref{lemma:Denominator of chi square} provides a lower bound on $\E_{\mu}[w_0(\Pa)]$ in the denominator of \eqref{eqn:L2 cancellation}. We will now obtain an upper bound on $\E_{\mu}[w_0^2(\Pa)]$ in the numerator of \eqref{eqn:L2 cancellation}. 
\begin{lemma}
\label{lemma:Numerator of chi square}
Let $\theta = \gamma^3_{\star}\phi^{2(k+1)}_{\star}\phi^2_{\max} e^{2T(\tilde{\delta} + 2\delta)}$. The following bound holds:
\begin{align*}
\E_{\mu}[w_0^2(\Pa)] \leq e^{2rT(\tilde{\delta} + 2\delta )} \E_{\mu}[\theta^{m(\Pa) - r}] \sum_{\s^r \in \SymS} \frac{\Phi^2(\s^r)}{r!}.
\end{align*}
\end{lemma}
Note that this bound involves an expectation of the form $\E_{\mu}[c^{m(\Pa)}]$ for some constant $c \in (0,\infty)$ that does not depend on $\Pa$, effectively the MGF of $m(\Pa)$; Lemma~\ref{lemma:expectation and pr bound} below provides an explicit upper bound on this expectation. 
\begin{proof}\let\qed\relax
The proof of Lemma~\ref{lemma:Numerator of chi square} is similar to that of Lemma~\ref{lemma:Denominator of chi square} in that we will decompose $\E_{\mu}[w_0^2(\Pa)]$ into a sum over the equivalence classes $\scrP_{[1]},\ldots,\scrP_{[r!]}$ and find constants $c, c^{\prime} \in (0,\infty)$ that depend on $\tQ$, and $\Q$ such that
\begin{align}
\E_{\mu}[w_0^2(\Pa)] &= \sum^{r!}_{v=1} \int_{\scrP_{[v]}}w_0^2(\Pa) \mu(d\Pa) \nonumber \\
&= 
\sum^{r!}_{v=1} \int_{\scrP_{[v]}}
\Phi^2(\s(\Pa), \bb(\Pa))
\me^{2 (\tpsi(\s(\Pa),\bb(\Pa) ) - \psi(\s(\Pa),\bb(\Pa) ))} 
\mu(d\Pa) 
\nonumber \\
& \leq 
\sum^{r!}_{v=1} \Phi^2(\s^r_{v})
\E_{\mu_{\mid \scrP_{[v]}}}\Big[\frac{\Phi^2(\s(\Pa),\bb(\Pa))}{\Phi^2(\s^r_{v})} e^{2m(\Pa)T (\tilde{\delta}+\delta)}\Big],
\label{eqn:Numerator of chi square overview}
\end{align}
with the inequality coming from Lemma~\ref{lemma: Uniform Bounds on Delta}. Now for $\Pa \in \Pa_{[v]}$ we have 
\begin{align}
\label{eqn:uniform ratio bound}
\Prob_{\mu}\left(\frac{\Phi(\s(\Pa), \bb(\Pa))}{\Phi(\s^r_{v})} \ind_{\scrP_{[v]}}(\Pa) \leq 
(\phi^{(k+1)}_{\star}
\phi_{\max})^{m(\Pa) - r}
\ind_{\scrP_{[v]}}(\Pa) \right) = 1.
\end{align}
To see this, note that each $m(\Pa) - r$ additional mutation along a path $\Pa$ has rate no larger than $\phi_{\max}$, and can change the context of at most $k+1$ sites under the DSM. Applying this to \eqref{eqn:Numerator of chi square overview} gives
\begin{equation}
\label{eqn:Chi square numerator eq 1}
\E_{\mu}[w_0^2(\Pa)] \leq e^{2rT(\tilde{\delta} + \delta )}\sum^{r!}_{v=1} \Phi^2(\s^r_{v} ) \int_{\scrP_{[v]}} \left[\phi^{(k+1)}_{\star}\phi_{\max} e^{T(\tilde{\delta} + \delta)}\right]^{2(m(\Pa) - r)} \mu(d\Pa).
\end{equation}
To complete the proof, we will need to bound the integral on the right-hand side
 of 
\eqref{eqn:Chi square numerator eq 1}; this is provided by the following additional Lemma. The integral in \eqref{eqn:Chi square numerator eq 1} bounds the increase in variance of $\Phi(\s(\Pa),\bb(\Pa))$ due to extra jumps
as discussed in Section~\ref{sec: approach} (see 
\eqref{eqn: approach expansion}). 
The following lemma is key to bounding this inflation factor. 
\end{proof}
%
\begin{lemma}
\label{lemma:Numerator of chi square p1}
Let $c \in (0,\infty)$ be a constant and $\gamma_{\star} := \gamma_{\max}/\gamma_{\min}$ the ratio of the largest to smallest ISM rates. Then
\begin{align*}
\int_{\scrP_{[i]}}c^{m(\Pa)} \mu(d\Pa) \leq \int_{\scrP_{[j]}}c^{m(\Pa)} \gamma^{3(m(\Pa) - r)}_{\star} e^{2m(\Pa)T\delta} \mu(d\Pa) \text{ for }i \neq j.
\end{align*}
\end{lemma}
\begin{proof}
Recall $\mu(\Pa) \propto \tP_{(T,\Q)}(\y,\Pa \mid \x)$ and
\begin{align*}
\tP_{(T,\Q)}(\y,\Pa \mid \x) = \left[\prod^{m(\Pa)}_{j=1} \gamma_{s^j}(b^j; x^{j-1}_{s^j})\right]
 \me^{-T\gamma(\cdot; \y) }
 \me^{\sum^{m(\Pa)}_{j=1} t^j\Delta^{\gamma}(j)},
\end{align*}
for $m(\Pa) \geq r$ and zero otherwise, where $ \me^{-T\gamma(\cdot; \y) }$ is a constant. Therefore, it suffices to show
\begin{align*}
\int_{\scrP_{[i]}}c^{m(\Pa)} f(\Pa)g(\Pa) d\Pa \leq \int_{\scrP_{[j]}}c^{m(\Pa)} \gamma^{3(m(\Pa) - r)}_{\star} e^{2m(\Pa)T\delta} f(\Pa)g(\Pa) d\Pa \text{ for }i \neq j,
\end{align*}
where 
\begin{align*}
f(\Pa) := \prod^{m(\Pa)}_{j=1} \gamma_{s^j}(b^j; x^{j-1}_{s^j}) \qquad\qquad g(\Pa) := \me^{\sum^{m(\Pa)}_{j=1} t^j\Delta^{\gamma}(j)} = e^{\psi(\s(\Pa),\mathbf{b}(\Pa))}.
\end{align*}
Let $\Gamma := f(\Pa^r)$ denote the value of $f$ for any length $r$ path (which is invariant under orderings). Since each of the $m(\Pa) - r$ additional mutations along a path $\Pa$ contributes a multiple of at least $\gamma_{\min}$ but at most $\gamma_{\max}$, and can change the context of at most one site (its own) under the ISM, we get
\begin{align}
\label{eqn:ISM max change bound}
\Prob_{\mu}\left((\gamma^{-1}_{\star}\gamma_{\min})^{m(\Pa)-r} \leq \frac{f(\Pa)}{\Gamma} \leq (\gamma_{\star}\gamma_{\max})^{m(\Pa)-r}\right) = 1.
\end{align}
%
%
Combining this with Lemma~\ref{lemma: Uniform Bounds on Delta} gives
%
\begin{align}
\label{eqn:Chi square upper bound p1 eq 1}
\Prob_{\mu}\left( 
(\gamma^{-1}_{\star}\gamma_{\min})^{m(\Pa)-r} e^{-m(\Pa)T\delta} \leq \frac{f(\Pa)g(\Pa)}{\Gamma} \leq 
(\gamma_{\star}\gamma_{\max})^{m(\Pa)-r}
e^{m(\Pa)T\delta} \right) = 1.
\end{align}
Applying \eqref{eqn:Chi square upper bound p1 eq 1} twice yields
\begin{align*}
\int_{\scrP_{[i]}}c^{m(\Pa)} f(\Pa)g(\Pa) \nu(d\Pa^m) &\leq \Gamma \sum^{\infty}_{m=r} c^m (\gamma_{\star}\gamma_{\max})^{m-r} e^{mT\delta} \int_{\scrP^m}  \ind_{\scrP_{[i]}}(\Pa^m)\nu(d\Pa^m)\\
&= \Gamma \sum^{\infty}_{m=r} c^m 
(\gamma_{\star}\gamma_{\max})^{m-r} e^{mT\delta} \int_{\scrP^m}  \ind_{\scrP_{[j]}}(\Pa^m)\nu(d\Pa^m) \\
& \leq \int_{\scrP_{[j]}}c^{m(\Pa)} \gamma_{\star}^{3(m(\Pa) - r)}e^{2m(\Pa)T\delta}f(\Pa)g(\Pa) \nu(d\Pa),
\end{align*}
%
where the middle equality $\int_{\scrP^m}  \ind_{\scrP_{[i]}}(\Pa^m)\nu(d\Pa^m) = \int_{\scrP^m}  \ind_{\scrP_{[j]}}(\Pa^m) \nu(d\Pa^m)$ follows by symmetry.
\end{proof}
With Lemma~\ref{lemma:Numerator of chi square p1} in hand, we are now ready to complete the proof of Lemma~\ref{lemma:Numerator of chi square}.
\begin{proof}
\textit{(Lemma~\ref{lemma:Numerator of chi square}, cont'd)}

Note that for any constant $c \in (0,\infty)$ we have by Lemma~\ref{lemma:Numerator of chi square p1}
\begin{align*}
\int_{\scrP_{[i]}} c^{m(\Pa)} \mu(d\Pa) &= \frac{1}{r!} \sum^{r!}_{j=1}  \int_{\scrP_{[i]}} c^{m(\Pa)} \mu(d\Pa) \\   &\leq \frac{1}{r!}\sum^{r!}_{j=1} \int_{\scrP_{[j]}} c^{m(\Pa)} \gamma^{3(m(\Pa) - r)}_{\star} e^{2m(\Pa)T\delta} \mu(d\Pa) \\
&= \frac{1}{r!} \E_{\mu}[c^{m(\Pa)} \gamma^{3(m(\Pa) - r)}_{\star} e^{2m(\Pa)T\delta}].
\end{align*}
Setting $c = \big(\phi^{(k+1)}_{\star}\phi_{\max} e^{T(\tilde{\delta} + \delta)}\big)^2$ and combining with \eqref{eqn:Chi square numerator eq 1}  gives the stated bound:
\begin{align*}
\E_{\mu}[w_0^2(\Pa)] 
& \leq e^{2rT(\tilde{\delta} + 2\delta )}  \E_{\mu}[\theta^{m(\Pa) - r}] \sum^{r!}_{v=1} \frac{\Phi^2(\s^r_{v})}{r!},
\end{align*}
where we recall that $\theta = \gamma^3_{\star}\phi^{2(k+1)}_{\star}\phi^2_{\max} e^{2T(\tilde{\delta} + 2\delta)}$.
\end{proof}
The following lemma, which is proved in Appendix~\ref{appdx: ISM expectation and pr bound}, will be used to obtain explicit lower and upper bounds on the quantities $p_r$ and $\E[\theta^{m(\Pa)-r}]$, respectively, that appear in Lemmas~\ref{lemma:Denominator of chi square} and \ref{lemma:Numerator of chi square}. (Recall from Lemma~\ref{lemma: Uniform Bounds on Delta} that $q=a-1$.)
\begin{lemma}
\label{lemma:expectation and pr bound}
Let $\theta \in \mathbb{R}$ and $c = \gmax^2/\gmin q^2 e^{Tq(\gmax - \gmin)}$. Then
%
\begin{align*}
\E_{\mu}[\theta^{m(\Pa)}] \leq \theta^r\exp\left(rT \theta c \exp(Tq \theta \gamma_{\max} ) + (n-r)T^2 \theta^2 c\gmin \exp(Tq\theta \gamma_{\max} )  \right).
\end{align*}
In addition, the following lower bound on $p_r$ given in \eqref{Eqn:Pr} holds
\begin{align*}
p_r \geq 
\exp\left(- rTc\exp(T q \gmax) - (n-r)T^2c \exp(Tq\gmax)\gmin \right).
\end{align*}
\end{lemma}
We now plug the bounds from Lemma~\ref{lemma:expectation and pr bound} into the bounds obtained in Lemmas~\ref{lemma:Denominator of chi square} and \ref{lemma:Numerator of chi square} to prove the following result relating $L^2(\pi, \mu)$ and $L^2(\tPhi, U_{\Sca})$. Theorem~\ref{thm:Main Thm Bound} follows as a special case. 
\begin{theorem}
\label{thm:main thm general}
Let $\theta = \gamma^3_{\star}\phi^{2(k+1)}_{\star}\phi^2_{\max} e^{2T(\tilde{\delta} + 2\delta)}$ and
\begin{align*}
c &= \frac{\gamma^2_{\max}}{\gmin}  q^2 e^{Tq(\gmax - \gmin)} [\theta e^{Tq\theta \gamma_{\max}}  +  2e^{Tq\gmax} ]  + 2(2\tilde{\delta} + 5\delta) \\
c^{\prime} &=  \gamma^2_{\max}  q^2 e^{Tq(\gmax - \gmin)}  [\theta^2  e^{Tq\theta \gamma_{\max}}  + 2e^{Tq\gmax}  ]   .
\end{align*}
Then the follow upper bound holds:
\begin{align*}
L^2(\pi,\mu) 
& \leq \exp\left(rT c + (n-r)T^2 c^{\prime} \right) L^2(\tPhi, U_{\Sca}). 
\end{align*}
In particular, under Assumption~\ref{assump: T assumption}, the importance sampling estimator with 
$N^{\star}(\epsilon) = \bigO(\epsilon^{-2}L^2(\tPhi, U_{\Sca}))$ for $\epsilon \in (0,1)$ provides a randomized approximation scheme for $p_{(T,\tQ)}(\y \mid \x)$.
\end{theorem}
\begin{proof}
Applying Lemmas~\ref{lemma:Denominator of chi square} and \ref{lemma:Numerator of chi square} together with \eqref{eqn:L2 cancellation} gives us
\begin{align*}
L^2(\pi,\mu) \leq e^{2rT(2\tilde{\delta} + 5\delta)} \frac{\E_{\mu}[\theta^{m(\Pa) - r}]}{p^2_r} L^2(\tPhi, U_{\Sca}).
\end{align*}
Applying Lemma~\ref{lemma:expectation and pr bound} gives us the stated bound:
\begin{align}
\label{eqn:main thm general proof eq 1}
e^{2rT(2\tilde{\delta} + 5\delta)} \frac{\E_{\mu}[\theta^{m(\Pa) - r}]}{p^2_r} L^2(\tPhi, U_{\Sca}) \leq  \exp\left(rT c + (n-r)T^2 c^{\prime} \right) L^2(\tPhi, U_{\Sca}).
\end{align}
\end{proof}
%
Theorem~\ref{thm:Main Thm Bound} follows immediately from Theorem~\ref{thm:main thm general} by taking a uniform bound over the context-dependent rates:
\begin{proof}(Theorem~\ref{thm:Main Thm Bound})
Apply Theorem~\ref{thm:main thm general} and note
\begin{align*}
L^2(\tPhi, U_{\Sca}) = \frac{\sum_{\s^r \in \SymS} \frac{\Phi^2(\s^r)}{r!}  }{(\sum_{\tilde{\s}^r \in \SymS} \frac{\Phi(\tilde{\s}^r)}{r!})^2} \leq \phi^{2r}_{\star}.
\end{align*}
\end{proof}

\subsection{Proof of Theorem~\ref{thm: main thm IS LB}}
\label{sec: LB}
\begin{figure}[h]
\centering
\begin{tikzpicture}[scale=0.5, baseline=(current bounding box.center)]  
    \node at (-1,0.5) {$\x^{\star}$ =};
    \foreach \x/\y in {0/$\tT$} {
        \draw (\x,0) rectangle (\x+1,1);
            \node at (\x+0.5,0.5) {\y};
        }
    \node at (1.5,0.5) {...};
    \foreach \x/\y in {2/$\tT$,3/$\tC$,4/$\tA$,5/$\tT$,6/$\tT$,7/$\tC$,8/$\tA$,9/$\tT$} {
        \draw (\x,0) rectangle (\x+1,1);
            \node at (\x+0.5,0.5) {\y};
        }
    \node at (10.5,0.5) {...};
    \foreach \x/\y in {11/$\tT$,12/$\tC$,13/$\tA$, 14/$\tT$ } {
        \draw (\x,0) rectangle (\x+1,1);
            \node at (\x+0.5,0.5) {\y};
        }
    \node at (15.5,0.5) {...};
    \foreach \x/\y in {16/$\tT$} {
        \draw (\x,0) rectangle (\x+1,1);
            \node at (\x+0.5,0.5) {\y};
        }
\draw [decorate,decoration={brace,mirror,amplitude=5pt}]
    (3,0) -- (5,0) node [black,midway,yshift=-0.4cm] {$1$};

    \draw [decorate,decoration={brace,mirror,amplitude=5pt}]
    (7,0) -- (9,0) node [black,midway,yshift=-0.4cm] {$2$};
   
    \draw [decorate,decoration={brace,mirror,amplitude=5pt}]
    (12,0) -- (14,0) node [black,midway,yshift=-0.4cm] {$r_{I}$};

\node at (-1,-3.5) {$\y^{\star}$ =};
\foreach \x/\y in {0/$\tT$} {
    \draw (\x,-4) rectangle (\x+1,-3);
        \node at (\x+0.5,-3.5) {\y};
    }
\node at (1.5,-3.5) {...};
            \foreach \x/\y in {2/$\tT$,3/$\tT$,4/$\tG$,5/$\tT$,6/$\tT$,7/$\tT$,8/$\tG$,9/$\tT$} {
                \draw (\x,-4) rectangle (\x+1,-3);
                \node at (\x+0.5,-3.5) {\y};
            }
            \node at (10.5,-3.5) {...};
            \foreach \x/\y in {11/$\tT$,12/$\tT$,13/$\tG$, 14/$\tT$ } {
                \draw (\x,-4) rectangle (\x+1,-3);
                \node at (\x+0.5,-3.5) {\y};
            }
            \node at (15.5,-3.5) {...};
            \foreach \x/\y in {16/$\tT$} {
                \draw (\x,-4) rectangle (\x+1,-3);
                \node at (\x+0.5,-3.5) {\y};
            }
\draw [decorate,decoration={brace,mirror,amplitude=5pt}]
    (3,-4) -- (5,-4) node [black,midway,yshift=-0.4cm] {$1$};

    \draw [decorate,decoration={brace,mirror,amplitude=5pt}]
    (7,-4) -- (9,-4) node [black,midway,yshift=-0.4cm] {$2$};
   
    \draw [decorate,decoration={brace,mirror,amplitude=5pt}]
    (12,-4) -- (14,-4) node [black,midway,yshift=-0.4cm] {$r_{I}$};
\end{tikzpicture}
\caption{Endpoint sequences used to establish lower bounds on KL$(\pi \| \mu)$. Note the $r_{I}$ islands corresponding to each $(\tC,\tA)$ pair mutating to a $(\tT,\tG)$ pair.} 
\label{fig: LB Sequences}
\end{figure}
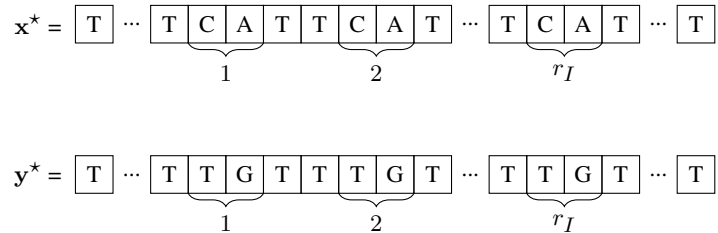

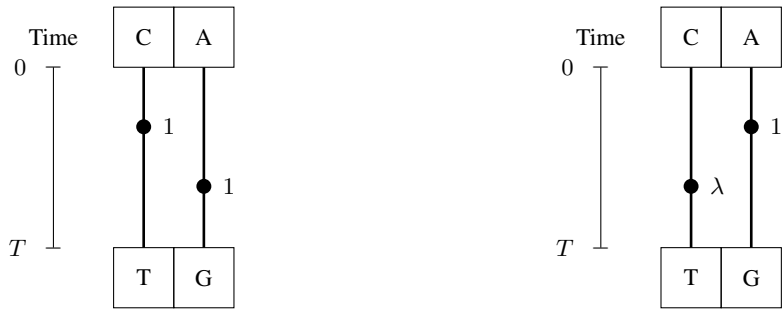
\begin{figure}[b]
\centering
\begin{subfigure}[b]{0.49\linewidth}
\centering
\begin{tikzpicture}[scale=0.8, baseline=(current bounding box.center)]     
    \foreach \x/\y in {0/$\tC$,1/$\tA$} {
                \draw (\x,0) rectangle (\x+1,1);
                \node at (\x+0.5,0.5) {\y};
            }
        
            \draw (-1,0) -- (-1,-3);
            \node [label=left:] at (-1,.5) {Time};
            \node [label=left:$0$] at (-1,0) {$-$};
            \node [label=left:$T$] at (-1,-3) {$-$};

            \draw [line width=1pt] (0.5,0) -- node [below,midway] {} (0.5,-3) node [right,midway] { } node [pos=0.33,circle,fill,inner sep=2pt,label={[label distance=.1mm]2.8: $1$ }] {};
            \draw [line width=1pt] (1.5,0) -- node [below,midway] {} (1.5,-3) node [right,midway] { } node [pos=0.66,circle,fill,inner sep=2pt,label={[label distance=.1mm]2.8: $1$ }] {};
        
            \foreach \x/\y in {0/$\tT$,1/$\tG$} {
                \draw (\x, -3) rectangle (\x+1, -4);
                \node at (\x+0.5,-3.5) {\y};
            }
        \end{tikzpicture}
        \caption{An example of an  island path  with the left site mutating before the right site.}
        \label{fig:subfig1}
\end{subfigure}
\hfill
\begin{subfigure}[b]{0.49\linewidth}
\centering
\begin{tikzpicture}[scale=0.8, baseline=(current bounding box.center)]     
            \foreach \x/\y in {0/$\tC$,1/$\tA$} {
                \draw (\x,0) rectangle (\x+1,1);
                \node at (\x+0.5,0.5) {\y};
            }
        
            \draw (-1,0) -- (-1,-3);
            \node [label=left:] at (-1,.5) {Time};
            \node [label=left:$0$] at (-1,0) {$-$};
            \node [label=left:$T$] at (-1,-3) {$-$};
        
            \draw [line width=1pt] (0.5,0) -- node [below,midway] {} (0.5,-3) node [right,midway] { } node [pos=0.66,circle,fill,inner sep=2pt,label={[label distance=.1mm]2.8: $\lambda$ }] {};
        
            \draw [line width=1pt] (1.5,0) -- node [below,midway] {} (1.5,-3) node [right,midway] { } node [pos=0.33,circle,fill,inner sep=2pt,label={[label distance=.1mm]2.8: $1$ }] {};
        
            \foreach \x/\y in {0/$\tT$,1/$\tG$} {
                \draw (\x, -3) rectangle (\x+1, -4);
                \node at (\x+0.5,-3.5) {\y};
            }
\end{tikzpicture}
\caption{An example of an island path with the right site mutating before the left site.}
\label{fig:subfig2}
\end{subfigure}
\caption{Possible orderings for a length 2 path within an island, with mutation events labeled by their corresponding rates. The sites immediately to the left and right of an island are both $\tT$'s (see Figure~\ref{fig: LB Sequences}).}
\label{fig:mainfig}
\end{figure}

We now prove Theorem~\ref{thm: main thm IS LB}. Our approach is to construct example sequences $\x^\star$ and $\y^\star$ such that the lower bound $\text{D}_{\text{KL}}(\pi \mid \mid \mu) = \Omega(r)$ holds; combining with Theorem~\ref{thm:Main Thm Bound} then yields Theorem~\ref{thm: main thm IS LB}. Supporting details are given in Appendix~\ref{Appdx:Importance Sampling Lower Bound}. 
Consider the sequences
$\x^{\star} = \text{T-(TCAT)}^{r_I}\text{-T}$ and 
$\y^{\star} = \text{T-(TTGT)}^{r_I}\text{-T}$
composed of repeating subsequences. As seen in Figure~\ref{fig: LB Sequences}, the observed mutations make up a set of $r_I(n) := r(n)/2$ repeating \textit{islands}, each of size 2.
%
%
\begin{definition*}[Island problem]
\label{def: island problem}
For the sequences 
given in Figure~\ref{fig: LB Sequences}, consider $\x^{\star}$ evolving to $\y^{\star}$ 
under the CpG model \eqref{eqn:CpG Island Rates} with context-dependent rates given by
\begin{align}
\label{eqn:CpG LB}
\tilde{\gamma}(b; \tx_i) = \gamma(b; x_i) \lambda^{\ind_{\text{CG}}(x_{i-1},x_i) + \ind_{\text{CG}}(x_i,x_{i+1})}
\end{align}
with $\gamma(b;b^{\prime}) \equiv 1$ for $b,b^{\prime} \in \{\tA,\tG,\tC,\tT \}$ and $\lambda \in (1,\infty)$. 
Approximate $p_{(T,\tQ)}(\y^{\star} \mid \x^{\star})$. 
\end{definition*}
For the island problem $\mu$ is the JC69 model \cite{Jukes:1969} with unit substitution rate, and  CpG pairs exhibit an increased mutation rate. In each island, the pair of observed mutations can occur in one of two orders in paths of length $r$: one order produces a CpG-type mutation event, while the other does not (see Figure~\ref{fig:mainfig}). Theorem~\ref{thm:Main Thm Bound} (under Assumption~\ref{assump: T assumption})
gives
\begin{align}
\label{eqn:Chisq LB}
\chi^2(\pi \mid \mid \mu) =
\bigO(\lambda^{2r(n)}).
\end{align}
Consequently, $N^{\star} = \mathcal{O}(\epsilon^{-2}\lambda^{2r(n)})$ samples are sufficient to obtain an $\epsilon$-relative error estimate of the marginal likelihood. We now establish a corresponding lower bound for this problem when $T(n) = \Theta(r(n)/n)$ and $r(n) \leq n^{\frac{1}{3}}$, by showing
\begin{align}
\label{eqn: KL LB Restate}
\text{D}_{\text{KL}}(\pi \mid \mid \mu) = \Omega(r(n)).
\end{align}
(Recall that this implies $\log(\chi^2(\pi \mid \mid \mu)) = \Theta(r(n))$ by \eqref{eqn:Chisq LB} and Jensen's inequality.) We will obtain \eqref{eqn: KL LB Restate} using the following lemma, which establishes a key connection between $\text{D}_{\text{KL}}(\pi \mid \mid \mu)$ and $\text{D}_{\text{KL}}(\tPhi \mid \mid U_{\calS})$.
\begin{restatable}{lemma}{LemmaKLs}
\label{lemma: Original KL and Sym KL}
 If $T(n) = \Theta(\frac{r(n)}{n})$ and $r(n) \leq n^{\frac{1}{3}}$, then for the island problem 
\begin{align*}
\text{D}_{\text{KL}}(\pi \mid \mid \mu) = \Omega\left( \text{D}_{\text{KL}}(\tPhi \mid \mid U_{\calS})  \right) .
\end{align*}
\end{restatable}
%
%
The proof of Lemma~\ref{lemma: Original KL and Sym KL} is given in Appendix~\ref{appdx:KL Bound for Lower Bound}.
We now compute $\text{D}_{\text{KL}}(\tPhi \mid \mid U_{\calS})$ exactly and establish \eqref{eqn: KL LB Restate}. First observe that since there are only two orderings at a given island (see Figure~\ref{fig:mainfig}), each occurring with equal probability under $U_{\calS}$, we have that $\Phi(\s^r) \in \{0,\lambda,\lambda^2,\ldots,\lambda^{r_I}\}$ and
\begin{align}
\label{eqn: Phi Probability}
\Prob_{U_{\calS}}(\Phi(\s^r) = \lambda^j) = \frac{1}{2^{r_I}} \binom{r_I}{j}.
\end{align}
Similarly, by the definition of $\tPhi$  
\begin{align}
\label{eqn:phi binom}
\Prob_{\tPhi}\big(\Phi(\s^r) = \lambda^j\big) = \binom{r_I}{j} \left(\frac{\lambda}{1+\lambda}\right)^j \left(\frac{1}{1+\lambda}\right)^{r_I - j}.
\end{align}
That is, the number of islands which are ordered as in Figure~\ref{fig:subfig2} is  binomially distributed under both the DSM and ISM models, with success probabilities $\lambda/(1+\lambda)$ and $1/2$, respectively. For $\lambda \in (1,\infty)$, these two binomial distributions diverge as $r_I$ grows. This is formalized in the following lemma: 
%
%
%
\begin{lemma}
\label{lemma:Problem Uniform Expectation LB}
For the island problem, there exists a constant $c(\lambda) > 0$ such that
\begin{align*}
\text{D}_{\text{KL}}(\tPhi \mid \mid U_{\calS}) > r(n) c(\lambda).
\end{align*}
\end{lemma}
\begin{proof}
By \eqref{eqn: Phi Probability} and the binomial theorem
\begin{align*}
\E_{U_{\calS}}[\Phi(\s^r)] =  \frac{1}{2^{r_I}}\sum^{r_I}_{j=0} \binom{r_I}{j}\lambda^j = \left( \frac{1+\lambda}{2}\right)^{r_I} .
\end{align*}
Next, we have again by \eqref{eqn: Phi Probability}
\begin{align*}
\text{D}_{\text{KL}}(\tPhi \mid \mid U_{\calS}) 
& 
= \E_{U_{\calS}}\left[ \frac{\tPhi(\s^r)}{U_{\calS}(\s^r)}\log\left(\frac{\tPhi(\s^r)}{U_{\calS}(\s^r)}\right) \right] \\
& 
= \E_{U_{\calS}}\left[ \log\left(\frac{\Phi(\s^r)}{\E_{U_{\calS}}[\Phi(\s^r)]}\right) \frac{\Phi(\s^r)}{\E_{U_{\calS}}[\Phi(\s^r)]}   \right] \\
&
= \sum^{r_I}_{i=0} \binom{r_I}{i}\log\left(\frac{\lambda^i}{\left( (1+\lambda)/2\right)^{r_I} }\right) \frac{\lambda^i}{(1+\lambda)^{r_I}} \\
&=\frac{\log(\lambda)}{(1+\lambda)^{r_I}}\sum^{r_I}_{i=0} i \lambda^i \binom{r_I}{i} - r_I\log((1+\lambda)/2).
\end{align*}
By the binomial theorem, we have $\sum^{r_I}_{i=0} \lambda^i \binom{r_I}{i} = (1+\lambda)^{r_I}$ and thus
\begin{align}
\label{eqn: Comb Identity}
    \sum^{r_I}_{i=0} i \lambda^i \binom{r_I}{i} = \lambda \cdot \frac{d}{d\lambda} (1+\lambda)^{r_I} = \lambda r_I (1+\lambda)^{r_I - 1}.
\end{align}
Finally, by \eqref{eqn: Comb Identity}:
\begin{align*}
\frac{\log(\lambda)}{(1+\lambda)^{r_I}}\sum^{r_I}_{i=0} i \lambda^i \binom{r_I}{i} - r_I\log\left(\frac{1+\lambda}{2}\right) = r_I\log(\lambda)\left[\frac{\lambda}{1+\lambda} -  \frac{\log\left(\frac{1+\lambda}{2}\right)}{\log(\lambda)} \right] > r c(\lambda)
\end{align*}
for some $c(\lambda) > 0$ since the function $\log(\lambda)(\frac{\lambda}{1+\lambda} -  \frac{\log\left(\frac{1+\lambda}{2}\right)}{\log(\lambda)})$ is strictly positive, continuous and monotone increasing in $\lambda \in (1,\infty)$.
\end{proof}
Theorem~\ref{thm: main thm IS LB} now follows by combining the upper bound on $\chi^2(\pi \mid \mid \mu)$ given in Theorem~\ref{thm:Main Thm Bound} with the lower bound on $\text{D}_{\text{KL}}(\pi \mid \mid \mu)$ given in Lemma~\ref{lemma: Original KL and Sym KL}.
\begin{proof}
(Theorem~\ref{thm: main thm IS LB}) First note that by Jensen's inequality
\begin{align*}
\text{D}_{\text{KL}}(\pi \mid \mid \mu) &= \int_{\scrP} \log(\pi(\Pa)/\mu(\Pa)) \pi(d\Pa) \\
&\leq \log\left(\int_{\scrP} (\pi(\Pa)/\mu(\Pa)) \pi(d\Pa)  \right) = \log(\chi^2(\pi \mid \mid \mu) +1).
\end{align*}
By assumption (Lemma~\ref{lemma: Original KL and Sym KL}), $T(n) = \Theta(r(n)/n)$ and $r(n) \leq n^{\frac{1}{3}}$, satisfying Assumption~\ref{assump: T assumption}. Consequently, by Theorem~\ref{thm:Main Thm Bound} there exists a constant $c > 0$ such that
\begin{align*}
\text{D}_{\text{KL}}(\pi \mid \mid \mu) \leq \log(\chi^2(\pi \mid \mid \mu) + 1 ) = \log(L^2(\pi,\mu)) \leq r(n) c = \bigO(r(n)).
\end{align*}
%
Conversely, Lemmas~\ref{lemma: Original KL and Sym KL} and \ref{lemma:Problem Uniform Expectation LB} imply that
\begin{align*}
\text{D}_{\text{KL}}(\pi \mid \mid \mu) = \Omega\left( \text{D}_{\text{KL}}(\tPhi \mid \mid U_{\calS})  \right) = \Omega(r(n)).
\end{align*}
Consequently, $\text{D}_{\text{KL}}(\pi \mid \mid \mu) = \Theta(r(n))$.
\end{proof}
\subsection{Model-Specific Improvements: the JC69+CpG Model}
\label{sec:Seq Bounds}
The bound on $\chi^2(\pi \mid \mid \mu)$ given in Theorem~\ref{thm:Main Thm Bound} holds for any pair of sequences under arbitrary context-dependence. However, in some cases the resulting bounds on $N^{\star}(\epsilon)$ can be impractically loose; here we show how the argument may be tightened by taking advantage of the context-dependency structure specific to a given DSM, yielding sharper bounds on $\chi^2(\pi \mid \mid \mu)$. We compare the bounds obtained with numerical results from simulation experiments.

\begin{figure}
\centering
\includegraphics[width=0.85\linewidth]{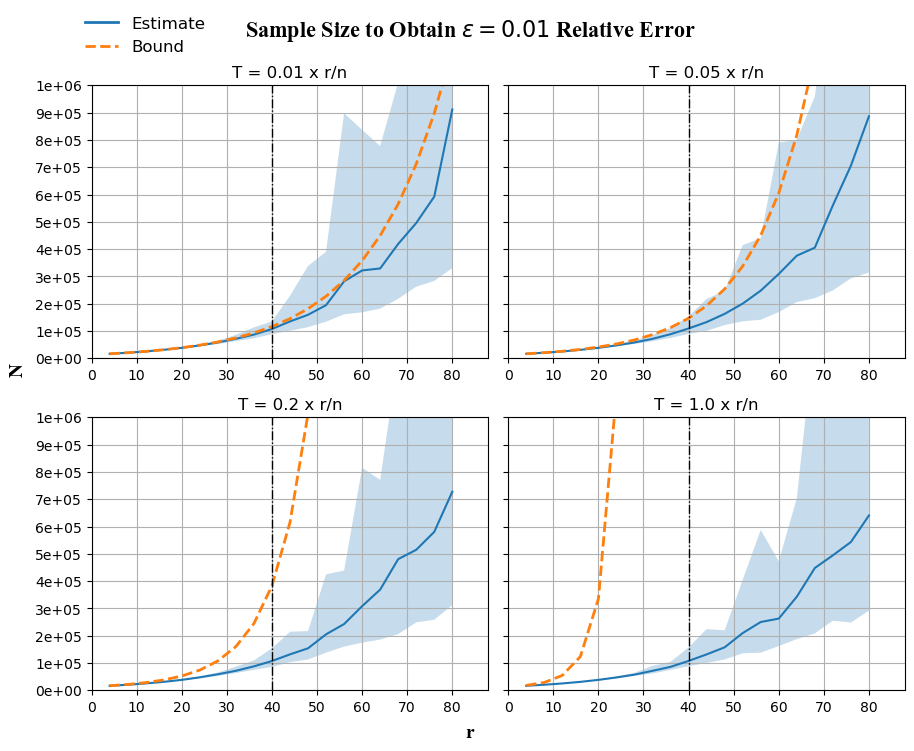}
\caption{Model-specific bounds (dashed lines) on $N^{\star}$ with $\epsilon = 0.01$ Chebychev's inequality using the  $L^2(\pi,\mu)$ bound of Proposition~\ref{prop:CG application}, for the JC69+CpG model with $n = 1600$ and $\lambda = 0.5$ and for various values of $r$ and $T$.  Solid lines and 95\% confidence intervals show sample sizes,  estimates obtained from the estimated squared coefficient of variation of $\hat{p}_{(T,\tQ)}(\y \mid \x)$ under importance sampling estimates (100 replicates using $N = 10000$ samples each). The vertical line marks $r = n^{\frac{1}{2}}$, where Assumption~\ref{assump: T assumption} breaks down.}
\label{Fig:NumericalResults}
\end{figure}

Consider the JC69+CpG model studied in \cite{Hobolth:2008}, where $\mu$ is given by the JC69 model:
%
\begin{align}
\label{eqn:CpG+JC69 main}
\tg_i(b; \tx_i) = \frac{1}{3}\lambda^{\ind_{\text{CG}}(x_{i - 1},x_i) + \ind_{\text{CG}}(x_i,x_{i + 1})},
\end{align}
so $(\Q_i)_{b,b} = - 1$ for all $i$ and all $b \in \scrA$. We now revisit the arguments in Section~\ref{sec:IS Results} to demonstrate how a tighter bound on $L^2(\pi,\mu)$ may be obtained for this model. We first show that constants appearing in the bounds in Lemmas~\ref{lemma:Denominator of chi square} and \ref{lemma:Numerator of chi square} can be improved:
\begin{proposition}
\label{prop:CG application}
Let $\theta_{\text{CG}} := \max(\lambda^{-2},\lambda^2) e^{4T\max(1-\lambda,\lambda-1)}$ and consider the JC69+CpG model \eqref{eqn:CpG+JC69 main}. Then
\begin{align*}
L^2(\pi, \mu) & \leq e^{8rT \max\{1-\lambda,\lambda-1\}  }\frac{\E_{\mu}[ \theta^{m(\Pa) - r}_{\text{CG}} ]}{p^2_r} L^2(\tPhi, U_{\calS} ).
\end{align*} 
\end{proposition}
The upper bound in Proposition~\ref{prop:CG application} can be significantly tighter than the bound in Theorem~\ref{thm:main thm general} (see \eqref{eqn:main thm general proof eq 1}). For example, $\theta_{\text{CG}}$ is much smaller than the quantity $\theta$ defined in Lemma~\ref{lemma:Numerator of chi square} since for the model \eqref{eqn:CpG+JC69 main} if $\lambda > 1$,
\begin{align*}
\theta =  \gamma^3_{\star}\phi^{2(k+1)}_{\star}\phi^2_{\max} e^{2T(\tilde{\delta} + 2\delta)} = \lambda^6 e^{12T(\lambda-1)  }.
\end{align*}
In Proposition~\ref{Prop:4} below we show how the upper bound in Proposition~\ref{prop:CG application} can be computed \textit{exactly} for this model and compare the bound to numerical estimates of $L^2(\pi,\mu)$.
\begin{proof}
Recall that $\delta = q(\gmax - \gmin)$ and $\tilde{\delta} = q(k+1)(\tilde{\gamma}_{\max}-\tilde{\gamma}_{\min})$ in Lemma~\ref{lemma: Uniform Bounds on Delta} are uniform bounds on the change in overall sequence mutation rate after a single substitution, under the ISM (JC69) and DSM (JC69+CpG) models, respectively. Under \eqref{eqn:CpG+JC69 main}, $\delta = 0$ since the JC69 rates are constant. A sharper $\tilde{\delta}$ can be obtained by noting that, in place of Lemma~\ref{lemma: Uniform Bounds on Delta}
\begin{align}
\label{eqn:Seq rate cpg+jc69 proof}
\Prob_{\mu}\left(|\tpsi(\s(\Pa),\bb(\Pa))| \leq  2m(\Pa)T \max\{1-\lambda,\lambda-1\}  \right) = 1, 
\end{align}
%
since the change in the number of CG pairs from $\x(t^j)$ to $\x(t^{j+1})$ is at most one (a single mutation can result in the net creation or removal of at most one CG pair). Thus the exit rate of at most two sites can change, leading to the uniform bound:
\begin{align*}
|\tilde{\gamma}(\cdot; \x(t^j)) - \tilde{\gamma}(\cdot; \x(t^{j+1}))| \leq 2\max\{1-\lambda,\lambda-1\}.
\end{align*}
Since this holds for ``extra" mutations as well, we also have
\begin{align}
\label{eqn:upper bound cpg+jc69 proof}
\Prob_{\mu}\left(\frac{\Phi(\s(\Pa), \bb(\Pa)) \ind_{\scrP_{[i]}}(\Pa)}{\Phi(\s^r_i)} \leq  \max\{\lambda^{-(m(\Pa) - r)},\lambda^{m(\Pa) - r}\} \ind_{\scrP_{[i]}}(\Pa)\right) = 1,
\end{align}
which provides an improvement on \eqref{eqn:uniform ratio bound}. 
Plugging \eqref{eqn:Seq rate cpg+jc69 proof} and \eqref{eqn:upper bound cpg+jc69 proof} into the proofs of Lemma~\ref{lemma:Denominator of chi square} and \ref{lemma:Numerator of chi square} in place of \eqref{eqn:Numerator of chi square overview} and \eqref{eqn:uniform ratio bound} yields the stated bound.
\end{proof}
\noindent The terms in the bound of Proposition~\ref{prop:CG application} can be computed exactly. By site independence we have
\begin{align}
\label{eqn:CG Example E and pr}
\E_{\mu}[\theta^{m(\Pa) - r}_{\text{CG}} ] = \frac{1}{\theta^r_{\text{CG}}} \prod^n_{i=1} \E_{\mu_i}[\theta^{m(\Pa_i)}_{\text{CG}} ] \quad\quad p_r = \prod_{i \in \calS} \Prob_{\mu_i}(m(\Pa_i) = 1) \prod_{i \notin \calS}\Prob_{\mu_i}(m(\Pa_i) = 0),
\end{align}
and under $\mu_i$, the distribution of $m(\Pa_i)$ is available in closed form:
\begin{align}
\label{eqn:m JC69}
\Prob_{\mu_i}(m(\Pa_i) = j) = \frac{e^{-T} \frac{T^j}{j!} \R^j_{x_i,y_i}}{p_{(T,\Q_i)}(y_i \mid x_i)},
\end{align}
%
where $\R$ is the jump chain corresponding to $\mu_i$ given by $\R_{b,b} = 0$ and $\R_{b,b^{\prime}} = 1/3$ for $b \neq b^{\prime}$. 
Hence, both quantities in \eqref{eqn:CG Example E and pr} can be computed exactly
using \eqref{eqn:m JC69}. Below, we consider the island problem of Section~\ref{sec: LB} and show how $ L^2(\tPhi, U_{\calS} )$ in Proposition~\ref{prop:CG application} can be computed exactly in this case. 
\begin{proposition}
\label{Prop:4}
For the island problem in Section~\ref{sec: LB}:
\begin{align*}
L^2(\tPhi, U_{\calS}) = \frac{2^{r_I} (1+\lambda^2)^{r_I} }{(1+\lambda)^{2r_I}}.
\end{align*}
\end{proposition}
\begin{proof}
Let $\C_i$ denote the set of sites lying in the context of site $i$.  Partition the set of observed mutations $\calS$ into $r_I$ \textit{islands} $\I_1,\ldots,\I_{r_I}$ defined as maximal subsets of observed mutation sites having overlapping contexts, so:
\begin{align*} 
\{\cup_{i \in \I_j}\C_i\} \cap  \{\cup_{i \in \I_{j^{\prime}}}\C_i\} = \emptyset \text{ for } j \neq j^{\prime}
\end{align*}
and let $r_j := |\I_j|$. For any path of length exactly $r$, the product of rates corresponding to the $r_j$ jumps in the $j$th island can take on at most $r_j!$ values $\Phi_{1,j},\ldots,\Phi_{r_j!,j}$. That is, we can write
\begin{align}
\label{eqn: L2 Island Rep}
L^2(\tPhi, U_{\calS})  = \frac{\frac{1}{r!}\sum_{\s^r \in \SymS}\Phi^2(\s^r) }{(\frac{1}{r!}\sum_{\s^r \in \SymS}\Phi(\s^r) )^2} =   \frac{r_1!\ldots r_{r_I}!\sum_{i_1,i_2,\ldots,i_{r_I}} \Phi^2_{i_1,1}\Phi^2_{i_2,2} \ldots \Phi^2_{i_{r_I},r_I} }{(\sum_{i_1,i_2,\ldots,i_{r_I}} \Phi_{i_1,1}\Phi_{i_2,2} \ldots \Phi_{i_{r_I},r_I}  )^2}
\end{align}
since there are $r_1!\ldots r_{r_I}!$ orderings of the mutations within each island and $\Phi(\s^r)$ is invariant over the $r! / r_1!\ldots r_{r_I}! = \binom{r}{r_1,\ldots,r_{r_I}}$ permutations that agree with the given island suborders. For the sequences $\x^{\star}$ and $\y^{\star}$ we have $r_j = 2$, $\Phi_{1,j} = 1$ and $\Phi_{2,j} = \lambda$ for all $j = 1,\ldots,r_I$ so the right-hand side simplifies: 
\begin{align*}
\frac{r_1!\ldots r_{r_I}!\sum_{i_1,i_2,\ldots,i_{r_I}} \Phi^2_{i_1,1}\Phi^2_{i_2,2} \ldots \Phi^2_{i_{r_I},r_I} }{(\sum_{i_1,i_2,\ldots,i_{r_I}} \Phi_{i_1,1}\Phi_{i_2,2} \ldots \Phi_{i_{r_I},r_I}  )^2} = \frac{2^{r_I}\prod^{r_I}_{i=1}(\Phi^2_{1,i} + \Phi^2_{2,i})}{\prod^{r_I}_{i=1}(\Phi_{1,i} + \Phi_{2,i})^2} = \frac{2^{r_I} (1+\lambda^2)^{r_I} }{(1+\lambda)^{2r_I}}.
\end{align*}
\end{proof}
%
Figure~\ref{Fig:NumericalResults} gives bounds (dashed lines) on $N^{\star}$ at a range of values of $T$ obtained by applying Chebychev's inequality for $\epsilon = 0.01$ using the upper bound on $L^2(\pi,\mu)$ given by combining Propositions~\ref{prop:CG application} and~\ref{Prop:4}, applied to the Island Problem (Definition~\ref{def: island problem}) with $n = 1600$ and $\lambda = 0.5$. Empirical estimates (solid lines) computed from the estimated relative variance of $\hat{p}_{(T,\tQ)}(\y \mid \x)$ under importance sampling are shown for comparison. Since 
$\epsilon^{-2} = 10000$, Chebychev's inequality requires $N \geq 10000$ even when $L^2(\pi,\mu) = 1$. We see that for $T$ small, our bound nearly matches the empirically observed values across the range of $r$ values shown, while as $T$ approaches $r/n$ our bound becomes quite loose. Observe that the empirical variances of the estimators grow exponentially beyond the $r = n^{\frac{1}{2}} = 40$ vertical line where Assumption~\ref{assump: T assumption} is violated, but remain quite small for $r \leq n^{\frac{1}{2}}$ despite our estimators being computed using $N = 10000$ instead of $N^{\star}$.
%

\section{Conclusion}
\label{sec:Conclusion} Calculation of marginal likelihoods under context-dependent evolutionary models is an important problem in phylogenetics and molecular evolution.  Given the extensive machinery available for independent site models, the idea of performing inference under DSMs by importance sampling from ISMs is an attractive one.   A potential drawback is the sample complexity of importance sampling, which often grows exponentially in the problem dimension.

Theorem~\ref{thm:Main Thm Bound} shows that as the length $n$ of the sequences $\x$ and $\y$ grows, the number of samples needed for the importance sampler to approximate the marginal sequence likelihood does not grow exponentially in the dimension $n$ of the space, but only in the number $r$ of observed mutations. Theorem~\ref{thm: main thm IS LB} however shows that in the worst case, $\bigO(e^r)$ samples are indeed \textit{required}. Our analysis in Section~\ref{sec:IS Results} shows that this exponential growth can be explained by the variability of the context-dependent rates across the \textit{orderings} of occurrence of these observed mutations when overlapping contexts occur. Section~\ref{sec:Seq Bounds} shows that the bound can be significantly sharpened and practical estimates on $N^{\star}$ may still be obtained for specific models (e.g. the CpG + JC69 model) using the techniques developed in  Section~\ref{sec: upper bound proof} to prove Theorem~\ref{thm:Main Thm Bound}. However, the practicality of the importance sampler will critically depend on whether the sites in $\calS$ have overlapping contexts; if \textit{all} observed mutations have non-overlapping contexts then $\bigO(\chi^2(\tPhi \mid \mid U_{\calS})) = \bigO(1)$, and the importance sampler provides a FPRAS for approximating the marginal sequence likelihood under Assumption~\ref{assump: T assumption}. Our results also suggest that identifying high probability orderings under $\tPhi$ may provide a way to to design better proposal distributions
for the importance sampler.


Finally, it is worth repeating that our results rely critically on the assumption that $T = \bigO(r/n)$
and $r^2 \leq n$
(Assumption~\ref{assump: T assumption}). The scaling assumption on $T$ was justified by the probability concentration tail bound on $T$ established in \citet{Mathews2:2025}, although that result differs from Assumption~\ref{assump: T assumption} by an unsatisfying additional factor of $\log(n)$. It is less clear whether it is possible to relax, or how to do so, the assumption that $r^2 \leq n$, as relaxing this assumption means that the $ \exp(rT c + (n-r)T^2 c^{\prime})$ term that appears in our bounds is no longer of constant order.


\bibliography{Bibliography.bib}

\bibliographystyle{imsart-nameyear.bst}

\appendix

\input{REVISION_CLEAN_AppendixB}
\input{REVISION_CLEAN_AppendixC}
\input{AppendixE}

\end{document}

%% file: REVISION_CLEAN_AppendixB.tex
\section{Results For Importance Sampling}
\label{Appdx:Importance Sampling Results}
\subsection{Proofs of Lemma~\ref{lemma: Uniform Bounds on Delta} and Lemma~\ref{lemma:uniform bound w}}\label{Appdx: weight bound}
%
\LemmaOne*
%
\begin{proof}
For the first claim, consider a path of length $m$, $\Pa^m$, and let $j \in \{1,\ldots,m\}$. Recall that exactly one mutation occurs at each $t^j\in \Pa$, occurring at site $s^j$, so all exit rates outside the context $\tilde{x}_{s^j}$  are constant over $t\in [t^{j-1},t^j)$. Since the exit rates of at most $k+1$ sites are affected, with each rate changing by at most $q(\tilde{\gamma}_{\max} - \tilde{\gamma}_{\min})$ in absolute value, we obtain 
\begin{align*}
\Prob_{\mu}\left(|\tpsi(\s(\Pa),\bb(\Pa))| \leq  m(\Pa)T \tilde{\delta}\right) = \Prob_{\mu}\Big(\big|\sum^{m(\Pa)}_{j=1} t^j \Delta^{\tg}(j) \big|\leq  m(\Pa)T \tilde{\delta}\Big) = 1,
\end{align*}
where we used $t^j \leq T$. This establishes the first stated bound. The second claim follows from the first by setting $k = 0$ and $\phi \equiv 1$. The third claim follows from the first two by the triangle inequality.
\end{proof}
\LemmaTwo*
\begin{proof}
Recall
\begin{align*}
w_0(\Pa) =
\prod^{m(\Pa)}_{j=1} \phi(b^j; \tilde{x}^{j-1}_{s^j}) 
\me^{\sum^{m(\Pa)}_{j=1} t^j (\tilde{\Delta}(j) - \Delta(j))} = \Phi(\s(\Pa),\bb(\Pa))\me^{\sum^{m(\Pa)}_{j=1} t^j (\tilde{\Delta}(j) - \Delta(j))} .
\end{align*}
%
%
By Lemma~\ref{lemma: Uniform Bounds on Delta} and the triangle inequality
\begin{align}
\label{eqn:Seq Mutation Rate Bound}
\Prob_{\mu}\Biggl(\biggl|\sum^{m(\Pa)}_{j=1} t^j (\tilde{\Delta}(j) - \Delta(j))\biggr| \leq m(\Pa)T  (\tilde{\delta} + \delta)\Biggr) = 1,
\end{align}
where we used that $t^j \leq T$. Hence,
\begin{align*}
\Prob_{\mu}\left(\Phi(\s(\Pa),\bb(\Pa)) e^{-m(\Pa)T  (\tilde{\delta} + \delta)} \leq w_0(\Pa) \leq \Phi(\s(\Pa),\bb(\Pa))e^{m(\Pa) T  (\tilde{\delta} + \delta)} \right) = 1.
\end{align*}
\end{proof}

\subsection{Proof of Lemma~\ref{lemma:expectation and pr bound}}\label{appdx: ISM expectation and pr bound}
The goal of this section is to derive an upper bound on expectations of the form $\E_{\mu}[\theta^{m(\Pa)}]$, where $\theta \in \mathbb{R}$ is a constant, $\mu$ is the joint distribution over paths under the ISM \eqref{Eqn:ISM}, and $m(\Pa)$ is the number of mutations in path $\Pa$, and to derive a lower bound on $p_r = \Prob_{\mu}(m(\Pa) = r)$. We first restate relevant notation from Section~\ref{sec: background} and provide an overview of our approach. We then state the main result of this section in Lemmas~\ref{lemma:appdx ex bound}~and~\ref{lemma:appdx pr bound}; Lemma~\ref{lemma:expectation and pr bound} follows as an immediate consequence. 

Recall that the path at the $i$th site is denoted by
\begin{align*}
\Pa_i = (m_i,t^1_i,\ldots,t_i^{m_i},b^1_i,\ldots,b^{m_i}_i).
\end{align*}
We let $\scrP^l_i \ni \Pa_i$ denote the set of all length $l$ paths at site $i$ and $\scrP_i = \cup^{\infty}_{l=0}\scrP^l_i$ all such paths. Under the independence assumption, the joint density of a path $\Pa$ factors as
\begin{align*}
\mu(\Pa \mid \x,\y) = \mu_1(\Pa_1\mid x_1,y_1) \times \ldots \times \mu_n(\Pa_n\mid x_n,y_n),
\end{align*}
where $\mu_i(\Pa_i \mid x_i,y_i) =   \tP_{(T,\Q_i)}(y_i,\Pa_i \mid x_i) /p_{(T,\Q_i)}(y_i \mid x_i)$ with
\begin{align}
\tP_{(T,\Q_i)}(y_i,\Pa_i \mid x_i) 
& = \left[ \prod^{m_i}_{j=1} \gamma_i(b^j_i; b^{j-1}_i) \right] \me^{\sum^{m_i}_{j=1}t^j_i (\gamma_i(\cdot; b^j_i) - \gamma_i(\cdot; b^{j-1}_i ))}
\me^{-T\gamma_i(\cdot; y_i)},
\label{eqn:appdx joint density}
\end{align}
if $x_i(T) = y_i$ and zero otherwise, and 
\begin{align}
\label{eqn:appdx marginalization}
p_{(T,\Q_i)}(y_i \mid x_i)  = (e^{T\Q_i})_{x_i,y_i}
= \int_{\scrP_i}\tP_{(T,\Q_i)}(y_i,\Pa_i \mid x_i) \nu_i(d\Pa_i)
\end{align}
%
is the transition probability from $x_i$ to $y_i$ at time $T$
obtained by marginalizing over all paths. For brevity, let $\bb^l_i = (b^1_i,\ldots,b^l_i)$ and $\bt^l_i = (t^1_i,\ldots,t^l_i)$ denote arbitrary length $l$ vectors of base changes and jump times, respectively. Let
\begin{align*}
\mathscr{T}^l = \{ \bt^l: 0 <t^1 < \ldots < t^l < T\}
\end{align*}
be the set of all $l$ jump time vectors that satisfy the ordering constraint, and 
\begin{align*}
\mathscr{B}^l_i = \{ \bb^l_i: b^l_i = y_i \text{ and } b_i^j\neq b_i^{j-1} \text{ for } j=1,\ldots,l\}
\end{align*}
be the set of valid base change vectors satisfying the endpoint constraints, where we let $b^{0}_i=x_i$. Below it will at times be useful to expand the integration over $\scrP_i$ using $\scrP_i = \cup_{l=0}^{\infty} \scrP^l_i$ into the following equivalent forms:
\begin{align}
\int_{\scrP_i}\theta^{m(\Pa_i)}
\tP_{(T,\Q_i)}(y_i,\Pa_i \mid x_i) \nu_i(d\Pa_i) &= \sum_{l=0}^{\infty} \theta^l \int_{\scrP^l_i}  
\tP_{(T,\Q_i)}(y_i,\Pa_i \mid x_i) \nu_i(d\Pa_i) \nonumber \\
\label{Eqn:ExpandedPathIntegral2}
&= \sum_{l=0}^{\infty}\theta^l 
\sum_{\bb^l_i \in \mathscr{B}^l_i} \int_{\mathscr{T}^l }
\tP_{(T,\Q_i)}(y_i,\Pa_i = (l, \bb^l_i,\bt^l_i) \mid x_i) d\bt_i^l \\
\label{Eqn:ExpandedPathIntegral3}
&= \sum_{l=0}^{\infty} \theta^l p^l_i,
\end{align}
where 
\begin{align}
\label{Eqn:Marginal Probability of L jumps}
p^l_i := \int_{\scrP^l_i} \tP_{(T,\Q_i)}(y_i,\Pa_i \mid x_i)\nu_i(d\Pa_i)
\end{align}
denotes the probability that $x_i(T) = y_i$ for a length $l$ path. Note that choosing $\theta = 1$ implies
\begin{align}
\label{Eqn:ExpandedPathIntegral4}
p_{(T,\Q_i)}(y_i \mid x_i)  &= \sum_{l=0}^{\infty}  
\sum_{\bb^l_i \in \mathscr{B}^l_i} \int_{\mathscr{T}^l }
\tP_{(T,\Q_i)}(y_i,\Pa_i = (l, \bb^l_i,\bt^l_i) \mid x_i) d\bt_i^l = \sum_{l=0}^{\infty}  p^l_i.
\end{align}
%
%
%
%
%
We will first derive an upper bound on
\begin{align}
\label{eqn:exp appdx approach}
\E_{\mu_i}[\theta^{m(\Pa_i)}] = \frac{\sum_{l=0}^{\infty} \theta^l p^l_i}{\sum_{l=0}^{\infty} p^l_i},
\end{align}
and lower bounds on
%
\begin{align}
\label{eqn:pr appdx approach 1}
\Prob_{\mu_i}(m(\Pa_i) = 1) &= \frac{p^1_i}{\sum^{\infty}_{l=1} p^l_i } \quad\quad \text{ for }x_i \neq y_i
\\
\Prob_{\mu_i}(m(\Pa_i) = 0) &= \frac{p^{0}_i}{\sum^{\infty}_{l=0} p^l_i } \quad\quad \text{ for }x_i = y_i,
\label{eqn:pr appdx approach 2}
\end{align}
where the equalities follow by the representations \eqref{Eqn:ExpandedPathIntegral3} and \eqref{Eqn:ExpandedPathIntegral4}. By appealing to site independence under $\mu$, we will then establish Lemma~\ref{lemma:expectation and pr bound}. 
%

%

Our first result establishes a lower bound on $p^1_i$ and $p^{0}_i$ appearing in \eqref{eqn:pr appdx approach 1} and \eqref{eqn:pr appdx approach 2}, respectively. Note that this 
also provides a lower bound on $p_{(T,\Q_i)}(y_i \mid x_i)$ since for any $j \in \{0,1,2,\ldots\}$ we have by \eqref{Eqn:ExpandedPathIntegral4}
\begin{align}\label{eqn:appdx approx bound transition}
p_{(T,\Q_i)}(y_i \mid x_i) = \sum_{l=0}^{\infty} p^l_i   \geq p^j_i.
\end{align}
In particular, if $m(\Pa_i)$ concentrates around $d_{\text{H}}(x_i,y_i)$ under $\mu$,  then choosing $j = d_{\text{H}}(x_i,y_i)$ in the inequality above will yield a sharp lower bound on $\sum_{l=0}^{\infty} p^l_i$.

\begin{lemma}\label{lemma:appdx p1 and po lower bound}
If $x_i = y_i$, then $p^{0}_i = e^{-T\gamma(\cdot; x_i)} \geq e^{-Tq\gamma_{\max}}$. 
%
%
If $x_i \neq y_i$, then
\begin{align*}
p^1_i &= \begin{cases}
\frac{\gamma(y_i; x_i)\big(e^{-\gamma(\cdot; y_i)T} - \me^{- \gamma(\cdot; x_i)T}\big)}{\gamma(\cdot; x_i) - \gamma(\cdot; y_i)} &\text{ if } \gamma(\cdot; x_i) \neq \gamma(\cdot; y_i)  \\
T\gamma(y_i; x_i) e^{-T\gamma(\cdot; y_i)} &\text{ if } \gamma(\cdot; x_i) =\gamma(\cdot; y_i)
\end{cases}
\quad 
\Biggr\} \;
\geq \; T\gamma_{\min}e^{-Tq\gamma_{\max}} .
\end{align*}
%
\end{lemma}
\begin{proof}
The first statement follows directly from \eqref{eqn:appdx joint density} and \eqref{Eqn:Marginal Probability of L jumps} by setting $m = 0$, the second by direct integration.
%
%
%
%
%
%
%
%
\end{proof}
As mentioned, the sharpness of these bounds depends on how concentrated $m(\Pa_i)$ is around $d_{\text{H}}(x_i,y_i)$. The following result bounds the probability that $m(\Pa_i)$ exceeds $l \in \{1,2,\ldots\}$ under $\mu$.
%
%
\begin{lemma}
\label{lemma:appdx pl upper bound}
Let $l \in \{1,2,\ldots \}$. Then
\begin{align*}
p_i^l \leq   e^{-q T \gamma_{\min}} \frac{(q T  \gamma_{\max} )^l}{l!} .
\end{align*}
\end{lemma}

%
\begin{proof}
Again using \eqref{eqn:appdx joint density}  and \eqref{Eqn:Marginal Probability of L jumps} note that $\prod^l_{j=1} \gamma_i(b^j_i; b^{j-1}_i) \leq \gamma^l_{\max}$ and
\begin{align*}
\me^{-\gamma_i(\cdot; y_i) (T - t^l_i)}\me^{-\sum^l_{j=1}(t^j_i - t^{j-1}_i) \gamma_i(\cdot; b^{j-1}_i)} &\leq e^{-q \gamma_{\min}\left( \sum^l_{j=1}(t^j_i - t^{j-1}_i) + (T - t^l_i) \right)} = e^{-q T \gamma_{\min}},
\end{align*}
providing the uniform upper bound $\tP_{(T,\Q_i)}(y_i,\Pa_i \mid x_i) \leq e^{-q T \gamma_{\min}} \gamma^l_{\max}$ for all $\Pa_i \in \scrP^l_i$. A simple counting argument gives $|\mathscr{B}^l_i| \leq q^l$, and using the representation \eqref{Eqn:ExpandedPathIntegral2} along with \eqref{Eqn:Marginal Probability of L jumps}, we obtain
\begin{align*}
p^l_i \leq e^{-q T \gamma_{\min}} \gamma^l_{\max} \int_{\scrP^l_i}  \nu_i(d\Pa_i) &= e^{-q T \gamma_{\min}}  \gamma_{\max}^l  \sum_{\mathscr{B}_i^l}  \int_{\mathscr{T}^l} 
d\bt^l_i \\
&=  e^{-q T \gamma_{\min}} \frac{(\gamma_{\max}T)^l}{l!} \sum_{\mathscr{B}_i^l} 1  \\
&\leq  e^{-q T \gamma_{\min}} \frac{(\gamma_{\max} q T)^l}{l!},
\end{align*} 
where the second equality follows by the identity $\int_{0 < t^1_i < \ldots < t^l_i < T}  dt^1_i \ldots t^l_i = \frac{T^l}{l!}$.
\end{proof}
\noindent The following result summarizes Lemma~\ref{lemma:appdx p1 and po lower bound} and Lemma~\ref{lemma:appdx pl upper bound}. For brevity we let
\begin{align*}
\lambda_0(\theta) := 
\gamma^2_{\max} \theta^2 q^2 e^{Tq((\theta+1)\gamma_{\max} - \gamma_{\min})}.
\end{align*}
\begin{lemma}
\label{lemma:appdx den and num combo}
Let $\theta \in \mathbb{R}$. Then
\begin{align*}
\sum^{\infty}_{l=2} \frac{\theta^lp^l_i}{p^{0}_i} \leq T^2\lambda_0(\theta) \;\; \text{ for } \ x_i = y_i  \quad\quad \sum^{\infty}_{l=2} \frac{\theta^lp^l_i}{p^1_i} \leq T\gamma^{-1}_{\min} \lambda_0(\theta) \;\; \text{ for } \ x_i \neq y_i.
\end{align*}
\end{lemma}
\begin{proof}
Applying Lemma~\ref{lemma:appdx p1 and po lower bound} and Lemma~\ref{lemma:appdx pl upper bound} we obtain for the  $x_i = y_i$ case
\begin{align*}
\sum^{\infty}_{l=2} \frac{\theta^lp^l_i}{p^{0}_i}  &\leq e^{Tq(\gamma_{\max} - \gamma_{\min})}  \sum^{\infty}_{l=2} \frac{(\gamma_{\max} \theta q T)^l}{l!} \leq T^2 \,\cdot \, \lambda_{0}(\theta).
\end{align*}
Similarly,
 \begin{align*}
\sum^{\infty}_{l=2} \frac{\theta^lp^l_i}{p^1_i} &\leq T^{-1}\gamma^{-1}_{\min} e^{Tq(\gamma_{\max} - \gamma_{\min})} \sum^{\infty}_{l=2} \frac{(\gamma_{\max}\theta q T)^l}{l!} \leq  T\gamma_{\min}^{-1}\,\cdot\, \lambda_0(\theta).
\end{align*} 
\end{proof}
\noindent The next result will be used to prove the first statement of Lemma~\ref{lemma:expectation and pr bound}.

\begin{lemma}\label{lemma:appdx ex bound}
For $\theta \in \mathbb{R}$, 
\begin{align*}
\E_{\mu}[\theta^{m(\Pa)}] \leq \theta^r\left(1 + T(\theta\gamma_{\min})^{-1} \lambda_0(\theta)\right)^r
\left( 1 + T^2 \lambda_0(\theta)\right)^{n-r}. 
\end{align*} 
\end{lemma}

\begin{proof}
First, consider a site without an observed mutation (where $x_i = y_i$). Then $p^1_i = 0$ and we have by \eqref{eqn:exp appdx approach} and Lemma~\ref{lemma:appdx den and num combo}  
\begin{align}
\label{eqn:appdx ub exp proof 1}
\E_{\mu_i}[\theta^{m(\Pa_i)}]  &\leq \frac{p^0_i + \sum^{\infty}_{l=2} \theta^l p^l_i  }{p^0_i}  \leq 1 + T^2 \lambda_0(\theta).
\end{align}
Similarly, for a site with an observed mutation (where $x_i \neq y_i$), $p^0_i = 0$ and again by \eqref{eqn:exp appdx approach} and Lemma~\ref{lemma:appdx den and num combo}
\begin{align}
\label{eqn:appdx ub exp proof 2}
\E_{\mu_i}[\theta^{m(\Pa_i)}]  \leq \frac{\theta p^1_i + \sum^{\infty}_{l=2} \theta^l p^l_i  }{p^1_i} &\leq \theta\left(1 +  T(\theta\gamma_{\min})^{-1} \lambda_0(\theta) \right).
\end{align}
Now by site independence:
\begin{align}
\label{eqn:exp site independence appdx}
\E_{\mu}[\theta^{m(\Pa)}] = \prod^n_{i=1} \E_{\mu_i}[\theta^{m(\Pa_i)}] =  \prod_{i \notin \calS} \E_{\mu_i}[\theta^{m(\Pa_i)}] \prod_{i \in \calS} \E_{\mu_i}[\theta^{m(\Pa_i)}] .
\end{align}
Recall that $\x$ and $\y$ are sequences of length $n$ with $\text{d}_{\text{H}}(\x,\y) = r$. Applying \eqref{eqn:appdx ub exp proof 1} and \eqref{eqn:appdx ub exp proof 2} to \eqref{eqn:exp site independence appdx} yields the stated bound.
\end{proof}

\noindent The next result follows similarly and will be used to prove the second statement of Lemma~\ref{lemma:expectation and pr bound}.
\begin{lemma}
\label{lemma:appdx pr bound}
The probability $p_r$ of exactly $r$ mutations under $\mu$ satisfies:
\begin{align*}
p_r \geq \left(\frac{1}{1 + T(\gamma_{\min})^{-1} \lambda_0(1)}\right)^r 
\left(\frac{1}{1 +  T^2 \lambda_0(1)} \right)^{n-r}.
\end{align*}
%
\end{lemma}
\begin{proof}
Note that the event $\{\Pa : m(\Pa)=r\}$ occurs when exactly one mutation occurs in each of the observed mutation sites $\calS$, and no others. Consider a site $i \not\in \calS$, so $x_i = y_i$. Then $p^1_i = 0$ and we again have by \eqref{eqn:pr appdx approach 2} and application of Lemma~\ref{lemma:appdx den and num combo} with $\theta = 1$
\begin{align}
\label{eqn:appdx lb pr proof 1}
\Prob_{\mu_i}(m(\Pa_i) = 0) 
&= \frac{1}{1 + \sum^{\infty}_{l=2} \frac{p^l_i}{p^0_i} } \geq \frac{1}{1 + T^2 \lambda_0(1) }.
\end{align}
Similarly, consider a  site $i \in \calS$ with observed mutation (so $x_i \neq y_i$). Then $p^{0}_i = 0$ and we have by \eqref{eqn:pr appdx approach 1} and application of Lemma~\ref{lemma:appdx den and num combo} with $\theta = 1$
\begin{align}
\label{eqn:appdx lb pr proof 2}
\Prob_{\mu_i}(m(\Pa_i) = 1)  = \frac{1}{1 + \sum^{\infty}_{l=2} \frac{p^l_i}{p^1_i} } \geq \frac{1}{1 +  T(\gamma_{\min})^{-1} \lambda_0(1)}.
\end{align}
%
By site independence
\begin{align}
\label{eqn:pr site independence appdx}
p_r = \Prob_{\mu}(m(\Pa) = r) = \prod_{i \notin \calS}\Prob_{\mu_i}(m(\Pa_i) = 0) \prod_{i \in \calS}\Prob_{\mu_i}(m(\Pa_i) = 1) .
\end{align}
Applying \eqref{eqn:appdx lb pr proof 1} and \eqref{eqn:appdx lb pr proof 2} to \eqref{eqn:pr site independence appdx} gives the stated bound.
\end{proof}
\noindent The proof of Lemma~\ref{lemma:expectation and pr bound} now follows directly by Lemma~\ref{lemma:appdx ex bound} and Lemma~\ref{lemma:appdx pr bound}.

\begin{proof}(Lemma~\ref{lemma:expectation and pr bound})
Let $c_1 = (\gamma^2_{\max}/\gamma_{\min})  q^2 e^{Tq(\gamma_{\max} - \gamma_{\min})}$. We obtain by Lemma~\ref{lemma:appdx ex bound} and the inequality $1 + x \leq e^{x}$
\begin{align*}
\E_{\mu}[\theta^{m(\Pa)}] &\leq \theta^r\left(1 + \theta Tc_1 \exp(Tq\theta \gamma_{\max} ) \right)^r\left( 1 + \theta^2T^2c_1 \gamma_{\min} \exp(Tq\theta \gamma_{\max})  \right)^{n-r} \\
&\leq \theta^r\exp\left(rT\theta c_1 \exp(Tq\theta \gamma_{\max}) + (n-r)T^2\theta^2 c_1\gamma_{\min} \exp(Tq\theta \gamma_{\max}) \right) .
\end{align*}
%
Similarly, letting
\begin{align*}
    c_2 = \frac{\gamma^2_{\max}}{\gamma_{\min}} q^2 e^{Tq(2\gamma_{\max} - \gamma_{\min})} \quad\quad c_3 = \gamma_{\min} c_2,
\end{align*}
we have by Lemma~\ref{lemma:appdx pr bound}
\begin{align*}
    p_r \geq \left( \frac{1}{1+T c_2} \right)^r\left( \frac{1}{1+T^2 c_3} \right)^{n-r} \geq \exp\left(-(rTc_2 + (n-r)T^2c_3) \right) .
\end{align*}
\end{proof}

%% file: REVISION_CLEAN_AppendixC.tex
\section{Importance Sampling Lower Bounds}
\label{Appdx:Importance Sampling Lower Bound}
%

%
\subsection{Lower Bound On $\pi(m(\Pa) = r \mid \x, \y)$}

\subsubsection{Overview and Notation}
We will first prove a lower bound on the probability of exactly $d_{\text{H}}(\x,\y) = r$ jumps occurring under the \textit{dependent site} model for sequences $\x$, $\y$ of length $n$, denoted
\begin{align}
\tp_r := \pi(m(\Pa) = r \mid \x, \y).
\label{Eqn:TildePr}
\end{align}
Specifically, we will show that there exists a constant $c > 0$, independent of $n$ and $r$, such that
\begin{align*}
\tp_r \geq \exp\left(-c \cdot (rT - (n-r)T^2 ) \right).
\end{align*}
 Our approach will be analogous to that used to obtain a lower bound on $p_r$
in Appendix~\ref{appdx: ISM expectation and pr bound}, but whereas those results were under the ISM $\mu$, here we do so with respect to the distribution of $\Pa_i$ (recall $\Pa_i$ denotes the path at site $i$ \eqref{eqn: ith path def}) under the DSM conditional on $\x$, $\y$, \textit{and} $\Pa_{-i}$ , where
\begin{align*}
\Pa_{-i} = (\Pa_1,\ldots,\Pa_{i-1},\Pa_{i+1},\ldots,\Pa_n).
\end{align*}
Inequalities similar to those in Appendix~\ref{appdx: ISM expectation and pr bound} will apply, but now hold almost surely as the resulting bounds will be functions of $\Pa_{-i}$. 

Let $\C_i$ denote the set of sites lying in the context of site $i$ (including site $i$) and let $\C^{\prime}_i =  \C_i \setminus \{i \}$. Write the number of jumps occurring among the paths $\{\Pa_{j}: j \in \C_i \}$ at the sites in $\C_i$ as $\bar{m}_i := m(\Pa_{\C_i}) =  m(\Pa_i) + \bar{m}_i^{\prime}$, where
\begin{align*}
\bar{m}_i^{\prime} := \sum_{j \in \C^{\prime}_i}m(\Pa_j),
\end{align*}
is the total number of jumps occurring among the paths at sites in $\C_i^{\prime}$. Denote the ordered times of the $\bar{m}_i$ jumps at sites in $\C_i$  by
\begin{align*}
\bar{t}_i^0 = 0 < \bar{t}_i^1 < \ldots < \bar{t}_i^{\bar{m}_i} < T,
\end{align*}
and let $\bar{b}^j_l = x_l(\bar{t}^j_i)$ denote the value of the $l$th site at time $\bar{t}_i^l$, and
$\bar{\Delta}^t_i(j) := \bar{t}_i^{j+1} - \bar{t}_i^j$ the time interval between successive jumps, with $\bar{\Delta}_i^t(\bar{m}_i) := T - \bar{t}_i^{\bar{m}_i}$. 
The joint density of $\y$ and $\Pa_i$ conditional on $\x$ \textit{and} $\Pa_{-i}$ is given by
\begin{align}
\label{eqn: FC joint density}
\tP_{(T,\tQ)}(\y, \Pa_i \mid \Pa_{-i}, \x) &=  \prod_{z \in \C_i} \prod^{\bar{m}_i}_{j=1}\left[ \left\{\tilde{\gamma}_z(\bar{b}^j_z; \tilde{x}^{j-1}_z) \right\}^{\ind(\bar{b}^j_z \neq \bar{b}^{j-1}_z)} e^{-\bar{\Delta}^t_i(j) \tilde{\gamma}_z(\cdot; \tilde{x}^{j-1}_z)}\right] \\
&\times e^{-\bar{\Delta}^t_i(\bar{m}_i) \tilde{\gamma}_z(\cdot; \tilde{y}_z )} \ind_{\x^m = \y}(\Pa) \nonumber,
\end{align}
where $\tilde{y}_z = \{y_i:  i \in \C_z \}$ is the context of site $z$ in the terminal sequence $\y$. The full conditional distribution of $\Pa_i$ under $\pi$ is 
\begin{align}
\label{eqn: EC FC distribution}
\pi(\Pa_i \mid \x, \y, \Pa_{-i}) = \frac{\tP_{(T,\tQ)}(\y, \Pa_i \mid \Pa_{-i}, \x)}{\int_{\scrP_i} \tP_{(T,\tQ)}(\y, \Pa_i \mid \Pa_{-i}, \x) \nu(d\Pa)}.
\end{align}
As in Appendix~\ref{appdx: ISM expectation and pr bound}, we define the marginal probability of $l$ jumps occurring at site $i$ resulting in $x_i(T) = y_i$, conditional on $x_i$ and $\Pa_{-i}$, by
\begin{align}
\label{eqn:marginal probability full conditional}
\tp^l_i  &:=  \sum_{\bb^l_i \in \mathscr{B}^l_i}\int_{\mathscr{T}^l} \tP_{(T,\tQ)}(\y, \Pa_i = (l,\bt^l_i, \bb^l_i) \mid \Pa_{-i}, \x)d\bt^l_i \quad \text{ for }\;  l \geq 1, 
\end{align}
and $\tp^0_i = \tP_{(T,\tQ)}(\y, \Pa_i \mid \Pa_{-i}, \x) \ind_{m_{i}(\Pa_i) = 0}(\Pa_i)$. Note that $\tp^l_i$, viewed as a function of $\Pa_{-i}$, is a random variable. Consequently, inequalities involving $\tp^l_i$ hold almost surely under $\pi$; we will not make this distinction explicit for brevity. 

\subsubsection{Proof of $\tp_r$ Bound}

We first derive uniform bounds on the joint density function $ \tP_{(T,\tQ)}(\y, \Pa_i \mid \Pa_{-i}, \x)$ given in \eqref{eqn: FC joint density}. Recall from Lemma~\ref{lemma: Uniform Bounds on Delta} the quantity $\tilde{\delta} := q(k+1)(\tgmax - \tgmin)$ corresponding to a uniform bound on the change in the sequence mutation rate $\Delta^{\tg}(j; \Pa) =  \tg(\cdot; \x^j) - \tg(\cdot; \x^{j-1})$ under the DSM.
\begin{lemma}
\label{lemma:FC Density Bound}
\begin{align*}
\Prob_{\pi}(\tilde{\gamma}^{\bar{m}_i}_{\min} e^{-T( \tilde{\delta}  \bar{m}_i + \sum_{j \in \C_i}\tilde{\gamma}_j(\cdot; \tilde{y}_j) )} \leq  \tP_{(T,\tQ)}(\y, \Pa_i \mid \Pa_{-i}, \x) &\leq  \tilde{\gamma}^{\bar{m}_i}_{\max}  e^{T( \tilde{\delta}\bar{m}_i - \sum_{j \in \C_i}\tilde{\gamma}_j(\cdot; \tilde{y}_j))}) = 1.
\end{align*}
\end{lemma}
\begin{proof}
For any $z \in \C_i$, note that
\begin{align*}
\tilde{\gamma}^{m(\Pa_z)}_{\min} \leq \prod^{\bar{m}_i}_{j=1} \left\{\tilde{\gamma}_z(\bar{b}^j_z; \tilde{x}^{j-1}_z ) \right\}^{\ind(\bar{b}^j_z \neq \bar{b}^{j-1}_z)} \leq \tilde{\gamma}^{m(\Pa_z)}_{\max}.
\end{align*}
In addition, the change in overall sequence mutation rate $|\tilde{\gamma}_z(\cdot; \tilde{x}^j_z ) - \tilde{\gamma}_z(\cdot; \tilde{x}^{j-1}_z)| \leq q( \tilde{\gamma}_{\max} - \tilde{\gamma}_{\min}) $ and so
%
\begin{align*}
-T \tilde{\gamma}_z(\cdot; \tilde{y}_z) + \sum^{\bar{m}_i}_{j=1}\bar{t}^j_i(\tilde{\gamma}_z(\cdot; \tilde{x}^j_z ) - \tilde{\gamma}_z(\cdot; \tilde{x}^{j-1}_z )) &\leq  T q \bar{m}_i  (\tilde{\gamma}_{\max} - \tilde{\gamma}_{\min}) -  T \tilde{\gamma}_z(\cdot; \tilde{y}_z).
\end{align*}
Consequently, for any $z \in \C_i$ rewriting the exponential term $e^{-\sum^{\bar{m}_{i}}_{j=1}\bar{\Delta}^t_i(j) \tilde{\gamma}_z\left(\cdot; \tilde{x}^{j-1}_z \right)}$ of the density \eqref{eqn: FC joint density} as a difference between sequence rates (see \eqref{eqn: delta t to t conversion}) gives:
\begin{align*}
&\prod^{\bar{m}_i}_{j=1}
\left[ 
\left\{\tilde{\gamma}_z(\bar{b}^j_z; \tilde{x}^{j-1}_z) \right\}^{\ind(\bar{b}^j_z \neq \bar{b}^{j-1}_z)} 
e^{-\bar{\Delta}^t_i(j) \tilde{\gamma}_z\left(\cdot; \tilde{x}^{j-1}_z \right)}\right] 
e^{-\bar{\Delta}^t_i(\bar{m}_i) \tilde{\gamma}_z(\cdot; \tilde{y}_z )}   \\
\leq &\tilde{\gamma}^{m(\Pa_z)}_{\max} e^{T q \bar{m}_i  (\tilde{\gamma}_{\max} - \tilde{\gamma}_{\min}) -  T \tilde{\gamma}_z(\cdot; \tilde{y}_z)}.
\end{align*}
Taking a product over all $z \in \C_i$ and recalling that $|\C_i| = k+1$ yields  
\begin{align*}
\tP_{(T,\tQ)}(\y, \Pa_i \mid \Pa_{-i}, \x) &\leq e^{-T \sum_{z \in \C_i}\tilde{\gamma}_z(\cdot; \tilde{y}_z)}\tilde{\gamma}^{\bar{m}_i}_{\max} e^{Tq (k+1)  \bar{m}_i(\tilde{\gamma}_{\max}-\tilde{\gamma}_{\min})} \\
&= \tilde{\gamma}^{\bar{m}_i}_{\max}  e^{T( \tilde{\delta}\bar{m}_i - \sum_{z \in \C_i}\tilde{\gamma}_z(\cdot; \tilde{y}_z))}.
\end{align*}
The lower bound is proved similarly. 
\end{proof}
The next result is the analog of Lemma~\ref{lemma:appdx p1 and po lower bound}, lower bounding the probability of zero or one jumps at site $i$, under the full conditional distribution of $\Pa_i$ under $\pi$. Recall that $\bar{m}_i^{\prime} = \sum_{j \in \C^{\prime}_i}m(\Pa_j)$ denotes the number of jumps along the paths at sites in $\C^{\prime}_i = \C_i \setminus \{i\}$, and see \eqref{eqn:marginal probability full conditional} for the definitions of $\tp^0_i$ and $\tp^1_i$. 
\begin{lemma}
\label{lemma:pl bound FC}
If $x_i \neq y_i$,
\begin{align*}
\tp^1_i \geq T \tilde{\gamma}^{\bar{m}^{\prime}_i + 1 }_{\min} e^{-T( \tilde{\delta}  (\bar{m}^{\prime}_i + 1) + \sum_{z \in \C_i}\tilde{\gamma}_z(\cdot; \tilde{y}_z) )}.
\end{align*}
If $x_i = y_i$,
\begin{align*}
\tp^0_i &\geq \tilde{\gamma}^{\bar{m}^{\prime}_i  }_{\min} e^{-T( \tilde{\delta}  \bar{m}^{\prime}_i + \sum_{z \in \C_i}\tilde{\gamma}_z(\cdot; \tilde{y}_z) )}.
\end{align*}
\end{lemma}
\begin{proof}
Consider a mutated site, so $x_i \neq y_i$. Then $|\mathscr{B}^1_i| = 1$ in \eqref{eqn:marginal probability full conditional}, and recalling  $\bar{m}_i = m(\Pa_i) + \bar{m}_i^{\prime}$, we have by Lemma~\ref{lemma:FC Density Bound},
\begin{align*}
\tp^1_i &\geq \tilde{\gamma}^{\bar{m}^{\prime}_i + 1 }_{\min} e^{-T( \tilde{\delta}  (\bar{m}^{\prime}_i + 1) + \sum_{z \in \C_i}\tilde{\gamma}_z(\cdot; \tilde{y}_z) )}  \int^{T}_0 1 \, dt^1_i = T \tilde{\gamma}^{\bar{m}^{\prime}_i + 1 }_{\min} e^{-T( \tilde{\delta}  (\bar{m}^{\prime}_i + 1) + \sum_{z \in \C_i}\tilde{\gamma}_z(\cdot; \tilde{y}_z) )}.
\end{align*}
Similarly, if $x_i = y_i$ we obtain by Lemma~\ref{lemma:FC Density Bound}
%
\begin{align*}
\tp^0_i &\geq \tilde{\gamma}^{\bar{m}^{\prime}_i  }_{\min} e^{-T( \tilde{\delta}  \bar{m}^{\prime}_i + \sum_{z \in \C_i}\tilde{\gamma}_z(\cdot; \tilde{y}_z) )}.
\end{align*}
\end{proof}
\noindent The next result, which also follows by Lemma~\ref{lemma:FC Density Bound}, bounds the probability that the number of jumps at site $i$ exceeds $\text{d}_{\text{H}}(x_i,y_i)$
under the full conditional distribution of $\Pa_i$. To simplify notation, we define the random variable
\begin{align*}
\upsilon(\bar{m}^{\prime}_i) := e^{2 T \tilde{\delta}  \bar{m}^{\prime}_i} (\tilde{\gamma}_{\max}/\tilde{\gamma}_{\min})^{\bar{m}^{\prime}_i}.
\end{align*}
The proof of the following result is again very similar to Lemma~\ref{lemma:appdx pl upper bound} but holds for the full conditional distribution of $\Pa_i$ under $\pi$.
\begin{lemma}
Let $l > d_{\text{H}}(x_i,y_i)$. If $x_i \neq y_i$,
\begin{align*}
\frac{\tp^l_i}{\tp^1_i} &\leq 
\frac{\upsilon(\bar{m}^{\prime}_i)  e^{T \tilde{\delta} } }{T \tilde{\gamma}_{\min}}  \frac{(T q \tilde{\gamma}_{\max} e^{T \tilde{\delta}} )^l}{l!}.
\end{align*}
If $x_i = y_i$,
\begin{align*}
\frac{\tp^l_i}{\tp^0_i} &\leq \upsilon(\bar{m}^{\prime}_i)  \frac{(T q \tilde{\gamma}_{\max} e^{T\tilde{\delta}} )^l}{l!} .
\end{align*}
\end{lemma}
\begin{proof}
If $\Pa_i \in \scrP^l_i$, then $\bar{m}_i = l + \bar{m}^{\prime}_i$. Applying Lemma~\ref{lemma:FC Density Bound} then gives
\begin{align*}
\tp^l_i &\leq  e^{T(\tilde{\delta} \cdot ( \bar{m}^{\prime}_i+l) -\sum_{z \in \C_i}\tilde{\gamma}_z(\cdot; \tilde{y}_z) ) } \tilde{\gamma}^{( \bar{m}^{\prime}_i+ l)}_{\max}\sum_{\bb^l_i \in \mathscr{B}^l_i}\int_{\mathscr{T}^l} 1 \, d\bt^l_i \\
&\leq e^{T(\tilde{\delta}   \bar{m}^{\prime}_i  -\sum_{z \in \C_i}\tilde{\gamma}_z(\cdot; \tilde{y}_z) ) } 
\tilde{\gamma}^{\bar{m}^{\prime}_i}_{\max}  \frac{(T q \tilde{\gamma}_{\max} e^{T\tilde{\delta}} )^l}{l!}.
\end{align*}
%
Recalling that $\upsilon(\bar{m}^{\prime}_i) = e^{2 T \tilde{\delta}  \bar{m}^{\prime}_i}(\tilde{\gamma}_{\max}/\tilde{\gamma}_{\min})^{\bar{m}^{\prime}_i}$, for an mutated site ($x_i \neq y_i$), by Lemma~\ref{lemma:pl bound FC},
\begin{align*}
\frac{\tp^l_i}{\tp^1_i} \leq
\frac{\upsilon(\bar{m}^{\prime}_i) e^{T \tilde{\delta}} }{T \tilde{\gamma}_{\min}}  \frac{(T q \tilde{\gamma}_{\max} e^{T \tilde{\delta}} )^l}{l!}.
\end{align*}
Similarly, for an unmutated site ($x_i = y_i$) by Lemma~\ref{lemma:pl bound FC} 
\begin{align*}
\frac{\tp^l_i}{\tp^0_i} &\leq \upsilon(\bar{m}^{\prime}_i)\frac{(T q \tilde{\gamma}_{\max} e^{T\tilde{\delta}} )^l}{l!}.
\end{align*}
\end{proof}
The next result is the penultimate step to completing the proof of the lower bound on $\tp_r$, providing the analogs of equations \eqref{eqn:appdx lb pr proof 1} and \eqref{eqn:appdx lb pr proof 2} in the proof of the lower bound for $p_r$ under the ISM $\mu$ in Appendix~\ref{Appdx:Importance Sampling Results}, for the conditional of $\Pa_i$ under the DSM $\pi$.
%
\begin{lemma}
\label{lemma:DSM FC Bounds}
Let
\begin{align*}
c = q^2 \tilde{\gamma}^2_{\max} e^{T ( q\tilde{\gamma}_{\max} e^{T\tilde{\delta}} + 2 \tilde{\delta} ) } \max\{1,\tilde{\gamma}^{-1}_{\min}\}. 
\end{align*}
Then for $x_i \neq y_i$
%
\begin{align*}
\Prob_{\pi}(m(\Pa_i) = 1 \mid  \Pa_{-i}) \geq \exp\left(-T \upsilon(\bar{m}^{\prime}_i)c \right).
\end{align*}
Similarly, for $x_i = y_i$,
\begin{align*}
\Prob_{\pi}(m(\Pa_i) = 0 \mid  \Pa_{-i}) \geq \exp\left(-T^2 \upsilon(\bar{m}^{\prime}_i)c \right).
\end{align*}
\end{lemma}
\begin{proof}
Suppose $x_i \neq y_i$ and write
\begin{align*}
\Prob_{\pi}(m(\Pa_i) = 1 \mid  \Pa_{-i}) = \frac{\tp^1_i}{\sum^{\infty}_{l=0} \tp^l_i} &= \frac{1}{1 + \sum^{\infty}_{l=2} \frac{\tp^l_i}{\tp^1_i}} .
\end{align*}
Let $c_1 := \left(\tilde{\gamma}^2_{\max}/\tilde{\gamma}_{\min}\right) q^2 e^{T (q \tilde{\gamma}_{\max} e^{T\tilde{\delta}} + 2 \tilde{\delta})}$.
We first have
\begin{align*}
\sum^{\infty}_{l=2} \frac{\tp^l_i}{\tp^1_i} < \frac{ \upsilon(\bar{m}^{\prime}_i) e^{T \tilde{\delta}} }{T \tilde{\gamma}_{\min}} \sum^{\infty}_{l=2}  
\frac{(T q \tilde{\gamma}_{\max} e^{T\tilde{\delta}} )^l}{l!} &\leq T \upsilon(\bar{m}^{\prime}_i) \left(\tilde{\gamma}^2_{\max}/\tilde{\gamma}_{\min}\right) q^2 e^{T (q \tilde{\gamma}_{\max} e^{T\tilde{\delta}} + 2 \tilde{\delta})} \\
&= T \upsilon(\bar{m}^{\prime}_i)c_1 .
\end{align*}
Hence,
\begin{align*}
\Prob_{\pi}(m(\Pa_i) = 1 \mid  \Pa_{-i})   \geq \frac{1}{1+T \upsilon(\bar{m}^{\prime}_i)c_1} \geq \exp\left(-T \upsilon(\bar{m}^{\prime}_i)c_1 \right).
\end{align*}
Now suppose $x_i = y_i$. Then
\begin{align*}
\sum^{\infty}_{l=2} \frac{\tp^l_i}{\tp^0_i} \leq \upsilon(\bar{m}^{\prime}_i)(T q \tilde{\gamma}_{\max} e^{T\tilde{\delta}})^2 \sum^{\infty}_{l=2} \frac{(T q \tilde{\gamma}_{\max} e^{T\tilde{\delta}} )^{l-2}}{l!} &\leq T^2 \upsilon(\bar{m}^{\prime}_i)q^2 \tilde{\gamma}^2_{\max} e^{T ( q\tilde{\gamma}_{\max} e^{T\tilde{\delta}} + 2 \tilde{\delta} ) } \\
&= T^2\upsilon(\bar{m}^{\prime}_i)\tilde{\gamma}_{\min} c_1.
\end{align*}
By the same argument, $\Prob_{\pi}(m(\Pa_i) = 0 \mid  \Pa_{-i}) \geq \exp\left(-T^2 \upsilon(\bar{m}^{\prime}_i)\tilde{\gamma}_{\min} c_1 \right)$. Setting $c = c_1 \cdot \max\{1 ,\tilde{\gamma}_{\min}\}$ yields the stated claim. 
\end{proof}
Finally, we now prove the desired lower bound on $\tp_r$, by repeatedly applying the tower rule (law of iterated expectation) and applying  Lemma~\ref{lemma:DSM FC Bounds}. Observe that for any length $r$ path, $\bar{m}^{\prime}_i \leq k$ with probability one under $\pi$ and so the following bound holds with probability one as well:
\begin{align}
\upsilon(\bar{m}^{\prime}_i) = e^{2 T \tilde{\delta}  \bar{m}^{\prime}_i} (\tilde{\gamma}_{\max}/\tilde{\gamma}_{\min})^{\bar{m}^{\prime}_i} \leq e^{2 T \tilde{\delta}k} (\tilde{\gamma}_{\max}/\tilde{\gamma}_{\min})^k =\upsilon(k) := \upsilon_k \qquad \text{ for } \quad \Pa \in \scrP^r.
\label{Eqn:LengthRPhiBound}
\end{align}
%
We will use this observation to establish the following result:
\begin{lemma}
\label{lemma: pr bound DSM}
%
We have the following lower bound:
\begin{align*}
\tp_r \geq \exp\left(- c \upsilon_k \cdot (rT - (n-r)T^2) \right).
\end{align*}
\end{lemma}
\begin{proof}
Recall that $\x$ and $\y$ are sequences of length $n$ with $\text{d}_{\text{H}}(\x,\y) = r$ and $\calS$ denotes the set of observed mutated sites. Write
%
\begin{align*}
\Prob_{\pi}(m(\Pa) = r) = \E_{\pi}\left[\prod^r_{i \in \calS} \ind_{\{m(\Pa_{i}) = 1 \}}(\Pa_i)    \prod^n_{i \notin \calS}\ind_{\{m(\Pa_i) = 0 \}}(\Pa_i) \right].
\end{align*}
Recall the constant defined in Lemma~\ref{lemma:DSM FC Bounds}
\begin{align*}
c = q^2 \tilde{\gamma}^2_{\max} e^{T ( q\tilde{\gamma}_{\max} e^{T\tilde{\delta}} + 2 \tilde{\delta} ) } \max\{1,\tilde{\gamma}^{-1}_{\min}\}.
\end{align*}
For brevity, write $\ind_{\{m(\Pa_i) = 0 \}}(\Pa_i) = \ind_0(\Pa_i)$ and $\ind_{\{m(\Pa_i) = 1 \}}(\Pa_i) = \ind_1(\Pa_i)$. Let $\ell_i \in \calS$ for $i = 1,\ldots, r$ denote the site index for the $i$th mutated site. Using the tower rule and Lemma~\ref{lemma:DSM FC Bounds} we have
\begin{align*}
\E_{\pi}\left[\prod^r_{i \in \calS} \ind_1(\Pa_i)    \prod^{n}_{i \notin \calS}\ind_0(\Pa_i) \right] &= \E_{\pi}\left[ \Prob_{\pi}(m(\Pa_{\ell_1}) = 1 \mid \Pa_{[-\ell_1]} ) \prod^r_{i=2} \ind_1(\Pa_{\ell_i})    \prod^n_{i \notin \calS}\ind_0(\Pa_i)  \right] \\
&\geq \E_{\pi}\left[ e^{-T \upsilon(\bar{m}^{\prime}_i)  c } \prod^r_{i=2} \ind_1(\Pa_{\ell_i})    \prod^n_{i \notin \calS}\ind_0(\Pa_i) \right]\\
& \geq e^{-T\upsilon_kc} \E_{\pi}\left[\prod^r_{i=2} \ind_1(\Pa_{\ell_i}) \prod^n_{i \notin \calS}\ind_0(\Pa_i) \right],
\end{align*}
%
where the last inequality comes from \eqref{Eqn:LengthRPhiBound}.
Repeatedly iterating the expectation and applying Lemma~\ref{lemma:DSM FC Bounds} yields
\begin{align*}
\Prob_{\pi}(m(\Pa) = r) &\geq e^{-rTc\upsilon_k} \E_{\pi}\left[\prod^n_{i \notin \calS}\ind_0(\Pa_i)\right] \geq \exp\left(-c \upsilon_k  \cdot  ( rT  + (n-r)T^2)  \right).
\end{align*}
\end{proof}

\subsection{Proof of Lemma~\ref{lemma: Original KL and Sym KL}}
\label{appdx:KL Bound for Lower Bound}
This section provides a proof of Lemma~\ref{lemma: Original KL and Sym KL}, which we restate here for convenience. Recall that we define for $\s^r \in \text{Sym}(\calS)$ (the symmetric group on $\calS$)
\begin{align*}
\Phi(\s^r) :=  \prod^r_{l=1}\phi_{s_l}(y_{s^l}; \tilde{x}^{l-1}_{s^l} ) \quad\quad \tilde{\Phi}(\s^{r}) := \frac{\Phi(\s^r)}{\sum_{\s^r \in \text{Sym}(\calS)}\Phi(\s^r)},
\end{align*}
and $U_{\calS}$ denotes the uniform measure over $\text{Sym}(\calS)$.
\LemmaKLs*
The proof of Lemma~\ref{lemma: Original KL and Sym KL} proceeds in two steps. The first step is to show that under the settings of Lemma~\ref{lemma: Original KL and Sym KL}
\begin{align}
\label{eqn:KL LB Overview 1}
\text{D}_{\text{KL}}(\pi \mid \mid \mu) = \Omega(\text{D}_{\text{KL}}(\pi_{\mid r} \mid \mid \mu_{\mid r})),
\end{align}
where the restrictions of $\pi$ and $\mu$ to the set of length $r$ paths are defined by
\begin{alignat*}{2}
& \pi_{\mid r}(\Pa) & := \frac{\pi(\Pa)\ind_{\scrP^r}(\Pa)}{\pi(\scrP^r)} &  =   
\frac{\tP_{(T,\tQ)}(\y, \Pa \mid \x)\ind_{ \{m(\Pa) = r \}}(\Pa)}{\int_{\scrP^r}\tP_{(T,\tQ)}(\y, \Pa \mid \x) \nu(d\Pa)} 
\\ \text{ and } \qquad\qquad\qquad\qquad 
& \mu_{\mid r}(\Pa) & := 
\frac{\mu(\Pa)\ind_{\scrP^r}(\Pa)}{\mu(\scrP^r)} &  = 
\frac{\tP_{(T,\Q)}(\y, \Pa \mid \x)\ind_{ \{m(\Pa) = r\} }(\Pa)}{\int_{\scrP^r}\tP_{(T,\Q)}(\y, \Pa \mid \x) \nu(d\Pa)}.
\end{alignat*}  
The next step is to show
\begin{align}
\label{eqn:KL LB Overview 3}
\text{D}_{\text{KL}}(\pi_{\mid r} \mid \mid \mu_{\mid r})  = \Omega(\text{D}_{\text{KL}}(\tilde{\Phi} \mid \mid U_{\calS})),
\end{align}
which will complete the proof of Lemma~\ref{lemma: Original KL and Sym KL} as \eqref{eqn:KL LB Overview 1} and \eqref{eqn:KL LB Overview 3} imply $\text{D}_{\text{KL}}(\pi \mid \mid \mu) = \Omega(\text{D}_{\text{KL}}(\tilde{\Phi} \mid \mid U_{\calS}))$.

\subsubsection{Supporting Results}
As previously mentioned, the first step is to establish \eqref{eqn:KL LB Overview 1}. To do so, we write
\begin{align}
\label{eqn:KL LB Overview 2}
\text{D}_{\text{KL}}(\pi \mid \mid \mu) = \underbrace{\int_{\scrP^r } \log\left(\frac{\pi(\Pa)}{\mu(\Pa)} \right)\pi(d\Pa) }_{(i)}
+ \underbrace{\int_{(\scrP^r)^{c} } \log\left(\frac{\pi(\Pa)}{\mu(\Pa)} \right)\pi(d\Pa)}_{(ii)}.
\end{align}
We first show that $(ii)$ is lower bounded by a constant (independent of $n$) in Lemma~\ref{lemma: KL Residual Bound}. We then show in Lemma~\ref{lemma: KL Lower Bound by Restriction} that $(i)$ is at least $\tp_r\text{D}_{\text{KL}}(\pi_{\mid r} \mid \mid \mu_{\mid r}) +  \tp_r\log(\tp_r )$ for $\tp_r$ given in \eqref{Eqn:TildePr}. Finally, the lower bound on $\tp_r$ obtained in Lemma~\ref{lemma: pr bound DSM} will allow us to establish \eqref{eqn:KL LB Overview 1}, which we state formally in Lemma~\ref{lemma: KL Lower Bound by Restriction Summary}. To show Lemma~\ref{lemma: KL Residual Bound}, we use the following result. 

\begin{lemma}
\label{lemma: Density Ratios IS LB}
Under the setting of Lemma~\ref{lemma: Original KL and Sym KL}, there exists a constant $c > 0$, independent of $n$ and $r$, such that
\begin{align*}
\frac{\pi(\Pa)}{\mu(\Pa)} \geq \lambda^{-r} \exp\left(-T(3n(\lambda - 1) + rc )  - (n-r)T^2c \right) .
\end{align*}
\end{lemma}
\begin{proof}
First note that since $\phi_{\min} = 1$ and $\gamma \equiv 1$, $\tilde{\gamma}(\cdot; \x^j) \geq 3n = \gamma(\cdot; \x^j)$ for any $j \in \{1,\ldots,m(\Pa)\}$ (recall $q = 3$ here since $\mathscr{A} = \{\tA, \tG, \tC, \tT$\}). Therefore for any length $m$ path
\begin{align}
\label{eqn:LB proof one eq 1}
e^{-\sum^m_{j=1}(t^j - t^{j-1}) \tilde{\gamma}(\cdot; \x^{j-1})} e^{-(T - t^{m}) \tilde{\gamma}(\cdot; \y)} \leq e^{-3Tn} = e^{-\sum^m_{j=1}(t^j - t^{j-1})\gamma(\cdot; \x^{j-1})} e^{-(T - t^{m}) \gamma(\cdot; \y)}.
\end{align}
%
Recalling that under the ISM $\gamma(b;x_i) = 1$ for all $b \in \mathscr{A} \setminus \{x_i\}$ and $x_i \in \mathscr{A}$, we then obtain by combining with \eqref{eqn:LB proof one eq 1}
\begin{align}
\label{eqn:LB proof one eq 2}
\tP_{(T,\tQ)}(\y, \Pa \mid \x) &=  \prod^{m(\Pa)}_{j=1} \left[\phi(b^j; \tilde{x}^{j-1}_{s_j}) e^{-(t^j - t^{j-1}) \tilde{\gamma}(\cdot; \x^{j-1})} \right] e^{-(T - t^{m(\Pa)}) \tilde{\gamma}(\cdot; \y)} \\
&\leq \lambda^{m(\Pa)} \tP_{(T,\Q)}(\y , \Pa \mid \x).
\end{align}
In addition, observe that the whole-sequence rate increase between the DSM and ISM is bounded by $\tilde{\gamma}(\cdot; \x^j) - \gamma(\cdot; \x^j) \leq 3n(\lambda - 1) $. Arguing as in \eqref{eqn:LB proof one eq 1} and using $\phi_{\min} = 1$, we see that the following lower bound holds as well:
\begin{align}
\label{eqn:LB proof one eq 3}
\tP_{(T,\tQ)}(\y, \Pa \mid \x) \geq \tP_{(T,\Q)}(\y , \Pa \mid \x) e^{-3Tn  (\lambda-1)}.
\end{align}
Using the upper bound on $\tP_{(T,\tQ)}(\y, \Pa \mid \x)$ obtained in \eqref{eqn:LB proof one eq 2} and applying Lemma~\ref{lemma:appdx ex bound}, we see that since $\gamma_{\max} = \gamma_{\min} = 1$ and $\lambda > 1$ we have for $c = (\lambda q)^{2}e^{Tq\lambda}$ 
\begin{align*}
\frac{p_{(T,\Q)}(\y \mid \x)}{p_{(T,\tQ)}(\y \mid \x)}  =
\frac{\int \tP_{(T,\Q)}(\y, \Pa \mid \x) \nu(d\Pa)}{\int \tP_{(T,\tQ)}(\y, \Pa \mid \x) \nu(d\Pa)} 
&\geq \frac{\int \tP_{(T,\Q)}(\y, \Pa \mid \x) \nu(d\Pa)}{\int \lambda^{m(\Pa)} \tP_{(T,\Q)}(\y, \Pa \mid \x) \nu(d\Pa) } \\
&=\frac{1}{\E_{\mu}[\lambda^{m(\Pa)}]} \\
&\geq \lambda^{-r} \exp\left(-c \cdot (rT + (n-r)T^2) \right).
\end{align*}
%
The stated bound follows by \eqref{eqn:LB proof one eq 3}:
\begin{align*}
\frac{\pi(\Pa)}{\mu(\Pa)} := \frac{\pi(\Pa \mid \x,\y)}{\mu(\Pa \mid \x,\y)} 
&= \frac{\tP_{(T,\tQ)}(\y, \Pa \mid \x) }{\tP_{(T,\Q)}(\y , \Pa \mid \x)}  \frac{p_{(T,\Q)}(\y \mid \x)}{p_{(T,\tQ)}(\y \mid \x)} \\
&\geq \lambda^{-r} \exp\left(-T(3n (\lambda - 1) + rc )  - (n-r)T^2c \right). 
\end{align*}
\end{proof}
The next lemma shows that the second term in \eqref{eqn:KL LB Overview 2} can be lower bounded by a constant that does not depend on $n$.
\begin{lemma}
\label{lemma: KL Residual Bound}
Under the setting of Lemma~\ref{lemma: Original KL and Sym KL}, there exist a constant $\alpha < 0$ such that
\begin{align*}
\int_{(\scrP^r)^{c} } \log\left(\frac{\pi(\Pa)}{\mu(\Pa)} \right)\pi(d\Pa)  > \alpha.
\end{align*}
\end{lemma}
\begin{proof}
Applying Lemma~\ref{lemma: Density Ratios IS LB} yields that for some $c > 0$
\begin{align}
\label{eqn:LB proof two eq 1}
\int_{(\scrP^r)^{c} } \log\left(\frac{\pi(\Pa)}{\mu(\Pa)} \right)\pi(d\Pa)  \geq \left[r\log(\lambda) + T(3n(\lambda - 1) + rc ) +  (n-r)T^2c\right] (\tp_r-1).
\end{align}
By Lemma~\ref{lemma: pr bound DSM}, there exists a constant $c^{\prime} > 0$ (see Lemma~\ref{lemma: pr bound DSM} for the explicit value) such that
\begin{align*}
\tp_r \geq \exp\left(-c^{\prime} \cdot (rT - (n-r) T^2) \right) \geq 1 - c^{\prime} \cdot (rT  - (n-r) T^2).
\end{align*}
Since $T = \bigO(r/n)$ and $r \leq n^{\frac{1}{3}}$ under 
the assumptions of Lemma~\ref{lemma: Original KL and Sym KL}, note that
\begin{align*}
    \max\{r^2 T, rn T^2, r^2 T^2, rn T^3, n^2 T^3, n^2 T^4 \} = \bigO(1).
\end{align*}
%
Hence, the following quantity is $\bigO(1)$ and there exists a constant $\alpha < 0$ sufficiently small such that
\begin{align}
\label{eqn:LB proof two eq 2}
\left[r\log(\lambda) + T(3n(\lambda - 1) + rc ) +   (n-r)T^2c\right] (\tp_r-1) > \alpha.
\end{align}
The stated bound follows by combining \eqref{eqn:LB proof two eq 1} and \eqref{eqn:LB proof two eq 2}.
\end{proof}
Having established a lower bound on $(ii)$ in \eqref{eqn:KL LB Overview 2}, we now turn to $(i)$ in \eqref{eqn:KL LB Overview 2} and connect this term to $\text{D}_{\text{KL}}(\pi_{\mid r} \mid \mid \mu_{\mid r}) $.
\begin{lemma}
\label{lemma: KL Lower Bound by Restriction}
We have the following bound
\begin{align*}
\int_{\scrP^r} \log\left(\frac{\pi(\Pa)}{\mu(\Pa)} \right)\pi(d\Pa)  \geq  \tp_r\text{D}_{\text{KL}}(\pi_{\mid r} \mid \mid \mu_{\mid r}) +  \tp_r \log(\tp_r ).
\end{align*}
\end{lemma}
\begin{proof}
Recall from \eqref{Eqn:Pr} and \eqref{Eqn:TildePr} that 
$$p_r = \mu(m(\Pa) = r \mid \x, \y) = \frac{\int_{\scrP^r}\tP_{(T,\Q)}(\y, \Pa \mid \x) \nu(d\Pa)}{\int_{\scrP}\tP_{(T,\Q)}(\y, \Pa \mid \x) \nu(d\Pa)} = 
\frac{\int_{\scrP^r}\tP_{(T,\Q)}(\y, \Pa \mid \x) \nu(d\Pa)}{p_{(T,\Q)}(\y \mid \x)},
$$
and similarly $\tp_r = (\int_{\scrP^r}\tP_{(T,\tQ)}(\y, \Pa \mid \x) \nu(d\Pa))/p_{(T,\tQ)}(\y \mid \x)$.
 Write
\begin{align*}
\pi(\Pa)
= \frac{\tP_{(T,\tQ)}(\y, \Pa \mid \x)}{\tp_r} \frac{\tp_r}{p_{(T,\tQ)}(\y \mid \x)} 
= \frac{ \tp_r \tP_{(T,\tQ)}(\y, \Pa \mid \x)}{\int_{\scrP^r}\tP_{(T,\tQ)}(\y, \Pa \mid \x) \nu(d\Pa)} .
\end{align*}
In addition, we have for any $\Pa \in \scrP$
\begin{align*}
\mu(\Pa)  = \frac{\tP_{(T,\Q)}(\y, \Pa \mid \x)}{\int_{\scrP}\tP_{(T,\Q)}(\y, \Pa \mid \x) \nu(d\Pa)} \leq \frac{  \tP_{(T,\Q)}(\y, \Pa \mid \x)}{\int_{\scrP^r}\tP_{(T,\Q)}(\y, \Pa \mid \x) \nu(d\Pa)} .
\end{align*}
Consequently,
\begin{align*}
\int_{\scrP^r} \log\left(\frac{\pi(\Pa)}{\mu(\Pa)} \right)\pi(d\Pa) 
&\geq \int_{\scrP^r} \log\left(\frac{\tp_r \pi_{\mid r}(\Pa) }{\mu_{\mid r}(\Pa)}\right)\pi(d\Pa) \\
&= \tp_r \int_{\scrP^r} \log\left(\frac{ \tp_r \pi_{\mid r}(\Pa) }{\mu_{\mid r}(\Pa)}\right)\pi_{\mid r}(d\Pa)   \\
& = 
\tp_r\text{D}_{\text{KL}}(\pi_{\mid r} \mid \mid \mu_{\mid r}) +  \tp_r\log(\tp_r ).
\end{align*}
%
\end{proof}
The next lemma combines Lemmas~\ref{lemma: pr bound DSM},  \ref{lemma: KL Residual Bound}, and \ref{lemma: KL Lower Bound by Restriction} to summarize the results we have obtained so far and states \eqref{eqn:KL LB Overview 1} formally.
\begin{lemma}
\label{lemma: KL Lower Bound by Restriction Summary}
Under the setting of Lemma~\ref{lemma: Original KL and Sym KL},
\begin{align*}
\text{D}_{\text{KL}}(\pi \mid \mid \mu) = \Omega\left(\text{D}_{\text{KL}}(\pi_{\mid r} \mid \mid \mu_{\mid r}) \right).
\end{align*}
\end{lemma}
\begin{proof}
By \eqref{eqn:KL LB Overview 2},  Lemma~\ref{lemma: KL Residual Bound}, and \ref{lemma: KL Lower Bound by Restriction}, there exists a constant $\alpha \in \mathbb{R}$ such that
%
\begin{equation*}
\text{D}_{\text{KL}}(\pi \mid \mid \mu) 
\geq \int_{\scrP^r } \log\left(\frac{\pi(\Pa)}{\mu(\Pa)} \right)\pi(d\Pa) 
+ \alpha
\geq \tp_r\text{D}_{\text{KL}}(\pi_{\mid r} \mid \mid \mu_{\mid r}) + \tp_r\log(\tp_r ) + \alpha.
\end{equation*}
Next, by Lemma~\ref{lemma: pr bound DSM} there exists a constant $c'= c \upsilon_k$ and hence $c^{\prime\prime}$ such that
\begin{align*}
\tp_r \geq \exp(-c^{\prime} \cdot (rT + (n-r)T^2)) \geq \exp(-c^{\prime \prime} )
\end{align*}
since $rT = \bigO(1)$ and $(n-r)T^2 = \bigO(1)$ by assumption, and so
\begin{equation*}
\text{D}_{\text{KL}}(\pi \mid \mid \mu) 
= \Omega\left(\text{D}_{\text{KL}}(\pi_{\mid r} \mid \mid \mu_{\mid r}) \right).
\end{equation*}
\end{proof}
We now obtain a lower bound on $\text{D}_{\text{KL}}(\pi_{\mid r} \mid \mid \mu_{\mid r})$. We do so by finding uniform lower and upper bounds on the restricted densities $\pi_{\mid r}(\Pa^r)$ and $\mu_{\mid r}(\Pa^r)$, respectively. We first establish uniform bounds the joint likelihoods $\tP_{(T,\tQ)}(\y,\Pa^r \mid \x)$ and $\tP_{(T,\Q)}(\y,\Pa^r \mid \x)$ for any length $r$ path $\Pa^r$.
\begin{lemma}
\label{lemma: Length r Density Bound}
\begin{align*}
e^{-rT\tilde{\delta}}\me^{-T\tilde{\gamma}(\cdot; \y)} \Phi(\s(\Pa^r))  \leq  &\tP_{(T,\tQ)}(\y,\Pa^r \mid \x) \leq  e^{rT\tilde{\delta}} \me^{-T\tilde{\gamma}(\cdot; \y)} \Phi(\s(\Pa^r))  \\
e^{-rT\delta}\me^{-T\gamma(\cdot; \y)}  \leq  &\tP_{(T,\Q)}(\y,\Pa^r \mid \x) \leq  e^{rT\delta} \me^{-T\gamma(\cdot; \y)} .  
\end{align*}
\end{lemma}
\begin{proof}
We prove the bound for the DSM; taking $\phi \equiv 1$ provides the proof for the ISM. By Lemma~\ref{lemma: Uniform Bounds on Delta}
\begin{align}
\label{eqn: Length r Density Bound}
e^{-rT\tilde{\delta}}  \leq e^{\sum^r_{j=1} t^j(\tilde{\gamma}(\cdot; \x^j) - \tilde{\gamma}(\cdot; \x^{j-1}))} \leq e^{rT\tilde{\delta}} .
\end{align}
Recalling $\Phi(\s(\Pa^r)) :=  \prod^r_{l=1}\phi_{s_l}(y_{s^l}; \tilde{x}^{l-1}_{s^l} ) $, we have
\begin{align*}
\tP_{(T,\tQ)}(\y,\Pa^r \mid \x) &= \left[\prod^r_{j=1}\phi(y_{s^j}; \tilde{x}^{j-1}_{s^j} ) e^{\sum^r_{j=1} t^j(\tilde{\gamma}(\cdot; \x^j) - \tilde{\gamma}(\cdot; \x^{j-1}))}\right] \me^{-T\tilde{\gamma}(\cdot; \y)} \\
&=  e^{\sum^r_{j=1} t^j(\tilde{\gamma}(\cdot; \x^j) - \tilde{\gamma}(\cdot; \x^{j-1}))} \me^{-T\tilde{\gamma}(\cdot; \y)}    \Phi(\s(\Pa^r)),
\end{align*}
and applying \eqref{eqn: Length r Density Bound} yields the stated bound.
\end{proof}
We now apply Lemma~\ref{lemma: Length r Density Bound} to obtain uniform bounds on the restricted densities $\pi_{\mid r}$ and $\mu_{\mid r}$.
\begin{lemma}
\label{lemma: Restricted Target Bound} 
The following bounds hold:
\begin{align*}
\pi_{\mid r}(\Pa^r) \geq  \frac{T^{-r} e^{-2rT \tilde{\delta}} \Phi(\s(\Pa^r))  }{\frac{1}{r!}\sum_{\s^r \in \text{Sym}(\calS)}\Phi(\s^r)  } \qquad\qquad \mu_{\mid r}(\Pa^r) \leq T^{-r}e^{2rT \delta}.
\end{align*} 
\end{lemma}
\begin{proof}
Applying Lemma~\ref{lemma: Length r Density Bound} yields
\begin{align*}
\pi_{\mid r}(\Pa) = \frac{\tP_{(T,\tQ)}(\y, \Pa^r \mid \x)}{\int_{\scrP^r}\tP_{(T,\tQ)}(\y, \Pa^r \mid \x) \nu(d\Pa^r)} \geq \frac{e^{-2rT \tilde{\delta}}  \Phi(\s(\Pa^r))   }{\int_{\scrP^r}   \Phi(\s(\Pa^r)) \nu(d\Pa^r) } 
&= \frac{e^{-2rT \tilde{\delta}} \Phi(\s(\Pa^r))  }{\sum_{\s^r \in \text{Sym}(\calS)}\Phi(\s^r) \int_{\mathscr{T}^r}1 \, d\bt^r  } \\
&= \frac{T^{-r} e^{-2rT \tilde{\delta}}  \Phi(\s(\Pa^r))  }{\frac{1}{r!}\sum_{\s^r \in \text{Sym}(\calS)} \Phi(\s^r)  }.
\end{align*}
An identical argument applies for upper bounding $\mu_{\mid r}$ by letting $\phi \equiv 1$.
\end{proof}
We can relate the lower bound on $\pi_{\mid r}(\Pa^r)$ in Lemma~\ref{lemma: Restricted Target Bound} to the uniform measure on $\text{Sym}(\calS)$, denoted $U_{\calS}$ by writing
\begin{align*}
\frac{1}{r!}\sum_{\s^r \in \text{Sym}(\calS)}\Phi(\s^r) = \E_{U_{\calS}}[\Phi(\s^r)],
\end{align*}
and noting that
\begin{align*}
\text{D}_{\text{KL}}(\tPhi_{\calS} \mid \mid U_{\calS} ) =  \E_{U_{\calS}}\left[ \log\left(\frac{\Phi(\s^r)}{\E_{U_{\calS}}[\Phi(\s^r)]}\right) \frac{\Phi(\s^r)}{\E_{U_{\calS}}[\Phi(\s^r)]}   \right] .
\end{align*}
This connection to $U_{\calS}$ is established in the next lemma.
\begin{lemma}
\label{lemma: Restricted KL Lower Bound}
\begin{align*}
\text{D}_{\text{KL}}(\pi_{\mid r} \mid \mid \mu_{\mid r}) \geq e^{-2rT \tilde{\delta}} \text{D}_{\text{KL}}(\Phi_{\calS} \mid \mid U_{\calS} ) - 2rT(\tilde{\delta} + \delta).
\end{align*}
\end{lemma}
\begin{proof}
Noting $\text{D}_{\text{KL}}(\pi_{\mid r} \mid \mid \mu_{\mid r}) = \int_{\scrP^r} \log\left(\frac{\pi_{\mid r}(\Pa)}{\mu_{\mid r}(\Pa)} \right)\pi_{\mid r}(d\Pa)$ and applying Lemma~\ref{lemma: Restricted Target Bound} twice gives
\begin{align}
\int_{\scrP^r} \log\left(\frac{\pi_{\mid r}(\Pa)}{\mu_{\mid r}(\Pa)} \right)\pi_{\mid r}(d\Pa)  &\geq \int_{\scrP^r} \log\left(\frac{\Phi(\s(\Pa^r)) }{\E_{U_{\calS}}[\Phi(\s^r)]   }  \right)\pi_{\mid r}(d\Pa^r)  - 2rT(\tilde{\delta} + \delta) \nonumber \\
& \geq T^{-r} e^{-2rT \tilde{\delta}} \int_{\scrP^r} \log\left(\frac{\Phi(\s(\Pa^r)) }{\E_{U_{\calS}}[\Phi(\s^r)]}  \right)\frac{ \Phi(\s(\Pa^r))  }{\E_{U_{\calS}}[\Phi(\s^r)]  } \nu (d\Pa^r) \label{Eqn:DKLrLowerBound1} \\
&- 2rT(\tilde{\delta} + \delta) \nonumber  .
\end{align} 
Next, using $\int_{\mathcal{T}^r} 1 \, d\bt^r = T^r/r!$ we see that
\begin{align}
&\int_{\scrP^r} \log\left(\frac{\Phi(\s(\Pa^r)) }{\E_{U_{\calS}}[\Phi(\s^r)]}  \right)\frac{ \Phi(\s(\Pa^r))  }{\E_{U_{\calS}}[\Phi(\s^r)] } \nu(d\Pa^r) \\
=  &\sum_{\s^r \in \text{Sym}(\calS)} \log\left(\frac{\Phi(\s^r) }{\E_{U_{\calS}}[\Phi(\s^r)]}  \right)\frac{ \Phi(\s^r)  }{\E_{U_{\calS}}[\Phi(\s^r)]} \int_{\mathcal{T}^r} 1 \, d\bt^r \nonumber \\
= &T^r \E_{U_{\calS}}\left[\log\left(\frac{\Phi(\s^r)}{\E_{U_{\calS}}[\Phi(\s^r)]}\right) \frac{\Phi(\s^r)}{\E_{U_{\calS}}[\Phi(\s^r)]}   \right].
\label{Eqn:DKLrLowerBound2}
\end{align}
Finally, combining \eqref{Eqn:DKLrLowerBound1} and \eqref{Eqn:DKLrLowerBound2} we obtain 
\begin{align*}
\text{D}_{\text{KL}}(\pi_{\mid r} \mid \mid \mu_{\mid r}) &\geq e^{-2rT \tilde{\delta}} \E_{U_{\calS}}\left[\log\left(\frac{\Phi(\s^r)}{\E_{U_{\calS}}[\Phi(\s^r)]}\right) \frac{\Phi(\s^r)}{\E_{U_{\calS}}[\Phi(\s^r)]}   \right] - 2rT(\tilde{\delta} + \delta).
\end{align*}
\end{proof}

The proof of Lemma~\ref{lemma: Original KL and Sym KL} now follows immediately by Lemma~\ref{lemma: Restricted KL Lower Bound}.
\begin{proof}(Lemma~\ref{lemma: Original KL and Sym KL})
Since $T = \bigO(r/n)$ and $r \leq n^{\frac{1}{3}}$ we have $\bigO(rT) = 1$. Hence, by Lemmas~\ref{lemma: KL Lower Bound by Restriction Summary} and \ref{lemma: Restricted KL Lower Bound} 
\begin{align*}
\text{D}_{\text{KL}}(\pi \mid \mid \mu) = \Omega\left(\text{D}_{\text{KL}}(\pi_{\mid r} \mid \mid \mu_{\mid r})\right) = \Omega\left( \text{D}_{\text{KL}}(\Phi_{\calS} \mid \mid U_{\calS} ) \right).
\end{align*}
\end{proof}

%% file: AppendixE.tex
\section{Example Where MGF of Weights Does Not Exist}\label{sec: Example MGF}
We will establish the following proposition, which states that the MGF of $w(\Pa)$ does not exist for the Island Problem (Definition~\ref{def: island problem}) with $r_I = 1$.
\begin{proposition}\label{prop: MGF DNE}
    Consider the Island Problem with $\x^{\star} = \text{TCAT}$ and 
$\y^{\star} = \text{TTGT}$ (i.e. $r_I = 1$). Then the MGF of $w(\Pa)$ for $\Pa \sim \mu$ does not exist, i.e., $\E_\mu[e^{t w(\Pa)}]$ is infinite for every $t \in (0,\infty)$. 
\end{proposition}
\begin{remark}
    The same statement holds for $w_0(\Pa)$; we make this fact clear in the proof of Proposition~\ref{prop: MGF DNE}. 
\end{remark}
\begin{proof}
    Let $B_L$ be the set of paths for which mutations occur in the following order:
\begin{enumerate}
    \item First, site $3$ mutates from $\tA$ to $\tG$, so that $\text{TCAT} \rightarrow \text{TCGT}$.
    \item Next, site $2$ repeatedly loops $\tC \rightarrow \tT \rightarrow \tC$, so that the transitions $\text{TCGT} \rightarrow \text{TTGT} \rightarrow \text{TCGT}$ occur $L$ times.
    \item Finally, site $2$ mutates from $\tC \rightarrow \tT$, so that $\text{TCGT} \rightarrow  \text{TTGT}$ and the endpoint constraint is satisfied.
\end{enumerate}
All paths in $B_L$ consist of $2L + 2$ mutations, $L+1$ of which are CpG-type mutations, and so
\begin{align}\label{eqn: mgf example phi}
    \Phi(\s(\Pa), \mathbf{b}(\Pa)) = \lambda^{L+1} \quad \text{ for } \quad \Pa \in B_L.
\end{align}
Next, notice that the exit rate under the ISM is $\gamma(\cdot; \x') \equiv 12$ for any sequence $\x'$ since $\gamma(b; x_i) \equiv 1$. Similarly, the exit rate under the DSM is 
\begin{align*}
    \tg(\cdot ; \x') =
    \begin{cases}
        12 & \text{ if } \x' \in \{\text{TCAT}, \text{TTGT}  \} \\
        6 + 6\lambda & \text{ if } \x' = \text{TCGT},
    \end{cases}
\end{align*}
where the bottom case follows since sites 2 and 3 form a CpG pair when $\x' =  \text{TCGT}$, and therefore exit their state at rate $3\lambda$. It follows that for any $\x' \in \{\text{TCAT}, \text{TTGT}, \text{TCGT}  \}$ (the only sequences visited along a path in $B_L$):
\begin{align}\label{eqn: mgf example psi}
     \gamma(\cdot; \x') - \tg(\cdot; \x') \geq 12 - (6 + 6\lambda) = -6(\lambda - 1).
\end{align}
By \eqref{eqn: mgf example phi}, \eqref{eqn: mgf example psi}, and the definition of $w(\Pa)$,
\begin{align}\label{eqn: mgf example w bound}
    \mathbbm{1}_{B_L}(\Pa)w(\Pa) 
    &\geq \mathbbm{1}_{B_L}(\Pa) \lambda^{L+1} e^{-6T(\lambda - 1)}.
\end{align}
Since there are $2L+2$ jumps along any path in $B_L$, and $\mu$ is the JC69 model with unit rate, 
\begin{align}\label{eqn: mgf example mu}
    \int_{B_L} \mu(d\Pa) =  \frac{e^{-12T}}{p_{(T,\Q)}(\y \mid \x)}  \frac{T^{2L + 2}}{(2L + 2)!}.
\end{align}
Suppose $t \in (0,\infty)$ and define $c = \lambda e^{-6T(\lambda - 1)}$. We obtain a lower bound on the MGF of $w(\Pa)$ by restricting to paths in $B_L$ and then applying \eqref{eqn: mgf example w bound} and \eqref{eqn: mgf example mu}. Note that $w(\Pa) =  w_0(\Pa)$ for $\Pa \in B_L$ since $\tg(\cdot; \y) = \gamma(\cdot; \y)$; hence, the following statements hold for $w_0(\Pa)$ as well. 
\begin{align}
    \E_{\mu}[e^{t w(\Pa)}] = \sum^{\infty}_{m=r}\int_{\mathscr{P}^{m}} e^{t w(\Pa)} \mu(d\Pa) &\geq \sum^{\infty}_{L=0} \int_{B_L} e^{t w(\Pa)} \mu(d\Pa) \\
    &\geq \frac{e^{-12T}}{p_{(T,\Q)}(\y \mid \x)} \sum^{\infty}_{L=0} e^{t c\lambda^{L}}   \frac{T^{2L + 2}}{(2L + 2)!} \label{eqn: series}.
\end{align}
Letting $a_L := e^{t c\lambda^{L}}   \frac{T^{2L + 2}}{(2L + 2)!}$, we see that 
\begin{align*}
    \frac{a_{L+1}}{a_L} = e^{tc \lambda^{L}(\lambda - 1)} \frac{T^2}{(2L+4)(2L+3)} \rightarrow \infty \text{ as } L \rightarrow \infty.
\end{align*}
Hence, the series \eqref{eqn: series} diverges for any $t \in (0,\infty)$, in which case the MGF of $w(\Pa)$ does not exist.
\end{proof}

%% file: REVISION_CLEAN_IS_Context_Dependent.bbl
\begin{thebibliography}{52}

\bibitem[\protect\citeauthoryear{Arndt and Hwa}{2005}]{Arndt:2005}
\begin{barticle}[author]
\bauthor{\bsnm{Arndt},~\bfnm{P.~F.}\binits{P.~F.}} \AND
  \bauthor{\bsnm{Hwa},~\bfnm{T.}\binits{T.}}
(\byear{2005}).
\btitle{Identification and Measurement of Neighbour-Dependent Nucleotide
  Substitution Processes}.
\bjournal{Bioinformatics}
\bvolume{21}
\bpages{2322--2328}.
\end{barticle}
\endbibitem

\bibitem[\protect\citeauthoryear{Brooks and Gelman}{1998}]{Brooks:1998}
\begin{barticle}[author]
\bauthor{\bsnm{Brooks},~\bfnm{S.~P.}\binits{S.~P.}} \AND
  \bauthor{\bsnm{Gelman},~\bfnm{A.}\binits{A.}}
(\byear{1998}).
\btitle{General Methods for Monitoring Convergence of Iterative Simulations}.
\bjournal{Journal of Computational and Graphical Statistics}
\bvolume{7}
\bpages{434--455}.
\end{barticle}
\endbibitem

\bibitem[\protect\citeauthoryear{Challis and Schmidler}{2012}]{Challis:2012}
\begin{barticle}[author]
\bauthor{\bsnm{Challis},~\bfnm{C.~J.}\binits{C.~J.}} \AND
  \bauthor{\bsnm{Schmidler},~\bfnm{S.~C.}\binits{S.~C.}}
(\byear{2012}).
\btitle{A Stochastic Evolutionary Model for Protein Structure Alignment and
  Phylogeny}.
\bjournal{Molecular Biology and Evolution}
\bvolume{29}
\bpages{3575 - 3587}.
\end{barticle}
\endbibitem

\bibitem[\protect\citeauthoryear{Chatterjee and
  Diaconis}{2018}]{Chaterjee:2018}
\begin{barticle}[author]
\bauthor{\bsnm{Chatterjee},~\bfnm{S.}\binits{S.}} \AND
  \bauthor{\bsnm{Diaconis},~\bfnm{P.}\binits{P.}}
(\byear{2018}).
\btitle{The sample size required in importance sampling}.
\bjournal{Annals of Applied Probability}
\bvolume{28}
\bpages{1099-1135}.
\end{barticle}
\endbibitem

\bibitem[\protect\citeauthoryear{Chib}{1995}]{Chib:1995}
\begin{barticle}[author]
\bauthor{\bsnm{Chib},~\bfnm{S.}\binits{S.}}
(\byear{1995}).
\btitle{Marginal Likelihood from the {G}ibbs Output}.
\bjournal{Journal of the American Statistical Association}
\bvolume{90}
\bpages{1313--1321}.
\end{barticle}
\endbibitem

\bibitem[\protect\citeauthoryear{Christensen, Hobolth and
  Jensen}{2005}]{Christensen:2005}
\begin{barticle}[author]
\bauthor{\bsnm{Christensen},~\bfnm{O.~F.}\binits{O.~F.}},
  \bauthor{\bsnm{Hobolth},~\bfnm{A.}\binits{A.}} \AND
  \bauthor{\bsnm{Jensen},~\bfnm{J.~L.}\binits{J.~L.}}
(\byear{2005}).
\btitle{Pseudo-Likelihood Analysis of Context-Dependent Codon Substitution
  Models}.
\bjournal{Journal of Computational Biology}
\bvolume{12}
\bpages{1166--1182}.
\end{barticle}
\endbibitem

\bibitem[\protect\citeauthoryear{Cowles and Carlin}{1996}]{Cowles:1996}
\begin{barticle}[author]
\bauthor{\bsnm{Cowles},~\bfnm{M.~K.}\binits{M.~K.}} \AND
  \bauthor{\bsnm{Carlin},~\bfnm{B.~P.}\binits{B.~P.}}
(\byear{1996}).
\btitle{{M}arkov Chain {M}onte {C}arlo Convergence Diagnostics: A Review}.
\bjournal{Journal of the American Statistical Association}
\bvolume{91}
\bpages{883--904}.
\end{barticle}
\endbibitem

\bibitem[\protect\citeauthoryear{Felsenstein}{1973}]{Felenstein:1973}
\begin{barticle}[author]
\bauthor{\bsnm{Felsenstein},~\bfnm{J.}\binits{J.}}
(\byear{1973}).
\btitle{Maximum Likelihood and Minimum-Steps Methods for Estimating
  Evolutionary Trees from Data on Discrete Characters}.
\bjournal{Systematic Zoology}
\bvolume{22}
\bpages{240-249}.
\end{barticle}
\endbibitem

\bibitem[\protect\citeauthoryear{Felsenstein}{1985}]{Felenstein:1985}
\begin{barticle}[author]
\bauthor{\bsnm{Felsenstein},~\bfnm{Joseph}\binits{J.}}
(\byear{1985}).
\btitle{Phylogenies and the Comparative Method}.
\bjournal{The American Naturalist}
\bvolume{125}
\bpages{1-15}.
\end{barticle}
\endbibitem

\bibitem[\protect\citeauthoryear{Fourment et~al.}{2020}]{Fourment:2020}
\begin{barticle}[author]
\bauthor{\bsnm{Fourment},~\bfnm{M.}\binits{M.}},
  \bauthor{\bsnm{Magee},~\bfnm{A.~F.}\binits{A.~F.}},
  \bauthor{\bsnm{Whidden},~\bfnm{C.}\binits{C.}},
  \bauthor{\bsnm{Bilge},~\bfnm{A.}\binits{A.}},
  \bauthor{\bsnm{Matsen},~\bfnm{F.~A.~.}\binits{F.~A.~.}} \AND
  \bauthor{\bsnm{Minin},~\bfnm{V.~N.}\binits{V.~N.}}
(\byear{2020}).
\btitle{19 Dubious Ways to Compute the Marginal Likelihood of a Phylogenetic
  Tree Topology}.
\bjournal{Systematic Biology}
\bvolume{69}
\bpages{209--220}.
\end{barticle}
\endbibitem

\bibitem[\protect\citeauthoryear{Gelman and Rubin}{1992}]{Gelman:1992}
\begin{barticle}[author]
\bauthor{\bsnm{Gelman},~\bfnm{A.}\binits{A.}} \AND
  \bauthor{\bsnm{Rubin},~\bfnm{D.~B.}\binits{D.~B.}}
(\byear{1992}).
\btitle{Inference from Iterative Simulation Using Multiple Sequences}.
\bjournal{Statistical Science}
\bvolume{7}
\bpages{457--472}.
\end{barticle}
\endbibitem

\bibitem[\protect\citeauthoryear{Halpern and Bruno}{1998}]{Halpern:1998}
\begin{barticle}[author]
\bauthor{\bsnm{Halpern},~\bfnm{A.~L.}\binits{A.~L.}} \AND
  \bauthor{\bsnm{Bruno},~\bfnm{W.~J.}\binits{W.~J.}}
(\byear{1998}).
\btitle{Evolutionary distances for protein-coding sequences: modeling
  site-specific residue frequencies}.
\bjournal{Molecular Biology and Evolution}
\bvolume{15}
\bpages{910--917}.
\end{barticle}
\endbibitem

\bibitem[\protect\citeauthoryear{Hobolth}{2008}]{Hobolth:2008}
\begin{barticle}[author]
\bauthor{\bsnm{Hobolth},~\bfnm{A.}\binits{A.}}
(\byear{2008}).
\btitle{A {M}arkov Chain {M}onte {C}arlo Expectation Maximization Algorithm for
  Statistical Analysis of {DNA} Sequence Evolution With Neighbor-Dependent
  Substitution Rates}.
\bjournal{Journal of Computational and Graphical Statistics}
\bvolume{17}
\bpages{138--162}.
\end{barticle}
\endbibitem

\bibitem[\protect\citeauthoryear{Hobolth and Stone}{2009}]{Hobolth:2009}
\begin{barticle}[author]
\bauthor{\bsnm{Hobolth},~\bfnm{A.}\binits{A.}} \AND
  \bauthor{\bsnm{Stone},~\bfnm{E.}\binits{E.}}
(\byear{2009}).
\btitle{Simulation from endpoint-conditioned, continuous-time {M}arkov chains
  on a finite state space, with applications to molecular evolution}.
\bjournal{Annals of Applied Statistics}
\bvolume{3}
\bpages{1204--1231}.
\end{barticle}
\endbibitem

\bibitem[\protect\citeauthoryear{Hobolth and Thorne}{2014}]{HobolthThorne:2014}
\begin{bincollection}[author]
\bauthor{\bsnm{Hobolth},~\bfnm{A.}\binits{A.}} \AND
  \bauthor{\bsnm{Thorne},~\bfnm{J.}\binits{J.}}
(\byear{2014}).
\btitle{Sampling and summary statistics of endpoint-conditioned paths in {DNA}
  sequence evolution}.
In \bbooktitle{Bayesian Phylogenetics: Methods Algorithms, and Applications}
(\beditor{\bfnm{M.~H.}\binits{M.~H.}~\bsnm{Chen}},
  \beditor{\bfnm{L.}\binits{L.}~\bsnm{Kuo}} \AND
  \beditor{\bfnm{P.}\binits{P.}~\bsnm{Lewis}}, eds.)
\bpages{247--273}.
\bpublisher{Chapman and Hall}.
\end{bincollection}
\endbibitem

\bibitem[\protect\citeauthoryear{Hwang and Green}{2004}]{Hwang:2004}
\begin{barticle}[author]
\bauthor{\bsnm{Hwang},~\bfnm{DG}\binits{D.}} \AND
  \bauthor{\bsnm{Green},~\bfnm{P}\binits{P.}}
(\byear{2004}).
\btitle{Bayesian {M}arkov chain {M}onte {C}arlo sequence analysis reveals
  varying neutral substitution patterns in mammalian evolution}.
\bjournal{Proceedings of the National Academy of Science}
\bvolume{101}
\bpages{13994-14001}.
\end{barticle}
\endbibitem

\bibitem[\protect\citeauthoryear{Jensen and Pedersen}{2000}]{Jensen:2000}
\begin{barticle}[author]
\bauthor{\bsnm{Jensen},~\bfnm{J.}\binits{J.}} \AND
  \bauthor{\bsnm{Pedersen},~\bfnm{A-MK}\binits{A.-M.}}
(\byear{2000}).
\btitle{Probabilistic Models of {DNA} Sequence Evolution with Context Dependent
  Rates of Substitution}.
\bjournal{Advances in Applied Probability}
\bvolume{32}
\bpages{499--517}.
\end{barticle}
\endbibitem

\bibitem[\protect\citeauthoryear{Jerrum, Valiant and
  Vazirani}{1986}]{Jerrum:1986}
\begin{barticle}[author]
\bauthor{\bsnm{Jerrum},~\bfnm{M.~R.}\binits{M.~R.}},
  \bauthor{\bsnm{Valiant},~\bfnm{L.~G.}\binits{L.~G.}} \AND
  \bauthor{\bsnm{Vazirani},~\bfnm{V.~V.}\binits{V.~V.}}
(\byear{1986}).
\btitle{Random Generation of Combinatorial Structures from a Uniform
  Distribution}.
\bjournal{Theoretical Computer Science}
\bvolume{43}
\bpages{169--188}.
\end{barticle}
\endbibitem

\bibitem[\protect\citeauthoryear{Jones and Hobert}{2001}]{Jones:2001}
\begin{barticle}[author]
\bauthor{\bsnm{Jones},~\bfnm{G.~L.}\binits{G.~L.}} \AND
  \bauthor{\bsnm{Hobert},~\bfnm{J.~P.}\binits{J.~P.}}
(\byear{2001}).
\btitle{Honest Exploration of Intractable Probability Distributions via
  {M}arkov Chain {M}onte {C}arlo}.
\bjournal{Statistical Science}
\bvolume{16}
\bpages{312--334}.
\end{barticle}
\endbibitem

\bibitem[\protect\citeauthoryear{Jukes and Cantor}{1969}]{Jukes:1969}
\begin{bincollection}[author]
\bauthor{\bsnm{Jukes},~\bfnm{T.~H.}\binits{T.~H.}} \AND
  \bauthor{\bsnm{Cantor},~\bfnm{C.~R.}\binits{C.~R.}}
(\byear{1969}).
\btitle{Evolution of protein molecules}.
In \bbooktitle{Mammalian Protein Metabolism}
(\beditor{\bfnm{H.~N.}\binits{H.~N.}~\bsnm{Munro}}, ed.)
\bpages{121--132}.
\bpublisher{Academic Press}, \baddress{New York}.
\end{bincollection}
\endbibitem

\bibitem[\protect\citeauthoryear{Kishino, Thorne and
  Bruno}{2001}]{Kishino:2001}
\begin{barticle}[author]
\bauthor{\bsnm{Kishino},~\bfnm{H.}\binits{H.}},
  \bauthor{\bsnm{Thorne},~\bfnm{J.~L.}\binits{J.~L.}} \AND
  \bauthor{\bsnm{Bruno},~\bfnm{W.~J.}\binits{W.~J.}}
(\byear{2001}).
\btitle{Performance of a Divergence Time Estimation Method under a
  Probabilistic Model of Rate Evolution}.
\bjournal{Molecular Biology and Evolution}
\bvolume{18}
\bpages{352-361}.
\end{barticle}
\endbibitem

\bibitem[\protect\citeauthoryear{Larson, Thorne and
  Schmidler}{2020}]{Larson:2020}
\begin{barticle}[author]
\bauthor{\bsnm{Larson},~\bfnm{Gary}\binits{G.}},
  \bauthor{\bsnm{Thorne},~\bfnm{Jeffrey~L.}\binits{J.~L.}} \AND
  \bauthor{\bsnm{Schmidler},~\bfnm{Scott~C.}\binits{S.~C.}}
(\byear{2020}).
\btitle{Incorporating Nearest-Neighbor Site Dependence into Protein Evolution
  Models}.
\bjournal{Journal of Computational Biology}
\bvolume{27}
\bpages{361-375}.
\end{barticle}
\endbibitem

\bibitem[\protect\citeauthoryear{Li, Mathews and Schmidler}{2025}]{Li:2024}
\begin{barticle}[author]
\bauthor{\bsnm{Li},~\bfnm{Y.}\binits{Y.}},
  \bauthor{\bsnm{Mathews},~\bfnm{J.}\binits{J.}} \AND
  \bauthor{\bsnm{Schmidler},~\bfnm{Scott~C.}\binits{S.~C.}}
(\byear{2025}).
\btitle{On Gibbs Sampling for Endpoint-Conditioned Neighbor-Dependent Sequence
  Evolution Models}.
\bjournal{Journal of Graphical and Computational Statistics}.
\bnote{(provisionally accepted)}.
\end{barticle}
\endbibitem

\bibitem[\protect\citeauthoryear{Li, Wiehe and Schmidler}{2025}]{Li:2025}
\begin{barticle}[author]
\bauthor{\bsnm{Li},~\bfnm{Yongkang}\binits{Y.}},
  \bauthor{\bsnm{Wiehe},~\bfnm{Kevin}\binits{K.}} \AND
  \bauthor{\bsnm{Schmidler},~\bfnm{Scott~C.}\binits{S.~C.}}
(\byear{2025}).
\btitle{Reconstructing B Cell Lineages in the Presence of Context-Dependent
  Somatic Hypermutation}.
\bjournal{(submitted)}.
\end{barticle}
\endbibitem

\bibitem[\protect\citeauthoryear{Lunter and Hein}{2004}]{Lunter:2004}
\begin{barticle}[author]
\bauthor{\bsnm{Lunter},~\bfnm{G.}\binits{G.}} \AND
  \bauthor{\bsnm{Hein},~\bfnm{J.}\binits{J.}}
(\byear{2004}).
\btitle{A nucleotide substitution model with nearest-neighbour interactions}.
\bjournal{Bioinformatics}
\bvolume{20 Suppl 1}
\bpages{i216--i223}.
\end{barticle}
\endbibitem

\bibitem[\protect\citeauthoryear{Mathews and Schmidler}{2025a}]{Mathews:2025}
\begin{barticle}[author]
\bauthor{\bsnm{Mathews},~\bfnm{J.}\binits{J.}} \AND
  \bauthor{\bsnm{Schmidler},~\bfnm{S.~C.}\binits{S.~C.}}
(\byear{2025}a).
\btitle{Approximating Marginal Likelihoods in Evolutionary Models Under
  Site-Dependence}.
\bnote{Unpublished Manuscript}.
\end{barticle}
\endbibitem

\bibitem[\protect\citeauthoryear{Mathews and Schmidler}{2025b}]{Mathews2:2025}
\begin{barticle}[author]
\bauthor{\bsnm{Mathews},~\bfnm{J.}\binits{J.}} \AND
  \bauthor{\bsnm{Schmidler},~\bfnm{S.~C.}\binits{S.~C.}}
(\byear{2025}b).
\btitle{Posterior bounds on divergence time of two sequences under
  dependent-site evolutionary models}.
\bjournal{arXiv preprint arXiv:2507.19659}.
\end{barticle}
\endbibitem

\bibitem[\protect\citeauthoryear{Mathews et~al.}{2023}]{Mathews:2023b}
\begin{barticle}[author]
\bauthor{\bsnm{Mathews},~\bfnm{Joseph}\binits{J.}},
  \bauthor{\bsnm{Itallie},~\bfnm{Elizabeth~Van}\binits{E.~V.}},
  \bauthor{\bsnm{Li},~\bfnm{Yongkang}\binits{Y.}},
  \bauthor{\bsnm{Wiehe},~\bfnm{Kevin}\binits{K.}} \AND
  \bauthor{\bsnm{Schmidler},~\bfnm{Scott~C.}\binits{S.~C.}}
(\byear{2023}).
\btitle{{Computing the Inducibility of {B} Cell Lineages Under a
  Context-Dependent Model of Affinity Maturation: {A}pplications to Sequential
  Vaccine Design}}.
\bjournal{\textit{The Journal of Immunology}}.
\bnote{(in press)}.
\end{barticle}
\endbibitem

\bibitem[\protect\citeauthoryear{Mossel}{2003}]{Mossel:2003}
\begin{barticle}[author]
\bauthor{\bsnm{Mossel},~\bfnm{E.}\binits{E.}}
(\byear{2003}).
\btitle{On the impossibility of reconstructing ancestral data and phylogenies}.
\bjournal{Journal of Computational Biology}
\bvolume{10}
\bpages{669 -- 676}.
\end{barticle}
\endbibitem

\bibitem[\protect\citeauthoryear{Mossel}{2004}]{Mossel:2004}
\begin{barticle}[author]
\bauthor{\bsnm{Mossel},~\bfnm{E.}\binits{E.}}
(\byear{2004}).
\btitle{Phase transitions in phylogeny}.
\bjournal{Transactions of the American Mathematical Society}
\bvolume{356}
\bpages{2379 -- 2404}.
\end{barticle}
\endbibitem

\bibitem[\protect\citeauthoryear{Pagel, Meade and Barker}{2004}]{Pagel:2004}
\begin{barticle}[author]
\bauthor{\bsnm{Pagel},~\bfnm{M.}\binits{M.}},
  \bauthor{\bsnm{Meade},~\bfnm{A.}\binits{A.}} \AND
  \bauthor{\bsnm{Barker},~\bfnm{D.}\binits{D.}}
(\byear{2004}).
\btitle{Bayesian Estimation of Ancestral Character States on Phylogenies}.
\bjournal{Systematic Biology}
\bvolume{53}
\bpages{673--684}.
\end{barticle}
\endbibitem

\bibitem[\protect\citeauthoryear{Pedersen, Wiuf and
  Christiansen}{1998}]{Pederson:1998}
\begin{barticle}[author]
\bauthor{\bsnm{Pedersen},~\bfnm{A.~K.}\binits{A.~K.}},
  \bauthor{\bsnm{Wiuf},~\bfnm{C.}\binits{C.}} \AND
  \bauthor{\bsnm{Christiansen},~\bfnm{F.~B.}\binits{F.~B.}}
(\byear{1998}).
\btitle{A codon-based model designed to describe lentiviral evolution}.
\bjournal{Molecular Biology and Evolution}
\bvolume{15}
\bpages{1069-1081}.
\end{barticle}
\endbibitem

\bibitem[\protect\citeauthoryear{Pederson and Jensen}{2001}]{Jensen:2001}
\begin{barticle}[author]
\bauthor{\bsnm{Pederson},~\bfnm{A-MK}\binits{A.-M.}} \AND
  \bauthor{\bsnm{Jensen},~\bfnm{J.}\binits{J.}}
(\byear{2001}).
\btitle{A dependent rates model and {MCMC} based methodology for the maximum
  likelihood analysis of sequences with overlapping reading frames}.
\bjournal{Molecular Biology and Evolution}
\bvolume{18}
\bpages{763--776}.
\end{barticle}
\endbibitem

\bibitem[\protect\citeauthoryear{Robinson et~al.}{2003}]{Robinson:2003}
\begin{barticle}[author]
\bauthor{\bsnm{Robinson},~\bfnm{D.}\binits{D.}},
  \bauthor{\bsnm{Jones},~\bfnm{D.}\binits{D.}},
  \bauthor{\bsnm{Kishino},~\bfnm{H.}\binits{H.}},
  \bauthor{\bsnm{Goldman},~\bfnm{N.}\binits{N.}} \AND
  \bauthor{\bsnm{Thorne},~\bfnm{J.}\binits{J.}}
(\byear{2003}).
\btitle{Protein Evolution with Dependence Among Codons Due to Tertiary
  Structure}.
\bjournal{Molecular Biology and Evolution}
\bvolume{20}
\bpages{1692--1704}.
\end{barticle}
\endbibitem

\bibitem[\protect\citeauthoryear{Rodrigue, Philippe and
  Lartillot}{2006}]{Rodrigue:2006}
\begin{barticle}[author]
\bauthor{\bsnm{Rodrigue},~\bfnm{N.}\binits{N.}},
  \bauthor{\bsnm{Philippe},~\bfnm{H.}\binits{H.}} \AND
  \bauthor{\bsnm{Lartillot},~\bfnm{N.}\binits{N.}}
(\byear{2006}).
\btitle{Assessing site-interdependent phylogenetic models of sequence
  evolution}.
\bjournal{Molecular Biology and Evolution}
\bvolume{23}
\bpages{1762-1775}.
\end{barticle}
\endbibitem

\bibitem[\protect\citeauthoryear{Rodrigue et~al.}{2005}]{Rodrigue:2005}
\begin{barticle}[author]
\bauthor{\bsnm{Rodrigue},~\bfnm{N.}\binits{N.}},
  \bauthor{\bsnm{Lartillot},~\bfnm{N.}\binits{N.}},
  \bauthor{\bsnm{Bryant},~\bfnm{D.}\binits{D.}} \AND
  \bauthor{\bsnm{Philippe},~\bfnm{H.}\binits{H.}}
(\byear{2005}).
\btitle{Site interdependence attributed to tertiary structure in amino acid
  sequence evolution}.
\bjournal{Gene}
\bvolume{347}
\bpages{207-217}.
\end{barticle}
\endbibitem

\bibitem[\protect\citeauthoryear{Rodríguez et~al.}{1990}]{Rodriguez:1990}
\begin{barticle}[author]
\bauthor{\bsnm{Rodríguez},~\bfnm{F.}\binits{F.}},
  \bauthor{\bsnm{Oliver},~\bfnm{J.~L.}\binits{J.~L.}},
  \bauthor{\bsnm{Marín},~\bfnm{A.}\binits{A.}} \AND
  \bauthor{\bsnm{Medina},~\bfnm{J.~R.}\binits{J.~R.}}
(\byear{1990}).
\btitle{The general stochastic model of nucleotide substitution}.
\bjournal{Journal of Theoretical Biology}
\bvolume{142}
\bpages{485--501}.
\end{barticle}
\endbibitem

\bibitem[\protect\citeauthoryear{Ronquist et~al.}{2012}]{MrBayes:2012}
\begin{barticle}[author]
\bauthor{\bsnm{Ronquist},~\bfnm{F.}\binits{F.}},
  \bauthor{\bsnm{Teslenko},~\bfnm{M.}\binits{M.}}, \bauthor{\bsnm{{van der
  Mark}},~\bfnm{P.}\binits{P.}},
  \bauthor{\bsnm{Ayres},~\bfnm{D.~L.}\binits{D.~L.}},
  \bauthor{\bsnm{Darling},~\bfnm{A.}\binits{A.}},
  \bauthor{\bsnm{H\"ohna},~\bfnm{S.}\binits{S.}},
  \bauthor{\bsnm{Larget},~\bfnm{B.}\binits{B.}},
  \bauthor{\bsnm{Liu},~\bfnm{L.}\binits{L.}},
  \bauthor{\bsnm{Suchard},~\bfnm{M.~A.}\binits{M.~A.}} \AND
  \bauthor{\bsnm{Huelsenbeck},~\bfnm{J.~P.}\binits{J.~P.}}
(\byear{2012}).
\btitle{{MrBayes} 3.2: Efficient {Bayesian} Phylogenetic Inference and Model
  Choice Across a Large Model Space}.
\bjournal{Systematic Biology}
\bvolume{61}
\bpages{539--542}.
\end{barticle}
\endbibitem

\bibitem[\protect\citeauthoryear{Rosenthal}{1995}]{Rosenthal:1995}
\begin{barticle}[author]
\bauthor{\bsnm{Rosenthal},~\bfnm{J.~S.}\binits{J.~S.}}
(\byear{1995}).
\btitle{Minorization Conditions and Convergence Rates for {M}arkov Chain
  {M}onte {C}arlo}.
\bjournal{Journal of the American Statistical Association}
\bvolume{90}
\bpages{558--566}.
\end{barticle}
\endbibitem

\bibitem[\protect\citeauthoryear{Sanderson}{1997}]{Sanderson:1997}
\begin{barticle}[author]
\bauthor{\bsnm{Sanderson},~\bfnm{MJ.}\binits{M.}}
(\byear{1997}).
\btitle{A Nonparametric Approach to Estimating Divergence Times in the Absence
  of Rate Constancy}.
\bjournal{Molecular Biology and Evolution}
\bvolume{14}
\bpages{1218}.
\end{barticle}
\endbibitem

\bibitem[\protect\citeauthoryear{Siepel and Haussler}{2004}]{Siepel:2004}
\begin{barticle}[author]
\bauthor{\bsnm{Siepel},~\bfnm{A.}\binits{A.}} \AND
  \bauthor{\bsnm{Haussler},~\bfnm{D.}\binits{D.}}
(\byear{2004}).
\btitle{Phylogenetic Estimation of Context-Dependent Substitution Rates by
  Maximum Likelihood}.
\bjournal{Molecular Biology and Evolution}
\bvolume{21}
\bpages{468--488}.
\end{barticle}
\endbibitem

\bibitem[\protect\citeauthoryear{Tavaré}{1986}]{Tavare:1986}
\begin{barticle}[author]
\bauthor{\bsnm{Tavaré},~\bfnm{Simon}\binits{S.}}
(\byear{1986}).
\btitle{Some Probabilistic and Statistical Problems in the Analysis of {DNA}
  Sequences}.
\bjournal{Lectures on Mathematics in the Life Sciences}
\bvolume{17}
\bpages{57--86}.
\end{barticle}
\endbibitem

\bibitem[\protect\citeauthoryear{Thorne, Kishino and
  Painter.}{1998}]{Thorne:1998}
\begin{barticle}[author]
\bauthor{\bsnm{Thorne},~\bfnm{J.~L.}\binits{J.~L.}},
  \bauthor{\bsnm{Kishino},~\bfnm{H.}\binits{H.}} \AND
  \bauthor{\bsnm{Painter.},~\bfnm{I.~S.}\binits{I.~S.}}
(\byear{1998}).
\btitle{Estimating the rate of evolution of the rate of molecular evolution}.
\bjournal{Molecular Biology and Evolution}
\bvolume{15}
\bpages{1647-1657}.
\end{barticle}
\endbibitem

\bibitem[\protect\citeauthoryear{VanDerwerken and
  Schmidler}{2013}]{Vandewerken:2013}
\begin{barticle}[author]
\bauthor{\bsnm{VanDerwerken},~\bfnm{D.}\binits{D.}} \AND
  \bauthor{\bsnm{Schmidler},~\bfnm{S.~C.}\binits{S.~C.}}
(\byear{2013}).
\btitle{Parallel {M}arkov Chain {M}onte {C}arlo}.
\bjournal{arXiv preprint}.
\end{barticle}
\endbibitem

\bibitem[\protect\citeauthoryear{VanDerwerken and
  Schmidler}{2017}]{Vandewerken:2017}
\begin{barticle}[author]
\bauthor{\bsnm{VanDerwerken},~\bfnm{D.}\binits{D.}} \AND
  \bauthor{\bsnm{Schmidler},~\bfnm{S.~C.}\binits{S.~C.}}
(\byear{2017}).
\btitle{Monitoring Joint Convergence of {MCMC} Samplers}.
\bjournal{Journal of Computational and Graphical Statistics}
\bvolume{26}
\bpages{558--568}.
\end{barticle}
\endbibitem

\bibitem[\protect\citeauthoryear{von Haeseler and
  Schöniger}{1998}]{VonHaeseler:1998}
\begin{barticle}[author]
\bauthor{\bparticle{von} \bsnm{Haeseler},~\bfnm{A.}\binits{A.}} \AND
  \bauthor{\bsnm{Schöniger},~\bfnm{M.}\binits{M.}}
(\byear{1998}).
\btitle{Evolution of {DNA} or amino acid sequences with dependent sites}.
\bjournal{Journal of Computational Biology}
\bvolume{5}
\bpages{149-163}.
\end{barticle}
\endbibitem

\bibitem[\protect\citeauthoryear{Wiehe et~al.}{2018}]{Wiehe:2018}
\begin{barticle}[author]
\bauthor{\bsnm{Wiehe},~\bfnm{K.}\binits{K.}},
  \bauthor{\bsnm{Bradley},~\bfnm{T.}\binits{T.}},
  \bauthor{\bsnm{Meyerhoff},~\bfnm{RR.}\binits{R.}},
  \bauthor{\bsnm{Hart},~\bfnm{C.}\binits{C.}},
  \bauthor{\bsnm{Williams},~\bfnm{WB.}\binits{W.}},
  \bauthor{\bsnm{Easterhoff},~\bfnm{D.}\binits{D.}},
  \bauthor{\bsnm{Faison},~\bfnm{WJ.}\binits{W.}},
  \bauthor{\bsnm{Kepler},~\bfnm{TB.}\binits{T.}},
  \bauthor{\bsnm{Saunders},~\bfnm{KO.}\binits{K.}},
  \bauthor{\bsnm{Alam},~\bfnm{SM.}\binits{S.}},
  \bauthor{\bsnm{Bonsignori},~\bfnm{M.}\binits{M.}} \AND
  \bauthor{\bsnm{Haynes},~\bfnm{BF.}\binits{B.}}
(\byear{2018}).
\btitle{Functional Relevance of Improbable Antibody Mutations for {HIV} Broadly
  Neutralizing Antibody Development}.
\bjournal{Cell Host Microbe}
\bvolume{23}
\bpages{759--765}.
\end{barticle}
\endbibitem

\bibitem[\protect\citeauthoryear{Wolpert and Schmidler}{2012}]{Wolpert:2012}
\begin{barticle}[author]
\bauthor{\bsnm{Wolpert},~\bfnm{R.~L.}\binits{R.~L.}} \AND
  \bauthor{\bsnm{Schmidler},~\bfnm{S.~C.}\binits{S.~C.}}
(\byear{2012}).
\btitle{$\alpha$-Stable Limit Laws for Harmonic Mean Estimators of Marginal
  Likelihoods}.
\bjournal{Statistica Sinica}
\bvolume{22}
\bpages{1233-1251}.
\end{barticle}
\endbibitem

\bibitem[\protect\citeauthoryear{Yang}{1994}]{Yang:1994}
\begin{barticle}[author]
\bauthor{\bsnm{Yang},~\bfnm{Z.}\binits{Z.}}
(\byear{1994}).
\btitle{Maximum likelihood phylogenetic estimation from DNA sequences with
  variable rates over sites: Approximate methods}.
\bjournal{Journal of Molecular Evolution}
\bvolume{39}
\bpages{306--314}.
\end{barticle}
\endbibitem

\bibitem[\protect\citeauthoryear{Yang}{2014}]{Yang:2014}
\begin{bbook}[author]
\bauthor{\bsnm{Yang},~\bfnm{Ziheng}\binits{Z.}}
(\byear{2014}).
\btitle{Molecular Evolution: A Statistical Approach}.
\bpublisher{Oxford University Press}.
\end{bbook}
\endbibitem

\bibitem[\protect\citeauthoryear{Yang, Kumar and Nei}{1995}]{Yang:1995}
\begin{barticle}[author]
\bauthor{\bsnm{Yang},~\bfnm{Z.}\binits{Z.}},
  \bauthor{\bsnm{Kumar},~\bfnm{S.}\binits{S.}} \AND
  \bauthor{\bsnm{Nei},~\bfnm{M.}\binits{M.}}
(\byear{1995}).
\btitle{A new method of inference of ancestral nucleotide and amino acid
  sequences}.
\bjournal{Genetics}
\bvolume{141}
\bpages{1641-1650}.
\end{barticle}
\endbibitem

\bibitem[\protect\citeauthoryear{Yang and Nielsen}{2008}]{Yang:2008}
\begin{barticle}[author]
\bauthor{\bsnm{Yang},~\bfnm{Z.}\binits{Z.}} \AND
  \bauthor{\bsnm{Nielsen},~\bfnm{R.}\binits{R.}}
(\byear{2008}).
\btitle{Mutation-selection models of codon substitution and their use to
  estimate selective strengths on codon usage}.
\bjournal{Molecular Biology and Evolution}
\bvolume{25}
\bpages{568--579}.
\end{barticle}
\endbibitem

\end{thebibliography}
